\documentclass[12pt,aps,prd,superscriptaddress,showpacs,longbibliography,floatfix,nofootinbib]{revtex4-1}

\usepackage[utf8]{inputenc}
\usepackage[OT2,T1]{fontenc}
\usepackage[russian,english]{babel}
\usepackage{slashed}

\pdfoutput=1

\usepackage{color}
\usepackage{graphicx}   
\usepackage{bm}
\usepackage{amsmath}
\usepackage{amsfonts}
\usepackage{hyperref}

\allowdisplaybreaks[4]

\def\be{\begin{equation}}
\def\ee{\end{equation}}
\def\ba{\begin{eqnarray}}
\def\ea{\end{eqnarray}}

\begin{document}

\title{Vertex-resummed kinetic theory of dipole-symmetric bosons from 2PI effective action }

\author{Chun-Wei Su}
\affiliation{Department of Physics, University of Colorado, Boulder, CO 80309, USA}

\date{\today}

\begin{abstract}
The formalism of non-equilibrium dynamics for the dipole-symmetric $U(N)$ model involving $N$-component charged scalar fields is studied on the Schwinger-Keldysh contour by a systematic $1/N$ expansion of two-particle-irreducible (2PI) effective action to the next-to-leading order (NLO). The differential equations describing non-equilibrium evolution equations of two-point functions and field expectation values related to the symmetry breaking in the dynamics are derived, which encompass scatterings between modes essential in the late-time evolution and thermalization beyond the scope of the mean field. At the late-time regime of the evolution, the leading-order (LO) gradient expansion is performed to construct the vertex-resummed kinetic theory by encoding non-perturbative effects in the vertex through the resummation, which is distinct from the standard kinetic theory written by the weak-coupling expansion.
\end{abstract}

\maketitle

\section{Introduction}

The exotic dipole symmetry in the quantum system has provided a natural way to realize the fracton phase \cite{Haah:2011drr,Nandkishore:2018sel,Pretko:2020cko}, which features immobile quasi-particles or excited bound states with limited mobility moving along a confined trajectory on a line/plane \cite{Vijay:2016phm,Pretko:2016kxt,Ma:2017aog}. The theory can be formulated involving interactions to higher-rank tensor gauge fields \cite{Pretko:2016kxt,Pretko:2016lgv,Pretko:2017fbf,Pretko:2017xar} and exhibits a duality to the elasticity theory \cite{Gromov:2017vir,Pretko_2018}.

From the field-theoretic perspective, these constraints can be made explicit by considering charged scalar fields transformed in the dipole symmetry, as developed in Ref.~\cite{Pretko:2018jbi,Pretko:2020cko}, in the continuum quantum field theory. Charged scalar fields in excitations are localized since the kinetic energy is forbidden by the dipole symmetry. A closely related consequence is that the spontaneous breaking of the dipole symmetry can lead to a large degeneracy linked to the regularization, e.g. cut-off of the system related to the number of sites on lattices, which becomes infinite in the continuum limit. This sensitivity on the upper bound of momenta is known as the UV-IR mixing \cite{Seiberg:2020bhn,Seiberg:2020wsg,Gorantla:2021bda}. In addition, in spite of the absence of kinetic term, charged scalar fields can still interact with each other by non-trivial quartic couplings up to four spatial derivatives enhanced with higher multiple symmetry. The field theory description and the kinetic theory become challenging since these spatial derivatives lead to a strongly-coupled system.

A controlled alternative is provided by the $1/N$ expansion for $N$-component charged scalar fields in the large $N$ limit. The field theory and the kinetic theory based on the $1/N$ expansion requires a resummation technique that sums an infinite number of diagrams at a specific order in powers of $1/N$. Leading-order (LO) corrections are considered at the mean-field level in the absence of collisions (modes do not interact with each other), where the kinetic theory has no collision kernel. The field theory description in terms of the $1/N$ expansion at the LO has been discussed in equilibrium \cite{Jensen:2022ibn} and non-equilibrium \cite{Islam:2023cmm} to study the phase diagram, while the investigation beyond the LO crucial in the evolution and the kinetic description at the late time has been still lacking so far. To address this issue, this study aims to extend the non-equilibrium dynamics to the next-to-leading-order (NLO) by the two-particle-irreducible (2PI) $1/N$ expansion. At NLO, non-local self-energies generate memory and collisional damping, providing controlled access to the collision process and the late-time relaxation.

The utility of the 2PI $1/N$ expansion has well been established in studies of non-equilibrium $N$-component scalar field theories by including scattering effects emerging at NLO \cite{Berges:2001fi,Aarts:2002dj}. This framework has been proved to be a legitimate access to several phenomena such as parametric resonance instability \cite{Berges:2002cz} and quasi-stationary states \cite{Berges:2010ez,Berges:2012us,Berges:2015ixa,PineiroOrioli:2015cpb,Walz:2017ffj,Berges:2017ldx,Mikheev:2018adp}. The scattering plays a crucial role in inducing interactions between modes relevant to thermalization at late time \cite{Bettencourt:1997nf,Mihaila:2000ib,Berges:2000ur}, which is essential in the early universe \cite{Khlebnikov:1996mc,Kofman:1997yn}. NLO corrections from the 2PI $1/N$ expansion extend the applicability of the non-equilibrium dynamics to a longer time scale \cite{Berges:2001fi,Aarts:2002dj} when fluctuations are believed to become sizable in the $O(N)$ model, where the evolution is started from a non-perturbatively large background field together with small fluctuations even if the standard quartic coupling is weak \cite{Berges:2002cz}. Furthermore, the 2PI $1/N$ expansion is applied in the case beyond the weak coupling in Ref.~\cite{Berges:2016nru}.

The scattering effects encoded in non-local self-energies by the $1/N$ expansion lead to a momentum-dependent coupling containing a series of loop diagrams in the vertex. Consequently, the kinetic theory describing the late-time behavior is modified by the 2PI $1/N$ expansion and becomes the vertex-resummed kinetic theory, which governs the non-perturbative quantum dynamics at the late time in several systems \cite{Berges:2010ez,Berges:2012us,Berges:2015ixa,PineiroOrioli:2015cpb,Walz:2017ffj,Berges:2017ldx}.

In this study, a simple model of $N$-component complex scalar fields with global $U(N)$ and dipole symmetry is considered. The system in $d$ space-time dimensions has spatial homogeneity, invariance of spatial rotation, and parity. Following the method employed in Ref.~\cite{Aarts:2002dj}, a systematic $1/N$ expansion is carried out to the NLO using the 2PI effective action. Non-equilibrium evolution equations for two-point functions, called the Kadanoff-Baym equations \cite{kadanoff1962quantum,PhysRevD.37.2878}, and for field expectation values $\phi_\textbf{i}^*(x^0)$ are derived and expressed in two different forms. In contrast to the large-$N$ mean-field treatment in Ref.~\cite{Jensen:2022ibn,Islam:2023cmm}, the time evolution at the NLO generates non-local self-energies and hence a genuine collision kernel in the kinetic description. Accordingly, vertex-resummed kinetic theory is derived by the LO gradient expansion of evolution equations at the late time, which encompasses three-vertex and four-vertex collision kernels identified in the phase with unbroken/broken global $U(N)$ symmetry.

The paper is organized as follows. In Sec.~\ref{sec:auxi}, the auxiliary field method is used to obtain mean-field results. The action is written in the form that contains NLO corrections by considering fluctuations of auxiliary fields. In Sec.~\ref{sec:effe}, self-energies are obtained by 2PI $1/N$ expansion of the effective action to the NLO. The evolution equations of two-point functions and the field expectation values of complex scalar fields are constructed in two forms involving scattering effects. In Sec.~\ref{sec:vere}, taking the limit at the late time, the LO gradient expansion reduce Kadanoff-Baym equations to the vertex-resummed kinetic theory, where the vertex function is calculated by the resummation. Some detailed expressions of functions in self-energies are shown in Appendix.~\ref{sec:fide}.

In this paper, any repeated indices denote summation, except those expressed with an underline. Specifically, repeated Latin indices shown in italics (bold and non-italics, e.g. $\textbf{i}$) sum over integers from $1$ to $2N$ ($1$ to $N$), while repeated primed indices in italics indicate $v^\prime\equiv2N+v$ and sum over only $v=1,2,3$. In addition, any repeated Greek indices in italics run through integers from 0 to $d-1$, where $d$ denotes the space-time dimension; repeated bold Greek indices, e.g. $\bm{\alpha}$, sum over integers from 1 to $2N+3$.

\section{Auxiliary-field technique} \label{sec:auxi}
The model considered in this study has a Lagrangian with a minimal coupling consisting of $N$-component complex scalar fields $\Phi_{\textbf{i}}$ with a global $U(N)$ and a dipole symmetry
\begin{align}
\mathcal{L}=&\,\partial_{x^0}\Phi_{\textbf{i}}^*\partial_{x^0}\Phi_{\textbf{i}}-m^2\Phi_{\textbf{i}}^*\Phi_{\textbf{i}}-\frac{\bar{\lambda}}{N}\bigg\{\Phi_{\textbf{i}}^*\Phi_{\textbf{j}}^*\bigg[\frac{1}{2}\Big(\Phi_{\textbf{i}}\partial_{\textbf{x}}^2\Phi_{\textbf{j}}+\Phi_{\textbf{j}}\partial_{\textbf{x}}^2\Phi_{\textbf{i}}\Big)-\partial_{\textbf{x}}\Phi_{\textbf{i}}\cdot\partial_{\textbf{x}}\Phi_{\textbf{j}} \bigg]+{\rm H.c.}\bigg\}
\label{eq:lagr}
\end{align}
where $x^0, \textbf{x}$ denote time and space variables, respectively, in the space-time $d$ dimensions; H.c. stands for the hermitian conjugate part. Both the charge $Q$ and the dipole $\textbf{D}$
\begin{align}
Q=&\,\int d^{d-1}\textbf{x} \,\rho(\textbf{x})\,,\quad\rho(\textbf{x})=-i\Big(\partial_{x^0}\Phi_\textbf{i}^*(x^0,\textbf{x})\Phi_\textbf{i}(x^0,\textbf{x})-\Phi_\textbf{i}^*(x^0,\textbf{x})\partial_{x^0}\Phi_\textbf{i}(x^0,\textbf{x})\Big)\,,\\
\textbf{D}=&\,\int d^{d-1}\textbf{x}\, \textbf{x}\rho(\textbf{x})
\end{align}
are conserved under the $U(1)$ transformations $\Phi_\textbf{i}(x^0,\textbf{x})\rightarrow e^{iq}\Phi_\textbf{i}(x^0,\textbf{x})$ and the dipole transformation $\Phi_\textbf{i}(x^0,\textbf{x})\rightarrow e^{i\bm{\ell}\cdot \textbf{x}}\Phi_\textbf{i}(x^0,\textbf{x})$, respectively. 

It is suggested to write down the action $S[\Phi_{\textbf{i}},\Phi_{\textbf{i}}^*]$ by the Lagrangian in Eq.~(\ref{eq:lagr}) in the momentum space
\begin{align}
iS[\Phi_{\textbf{i}},\Phi_{\textbf{i}}^*]
=&\,i\int_{x^0,\mathcal{C}} \int_\textbf{k}\mathcal{L}\nonumber\\
=&\,i\int_{x^0,\mathcal{C}} \int_\textbf{k} \Phi^*(x^0,\textbf{k})\big(-\partial_{x^0}^2-m^2\big)\Phi(x^0,\textbf{k})+i\int_{x^0,\mathcal{C} } \prod_{\ell=1}^{4}\int_{\textbf{k}_\ell}\Phi_{\textbf{i}}^*(x^0,\textbf{k}_1)\Phi_{\textbf{j}}^*(x^0,\textbf{k}_2)\nonumber\\
&\,\times\Phi_{\textbf{j}}(x^0,\textbf{k}_3)\Phi_{\textbf{i}}(x^0,\textbf{k}_4)V_4(\textbf{k}_1,\textbf{k}_2,\textbf{k}_3,\textbf{k}_4)\,,
\label{eq:accc}
\end{align}
where $\int_\textbf{k}\equiv\int_{-\infty}^\infty d^{d-1}\textbf{k}/(2\pi)^{d-1}$; the time integral $\int_{x^0,\mathcal{C}}$ is integrated along a closed time contour $\mathcal{C}=\mathcal{C}^+\cup\mathcal{C}^-$ so called the Schwinger-Keldysh contour \cite{Schwinger:1960qe,Keldysh:1964ud} on the real axis of time from the initial time $t^0$ to $t=\infty$ along $\mathcal{C}^+$ and back to $t^0$ along $\mathcal{C}^-$ in the description of non-equilibrium dynamics on the basis of the path integral formulation. In the momentum space, the dipole transformation shifts momenta by $\bm{\ell}$, leading to $\Phi_\textbf{i}(x^0,\textbf{k})\rightarrow \Phi_\textbf{i}(x^0,\textbf{k}+\bm{\ell})$. The spatial derivatives in quartic interactions in Eq.~(\ref{eq:accc}) become related to the momentum-dependent coupling
\begin{align}
V_4(\textbf{k}_1,\textbf{k}_2,\textbf{k}_3,\textbf{k}_4)\equiv&\,\bar{V}_4(\textbf{k}_1,\textbf{k}_2,\textbf{k}_3,\textbf{k}_4)(2\pi)^{d-1}\delta^{d-1}(\textbf{k}_1+\textbf{k}_2-\textbf{k}_3-\textbf{k}_4)\,,\\
\bar{V}_4(\textbf{k}_1,\textbf{k}_2,\textbf{k}_3,\textbf{k}_4)\equiv&\,\frac{\bar{\lambda}}{2N}\Big(|\textbf{k}_1-\textbf{k}_2|^2+|\textbf{k}_3-\textbf{k}_4|^2\Big).
\end{align}

After integrating over $\textbf{k}_4$ and applying the re-definition of the variable $\textbf{q}\equiv \textbf{k}_2-\textbf{k}_3$, the action becomes
\begin{align} 
iS[\Phi_{\textbf{i}},\Phi_{\textbf{i}}^*]
=&\,i\int_{x^0,\mathcal{C}}\bigg[\int_\textbf{k}\Phi_{\textbf{i}}^*(x^0,\textbf{k})\big(-\partial_{x^0}^2-m^2\big)\Phi_{\textbf{i}}(x^0,\textbf{k})+\prod_{\ell=1}^2\int_{\textbf{k}_\ell}\int_\textbf{q}\Phi_{\textbf{i}}^*(x^0,\textbf{k}_1)\Phi_{\textbf{j}}^*(x^0,\textbf{k}_2)\nonumber\\
&\,\times\Phi_{\textbf{j}}(x^0,\textbf{k}_2-\textbf{q})\Phi_{\textbf{i}}(\textbf{k}_1+\textbf{q})\frac{\bar{\lambda}}{2N}\Big(|\textbf{k}_1-\textbf{k}_2|^2+|\textbf{k}_2-\textbf{k}_1-2\textbf{q}|^2\Big)\bigg]\,.\label{eq:Sshu}
\end{align}
It is noticed that the momentum integration $\int_{\textbf{k}_1}$ only acts on $\Phi_\textbf{i}^*\Phi_\textbf{i}$ while $\int_{\textbf{k}_2}$ on $\Phi_\textbf{j}^*\Phi_\textbf{j}$ in Eq.~(\ref{eq:Sshu}). That is, these two integrals become decoupled. Then, one can apply the Hubbard-Stratonovich transformation by introducing several auxiliary fields $\sigma$, $\bm{\rho}=\rho^\alpha$, $\zeta$, $\Delta$, $\bm{\chi}=\chi^\alpha$ and $\zeta$ in the action
\begin{align}
&\,iS[\Phi_{\textbf{i}},\Phi_{\textbf{i}}^*,\Upsilon_{v^\prime}^\alpha,\Theta_{v^\prime}^\alpha]\nonumber\\
=&\, i\int_{x^0,\mathcal{C} } \int_\textbf{q}\bigg[\Phi_\textbf{i}^*(x^0,\textbf{q})\big(-\partial_{x^0}^2-m^2\big)\Phi_\textbf{i}(x^0,\textbf{q})+\frac{\bar{\lambda}}{2N}\Upsilon_{s^\prime}^\alpha(x^0,-\textbf{q})A_{s^\prime v^\prime}^{\alpha\beta}(\textbf{q})\Upsilon_{v^\prime}^\beta(x^0,\textbf{q})\nonumber\\
&\,+\Theta_{s^\prime}^\alpha(x^0,-\textbf{q})\Upsilon_{s^\prime}^\alpha(x^0,\textbf{q})-\Delta(x^0,-\textbf{q})\int_\textbf{k}\Phi_{\textbf{i}}^*(x^0,\textbf{k})\Phi_{\textbf{i}}(x^0,\textbf{k}+\textbf{q}) \nonumber\\
&\,-\bm{\chi}(x^0,-\textbf{q})\cdot\int_\textbf{k} \textbf{k}\,\Phi_{\textbf{i}}^*(x^0,\textbf{k})\Phi_{\textbf{i}}(x^0,\textbf{k}+\textbf{q})-\xi(x^0,-\textbf{q})\int_\textbf{k} \textbf{k}^2\Phi_{\textbf{i}}^*(x^0,\textbf{k})\Phi_{\textbf{i}}(x^0,\textbf{k}+\textbf{q})\bigg]\,,\label{eq:Hust}
\end{align}
where 
\begin{align}
\Upsilon_{v^\prime}^\alpha(x^0,\textbf{q})\equiv
\begin{pmatrix}
\sigma(x^0,\textbf{q}) \\
\rho^\alpha(x^0,\textbf{q}) \\
  \zeta(x^0,\textbf{q})
\end{pmatrix}
\,,\quad
&\,\Theta_{v^\prime}^\alpha(x^0,\textbf{q})=
\begin{pmatrix}
\Delta(x^0,\textbf{q}) \\
\chi^\alpha(x^0,\textbf{q}) \\
\xi(x^0,\textbf{q}) \\
\end{pmatrix}
\,,
\end{align}
and I define the matrix $A_{v^\prime s^\prime}^{\alpha\beta}(\textbf{q})$ and its inverse $(A^{-1})_{v^\prime s^\prime}^{\alpha\beta}(\textbf{q})$ 
\begin{align}
&\,A_{v^\prime s^\prime}^{\alpha\beta}(\textbf{q})=
\begin{pmatrix}
4\textbf{q}^2 & 4 q^\beta & 2\\
-4q^\alpha & -4\delta^{\alpha\beta}& 0 \\
2& 0 & 0
\end{pmatrix}
_{v^\prime s^\prime}
\Rightarrow&\, 
(A^{-1})_{v^\prime s^\prime}^{\alpha\beta}(\textbf{q})=\frac{1}{16}
\begin{pmatrix}
0 & 0 & 8\\
0 & -4 \delta^{\alpha\beta} & -8q^\alpha \\
8 & 8q^\beta & 0
\end{pmatrix}
_{v^\prime s^\prime}
\label{eq:maaa}
\end{align}
in order to perform the Gaussian integral. The components labeled with $1\leq \alpha \leq d-1$ are non-trivial when the index $v^\prime$ right beneath $\alpha$ is $v^\prime=2^\prime$ (element labled by $\alpha=0$, $v^\prime=2^\prime$ is zero). On the other hand, components labeled with $v^\prime=1^\prime$ or $3^\prime$ are absent, except for $\alpha=0$. The above rules used to assign non-trivial components are also employed in other functions with a similar index structure present in the following discussions. The last four terms in Eq.~(\ref{eq:Hust}) play a role of imposing the integral form of Dirac delta functions on the generating functional such that the quartic interaction can be re-expressed as a quadratic form by $\Upsilon_{v^\prime}^\alpha$. Hence, the action in Eq.~(\ref{eq:Hust}) becomes
 \begin{align}
&\, iS[\Phi_{\textbf{i}},\Phi_{\textbf{i}}^*,\Delta,\bm{\chi},\xi]\nonumber\\
 =&\, i\int_{x^0,\mathcal{C}}\int_\textbf{q}\bigg[\Phi_\textbf{i}^*(x^0,\textbf{q})\big(-\partial_{x^0}^2-m^2\big)\Phi_\textbf{i}(x^0,\textbf{q}) -\frac{1}{2}\frac{N}{\bar{\lambda}}\Theta^*(x^0,\textbf{q})A^{-1}(\textbf{q})\Theta(x^0,\textbf{q})\nonumber\\
 &\,-\Delta(x^0,-\textbf{q})\int_\textbf{k}\Phi_{\textbf{i}}^*(x^0,\textbf{k})\Phi_{\textbf{i}}(x^0,\textbf{k}+\textbf{q})-\bm{\chi}(x^0,-\textbf{q})\cdot\int_\textbf{k} \textbf{k}\,\Phi_{\textbf{i}}^*(x^0,\textbf{k})\Phi_{\textbf{i}}(x^0,\textbf{k}+\textbf{q})\nonumber\\
 &\,-\xi(x^0,-\textbf{q})\int_\textbf{k} \textbf{k}^2\Phi_{\textbf{i}}^*(x^0,\textbf{k})\Phi_{\textbf{i}}(x^0,\textbf{k}+\textbf{q})\bigg]\,,
\label{eq:SSma}
\end{align}
after integrating out three types of auxiliary fields including $\sigma$, $\rho^\alpha$ and $\zeta$.

Furthermore, one can derive classical equations of motion of these auxiliary fields by varying the action with respect to fields $\Delta$, $\chi$ and $\xi$ 
\begin{align}
\frac{\delta S[\Phi_{\textbf{i}},\Phi_{\textbf{i}}^*,\Delta,\bm{\chi},\xi]}{\delta \Delta(x^0,\textbf{k})}=0: &\,\,\,\xi(x^0,\textbf{q})=-2\frac{\bar{\lambda}}{N}\int_\textbf{k} \Phi_{\textbf{i}}^*(x^0,\textbf{k}) \Phi_{\textbf{i}}(x^0,\textbf{k}+\textbf{q}) \,,\label{eq:quau}\\
\frac{\delta S[\Phi_{\textbf{i}},\Phi_{\textbf{i}}^*,\Delta,\bm{\chi},\xi]}{\delta \bm{\chi}(x^0,\textbf{k})}=0 :&\,\,\,
\bm{\chi}(x^0,\textbf{q})=-2\textbf{q}\,\xi(x^0,\textbf{q})+4\frac{\bar{\lambda}}{N}\int_\textbf{k} \textbf{k}\,\Phi_{\textbf{i}}^*(x^0,\textbf{k})\Phi_{\textbf{i}}(x^0,\textbf{k}+\textbf{q})\,,\label{eq:a1au}\\
\frac{\delta S[\Phi_{\textbf{i}},\Phi_{\textbf{i}}^*,\Delta,\bm{\chi},\xi]}{\delta \xi(x^0,\textbf{k})}=0:
&\, \,\, \Delta(x^0,\textbf{q})=-\textbf{q}\cdot\bm{\chi}(x^0,\textbf{q})-2\frac{\bar{\lambda}}{N}\int_\textbf{k} \textbf{k}^2\,\Phi_{\textbf{i}}^*(x^0,\textbf{k})\Phi_{\textbf{i}}(x^0,\textbf{k}+\textbf{q})\,.\label{eq:a2au}
\end{align}
Subsequently, they can be decoupled as follows
\begin{align}
\begin{cases}
\xi(x^0,\textbf{q})=&-2\frac{\bar{\lambda}}{N}\int_\textbf{k} \Phi_{\textbf{i}}^*(x^0,\textbf{k}) \Phi_{\textbf{i}}(x^0,\textbf{k}+\textbf{q}) \\
\bm{\chi}(x^0,\textbf{q})=&4\frac{\bar{\lambda}}{N}\int_\textbf{k} \textbf{q}\,\Phi_{\textbf{i}}^*(x^0,\textbf{k}) \Phi_{\textbf{i}}(x^0,\textbf{k}+\textbf{q}) +4\frac{\bar{\lambda}}{N}\int_\textbf{k} \textbf{k}\,\Phi_{\textbf{i}}^*(x^0,\textbf{k})\Phi_{\textbf{i}}(x^0,\textbf{k}+\textbf{q}) \\
\Delta(x^0,\textbf{q})=&-4\textbf{q}^2\frac{\bar{\lambda}}{N}\int_\textbf{k} \Phi_{\textbf{i}}^*(x^0,\textbf{k}) \Phi_{\textbf{i}}(x^0,\textbf{k}+\textbf{q}) -4\frac{\bar{\lambda}}{N}\int_\textbf{k} \textbf{q}\cdot\textbf{k}\,\Phi_{\textbf{i}}^*(x^0,\textbf{k})\Phi_{\textbf{i}}(x^0,\textbf{k}+\textbf{q}) \\
&-2\frac{\bar{\lambda}}{N}\int_\textbf{k} \textbf{k}^2\,\Phi_{\textbf{i}}^*(x^0,\textbf{k})\Phi_{\textbf{i}}(x^0,\textbf{k}+\textbf{q})\,.
\end{cases}
\label{eq:Heeq}
\,.
\end{align}
The field expectation values of $\xi$, $\bm{\chi}$ and $\Delta$ expressed as $\bar{\xi}$, $\bar{\bm{\chi}}$ and $\bar{\Delta}$, respectively, are obtained by taking expectation values of Eq.~(\ref{eq:Heeq})
\begin{align}
\bar{\xi}(x^0,-\textbf{q})=&\,-2(2\pi)^{d-1}\delta^{d-1}(\textbf{q})\frac{\bar{\lambda}}{N}\bigg[\phi_{\textbf{i}}^*(x^0)\phi_{\textbf{i}}(x^0)+\int_{\textbf{k}^\prime} G_{(2\textbf{i}-1)(2\textbf{i}-1)}(x^0,x^0,\textbf{k}^\prime-\textbf{q})\bigg]\,,\label{eq:xiba}\\
\bar{\Delta}(x^0,-\textbf{q})=&\,(2\pi)^{d-1}\delta^{d-1}(\textbf{q})\frac{\bar{\lambda}}{N} \bigg[-4\textbf{q}^2\phi_{\textbf{i}}^*(x^0)\phi_{\textbf{i}}(x^0)-2\int_{\textbf{k}^\prime} \textbf{k}^{\prime 2} G_{(2\textbf{i}-1)(2\textbf{i}-1)}(x^0,x^0,\textbf{k}^\prime-\textbf{q})\bigg]\,,\label{eq:Deba}\\
\bar{\bm{\chi}}(x^0,-\textbf{q})=&\,(2\pi)^{d-1}\delta^{d-1}(\textbf{q})\frac{\bar{\lambda}}{N}\bigg[-4\textbf{q}\,\phi_{\textbf{i}}^*(x^0)\phi_{\textbf{i}}(x^0)+4\int_{\textbf{k}^\prime} \textbf{k}^\prime G_{(2\textbf{i}-1)(2\textbf{i}-1)}(x^0,x^0,\textbf{k}^\prime-\textbf{q})\bigg]\,,
\label{eq:chba}
\end{align}
where the complex scalar fields are shifted  $\Phi_\textbf{i}\rightarrow\phi_\textbf{i}+\Phi_\textbf{i}$ by the field expectation values $\phi_\textbf{i}$ (therefore, the expectation value $\langle\hat{\Phi}_\textbf{i}\rangle$ becomes zero after the shift). The spatial homogeneity is assumed and applied in calculations of Eq.~(\ref{eq:xiba}), (\ref{eq:Deba}) and (\ref{eq:chba}), where the field expectation values $\phi_{\textbf{i}}$ are  
\begin{align}
\phi_{\textbf{i}}(x^0,\textbf{k})=&\,(2\pi)^{d-1} \phi_{\textbf{i}}(x^0)\,\delta^{d-1}(\textbf{k}).
\end{align}
and the two-point function $G_{nm}(x^0,y^0,\textbf{k})$ of fluctuations constructed by field operators $\hat{\Phi}_\textbf{i}$ and $\hat{\Phi}_\textbf{i}^*$ contour-ordered on the closed-time contour $\mathcal{C}$ denoted by $\mathcal{T}_\mathcal{C}$ is 
\begin{align}
&\,(2\pi)^{d-1}\delta^{d-1}(\textbf{k}-\textbf{k}^\prime)G_{nm}(x^0,y^0,\textbf{k})\nonumber\\
=&\,
\begin{pmatrix}
\big\langle \mathcal{T}_\mathcal{C}\,\hat{\Phi}_{\textbf{i}}(x^0,\textbf{k})\hat{\Phi}_{\textbf{j}}^*(y^0,\textbf{k}^\prime)\big\rangle & \big\langle \mathcal{T}_\mathcal{C}\,\hat{\Phi}_{\textbf{i}}(x^0,\textbf{k})\hat{\Phi}_{\textbf{j}}(y^0,-\textbf{k}^\prime)\big\rangle \\
\big\langle \mathcal{T}_\mathcal{C}\,\hat{\Phi}_{\textbf{i}}^*(x^0,-\textbf{k})\hat{\Phi}_{\textbf{j}}^*(y^0,\textbf{k}^\prime)\big\rangle   &   \big\langle \mathcal{T}_\mathcal{C}\,\hat{\Phi}_{\textbf{i}}^*(x^0,-\textbf{k})\hat{\Phi}_{\textbf{j}}(y^0,-\textbf{k}^\prime)\big\rangle
\end{pmatrix}
\,,\label{eq:Gtwo}\\
&\,\textbf{i}=
\begin{cases}
\frac{n+1}{2}\,,\quad &n\,\,\, {\rm odd} \\
\frac{n}{2}\,,\quad &n\,\,\, {\rm even}
\end{cases}
\,,\quad 
\textbf{j}=\,
\begin{cases}
\frac{m+1}{2}\,,\quad &m\,\, \,{\rm odd} \\
\frac{m}{2}\,,\quad &m\,\, \,{\rm even}
\end{cases}
\,,\nonumber
\end{align}
with integers $n,m=1,\cdots,2N$. The expectation values in Eq.~(\ref{eq:xiba}), (\ref{eq:Deba}) and (\ref{eq:chba}) are proved to be spatially homogeneous as well since they all share the same factor of $(2\pi)^{d-1}\delta^{d-1}(\textbf{q})$, which eliminates terms proportional to $\textbf{q}$ and $\textbf{q}^2$.

Collecting Eq.~(\ref{eq:xiba}), (\ref{eq:Deba}) and (\ref{eq:chba}) together, I obtain the LO local self-energy defined by $M_{mn}(x^0,x^0,\textbf{k})$ in the $1/N$ expansion that contains non-trivial components
\begin{align}
&\,M_{(2\underline{\textbf{i}}-1)(2\underline{\textbf{i}}-1)}(x^0,x^0,\textbf{k})\equiv \tilde{M}^2(x^0,x^0,\textbf{k})\nonumber\\
=&\,-\frac{2\bar{\lambda}}{N} \bigg[\phi_{\textbf{j}}^*(x^0)\phi_{\textbf{j}}(x^0) \textbf{k}^2+\int_{\textbf{k}^\prime}G_{(2\textbf{j}-1)(2\textbf{j}-1)}(x^0,x^0,\textbf{k}^\prime)\Big(\textbf{k}^2-2\textbf{k}\cdot\textbf{k}^\prime+\textbf{k}^{\prime 2}\Big)\bigg]\,,\label{eq:sig0}\\
&\,M_{(2\underline{\textbf{i}})(2\underline{\textbf{i}})}(x^0,x^0,\textbf{k})\nonumber\\
=&\,-\frac{2\bar{\lambda}}{N} \bigg[\phi_{\textbf{j}}^*(x^0)\phi_{\textbf{j}}(x^0) \textbf{k}^2+\int_{\textbf{k}^\prime}G_{(2\textbf{j})(2\textbf{j})}(x^0,x^0,-\textbf{k}^\prime)\Big(\textbf{k}^2+2\textbf{k}\cdot\textbf{k}^\prime+\textbf{k}^{\prime 2}\Big)\bigg]\,.
\label{eq:sig1}
\end{align}
To see its relation with the dipole symmetry, let us consider the dipole transformation
\begin{align}
\phi_\textbf{i}(x^0,\textbf{k})\rightarrow\phi_\textbf{i}(x^0,\textbf{k}+\bm{\ell})\,,\quad G_{nm}(x^0,y^0,\textbf{k})\rightarrow G_{nm}(x^0,y^0,\textbf{k}+\bm{\ell})\,,
\end{align}
which are characterized as charge and dipole condensate \cite{Jensen:2022ibn}, respectively. Momenta $\textbf{k}_{1,2,3,4}$ in Eq.~(\ref{eq:Sshu}) and $\textbf{k}$ in Eq.~(\ref{eq:Hust}) are all shifted by $\bm{\ell}$ under the dipole transformation (\textbf{q} is invariant since $\textbf{q}=\textbf{k}_2-\textbf{k}_3$). It is noticed that Eq.~(\ref{eq:sig0}) and (\ref{eq:sig1}) still hold after applying the dipole transformation by $\textbf{k}\rightarrow \textbf{k}+\bm{\ell}$ and $\textbf{k}^\prime\rightarrow\textbf{k}^\prime+\bm{\ell}$.  In addition, symmetries are maintained by the invariance $\tilde{M}(x^0,x^0,\textbf{k})=\tilde{M}(x^0,x^0,\textbf{k}+\bm{\ell})$. Any non-zero $\phi_\textbf{i}(x^0)$ in Eq.~(\ref{eq:sig0}) and (\ref{eq:sig1}) is found to spontaneously break both $U(N)$ and the dipole symmetry simultaneously since $\textbf{k}^2\neq(\textbf{k}+\bm{\ell})^2$. On the other hand, for the $U(N)$-symmetric system with vanishing $\phi_\textbf{i}(x^0)$, the dipole symmetry is spontaneously broken  by the non-invariant integral $\int_{\textbf{k}^\prime}$ achieved by the condition $G_{nm}(x^0,x^0,\textbf{k})\neq G_{nm}(x^0,x^0,\textbf{k}+\bm{\ell})$.

To go beyond the LO in the $1/N$ expansion, auxiliary fields are re-defined in the action as $\Delta\rightarrow \bar{\Delta}+\Delta$, $\chi^\alpha\rightarrow \bar{\chi}^\alpha+\chi^\alpha$ and $\xi\rightarrow \bar{\xi}+\xi$. Accordingly, the action consists of
\begin{align}
iS[\phi_\textbf{i},\phi_\textbf{i}^*,\bar{\Theta}_{s^\prime}^\alpha,\Phi_\textbf{i},\Phi_\textbf{i}^*,\Theta_{s^\prime}^\alpha]=&\,
iS[\phi_{\textbf{i}},\phi_{\textbf{i}}^*,\bar{\Theta}_{v^\prime}^\alpha]-\frac{1}{2}\int_{x^0,y^0,\mathcal{C} }\int_{\textbf{k}}\Xi_{\bm{\mu}}^{\alpha*}(x^0,\textbf{k})(D^{-1})_{\bm{\mu}\bm{\nu}}^{\alpha\beta}(x^0,y^0,\textbf{k})\Xi_{\bm{\nu}}^\beta(y^0,\textbf{k})\,,\label{eq:shac}\\
iS[\phi_{\textbf{i}},\phi_{\textbf{i}}^*,\bar{\Theta}]=&\,i V\int_{x^0,\mathcal{C}}\phi_\textbf{i}^*(x^0)\Big(-\partial_{x^0}^2-m^2-\tilde{M}^2(x^0,x^0,0)\Big)\phi_\textbf{i}(x^0)\nonumber\\
&\,-i\frac{N}{\bar{\lambda}}\int_{x^0,\mathcal{C}}\int_{\textbf{k}}\,\frac{1}{2}\, \bar{\Theta}_{s^\prime}^{\alpha*}(x^0,\textbf{k}) (A^{-1})_{s^\prime v^\prime}^{\alpha\beta}(\textbf{k})\bar{\Theta}_{v^\prime}^\beta(x^0,\textbf{k})\,,
\end{align}
where $V$ is the volume of the system identified as $(2\pi)^{d-1}\delta^{d-1}(0)$ and
\begin{align}
\bar{\Theta}_{v^\prime}^\alpha(x^0,\textbf{q})=
\begin{pmatrix}
\bar{\Delta}(x^0,\textbf{q}) \\
\bar{\chi}^\alpha(x^0,\textbf{q}) \\
\bar{\xi}(x^0,\textbf{q}) \\
\end{pmatrix}
\,,\quad
\Xi_{\bm{\mu}}^\alpha(x^0,\textbf{k})\equiv&\,
\begin{pmatrix}
\Phi_{\textbf{i}}(x^0,\textbf{k}) \\
\Phi_{\textbf{i}}^*(x^0,-\textbf{k}) \\
\Delta(x^0,\textbf{k}) \\
\chi^\alpha(x^0,\textbf{k}) \\
\xi(x^0,\textbf{k})
\end{pmatrix}
\,.
\end{align}
The fields $\Xi_{\bm{\mu}}^\alpha$ contain $N$-component complex scalar fields and three kinds of auxiliary fields labeled with bold Greek letters as integers $\bm{\mu}=1,\cdots,2N+3$ shown in lower indices \footnote{The bold Greek letter as a lower index includes components labeled with $v^\prime$, so the rules mentioned below Eq.~(\ref{eq:maaa}) still apply.}; $\alpha$ is 0 or a given spatial dimension from 1 to $d-1$ for the auxiliary field $\chi^\alpha$. The inverse propagator parametrized as
\begin{align}
(D^{-1})_{\bm{\mu}\bm{\nu}}^{\alpha\beta}(x^0,y^0,\textbf{k})\equiv&\,(D_0^{-1})_{\bm{\mu}\bm{\nu}}^{\alpha\beta}(x^0,y^0,\textbf{k})
-
\begin{pmatrix}
\Sigma_{mn}(x^0,y^0,\textbf{k})  & \tilde{\Lambda}_{m s^\prime}^\beta(x^0,y^0,\textbf{k}) \\
\Lambda_{v^\prime n}^\alpha(x^0,y^0,\textbf{k}) & \Pi_{v^\prime s^\prime}^{\alpha\beta}(x^0,y^0,\textbf{k})
\end{pmatrix}
\,,\label{eq:DiR2} \\
(D_0^{-1})_{\bm{\mu}\bm{\nu}}^{\alpha\beta}(x^0,y^0,\textbf{k})\equiv&\,i
\begin{pmatrix}
-iG_{0\, mn}^{-1}(x^0,y^0,\textbf{k}) & \quad \tilde{\Omega}_{m s^\prime}^\beta(x^0,\textbf{k}) \delta_\mathcal{C}(x^0-y^0)\\
\Omega_{v^\prime n}^\alpha(x^0,\textbf{k}) \delta_\mathcal{C}(x^0-y^0) &\quad  \frac{N}{\bar{\lambda}} (A^{-1})_{v^\prime s^\prime}^{\alpha\beta}(\textbf{k}) \delta_\mathcal{C}(x^0-y^0)
\end{pmatrix}
\end{align}
has the classical part including 
\begin{align}
&\,G_{0\,mn}^{-1}(x^0,y^0,\textbf{k})\,, \quad \frac{N}{\bar{\lambda}} (A^{-1})_{v^\prime s^\prime}^{\alpha\beta}(\textbf{k})\,i\delta_\mathcal{C}(x^0-y^0)\,, \\
&\,i\delta_\mathcal{C}(x^0-y^0)\Omega_{v^\prime n}^\alpha(x^0,\textbf{k})\,,\quad i\delta_\mathcal{C}(x^0-y^0)\tilde{\Omega}_{ms^\prime}^\beta(x^0,\textbf{k})\,,
\end{align} 
where the inverse propagator $G_{0\,mn}^{-1}$ is expressed as
\begin{align}
G_{0\,mn}^{-1}(x^0,y^0,\textbf{k})\equiv&\, i\Big(\big(\partial_{x^0}^2+m^2\big)\delta_{mn}+M_{mn}(x^0,x^0,\textbf{k})\Big)\delta_\mathcal{C}(x^0-y^0)
\end{align}
and $\tilde{\Omega}_{ms^\prime}^b$, $\Omega_{v^\prime n}$ are determined by considering the interactions shown in the last three terms of Eq.~(\ref{eq:SSma}), and self-energies. The above expressions are based on the Dirac delta function $\delta_\mathcal{C}(x^0-y^0)$ satisfying $\int_{y^0,\mathcal{C}}\delta_\mathcal{C}(x^0-y^0)f(y^0)=f(x^0)$ defined along the contour $\mathcal{C}$, where the non-vanishing case occurs only when $x^0$ is equal to $y^0$ on a branch.

By choosing either $\Phi_{\textbf{i}}$ or $\Phi_{\textbf{i}}^*$ to be replaced by $\phi_{\textbf{i}}$ or $\phi_{\textbf{i}}^*$ in the last three terms of Eq.~(\ref{eq:SSma}) due to the field re-definition
\begin{align}
&\,-i\int_{x^0,\mathcal{C}}\int_\textbf{q}\Delta(x^0,-\textbf{q})\phi_{\textbf{i}}^*(x^0)\Phi_{\textbf{i}}(x^0,\textbf{q})\nonumber\\
=&\,-\frac{i}{2}\int_{x^0,\mathcal{C}}\bigg[\Delta(x^0,-\textbf{q})\phi_{\textbf{i}}^*(x^0)\Phi_{\textbf{i}}(x^0,\textbf{q})+\Delta(x^0,\textbf{q})\phi_{\textbf{i}}^*(x^0)\Phi_{\textbf{i}}(x^0,-\textbf{q})\bigg]\,,\\
\nonumber\\
&\,-i\int_{x^0,\mathcal{C}}\int_\textbf{q}\Big(\Delta(x^0,\textbf{k})+\textbf{k}\cdot\bm{\chi}(x^0,\textbf{k})+\xi(x^0,\textbf{k})\textbf{k}^2\Big)\Phi_{\textbf{i}}^*(x^0,\textbf{k})\phi_{\textbf{i}}(x^0)\nonumber\\
=&\,-\frac{i}{2}\int_{x^0,\mathcal{C}}\int_\textbf{q}\bigg[\Big(\Delta(x^0,\textbf{k})+\textbf{k}\cdot\bm{\chi}(x^0,\textbf{k})+\xi(x^0,\textbf{k})\textbf{k}^2\Big)\phi_{\textbf{i}}(x^0)\Phi_{\textbf{i}}^*(x^0,\textbf{k})\nonumber\\
&\,+\Big(\Delta(x^0,-\textbf{k})-\textbf{k}\cdot\bm{\chi}(x^0,-\textbf{k})+\xi(x^0,-\textbf{k})\textbf{k}^2\Big)\phi_{\textbf{i}}(x^0)\Phi_{\textbf{i}}^*(x^0,-\textbf{k})\bigg]\,,
\end{align}
functions $\Omega_{v^\prime n}^\alpha$ and $\tilde{\Omega}_{mv^\prime}^\alpha$ are accordingly obtained as
\begin{align}
\Omega_{v^\prime n}^\alpha(x^0,\textbf{k})
=&\,
\begin{pmatrix}
\phi_{\textbf{i}}^*(x^0)  & \phi_{\textbf{i}}(x^0)\\
0  & -k^\alpha\,\phi_{\textbf{i}}(x^0) \\
0  &  \textbf{k}^2\phi_{\textbf{i}}(x^0)
\end{pmatrix}
\otimes
\begin{pmatrix}
\delta_{\textbf{i}1} & \delta_{\textbf{i}2} & \cdots & \delta_{\textbf{i}N}
\end{pmatrix}
_{1\times N}
\,, \label{eq:Omeg}\\ 
\tilde{\Omega}_{n v^\prime}^\alpha(x^0,\textbf{k})
=&\,
\begin{pmatrix}
\phi_{\textbf{i}}(x^0) & k^\alpha\,\phi_{\textbf{i}}(x^0) & \textbf{k}^2\phi_{\textbf{i}}(x^0) \\
\phi_{\textbf{i}}^*(x^0) &  0  & 0
\end{pmatrix}
\otimes
\begin{pmatrix}
\delta_{\textbf{i}1} \\
\delta_{\textbf{i}2}\\ 
\vdots \\
\delta_{\textbf{i}N}
\end{pmatrix}
_{N\times 1}
\,.\label{eq:tOme}
\end{align}
Accordingly, the relations between Eq.~(\ref{eq:Omeg}) and (\ref{eq:tOme}) are summarized by
\begin{align}
\Omega_{1^\prime (2\textbf{i}-1)}^\alpha(x^0,\textbf{k})=&\,\tilde{\Omega}_{(2\textbf{i})1^\prime}^\alpha(x^0,\textbf{k})=\tilde{\Omega}_{(2\textbf{i})1^\prime}^\alpha(x^0,-\textbf{k})\,,\label{eq:Ome1}\\
\Omega_{v^\prime (2\textbf{i})}^\alpha(x^0,\textbf{k})=&\,\tilde{\Omega}_{(2\textbf{i}-1)v^\prime}^\alpha(x^0,-\textbf{k})\,,\label{eq:Ome2}\\
\Omega_{v^\prime (2\textbf{i})}^\alpha(x^0,\textbf{k})=&\,\tilde{\Omega}_{(2\textbf{i}-1)v^\prime}^\alpha(x^0,\textbf{k})\,,\quad{\rm if} \,\, v=1\,\,{\rm or} \,\, 3\,. \label{eq:Ome3}
\end{align}

The Schwinger-Dyson equation for the two-point function $D_{\bm{\mu}\bm{\nu}}^{\alpha\beta}(x^0,y^0,\textbf{k})$ on the closed-time path has the following expression
\begin{align}
\int_{z^0,\mathcal{C}}  (D^{-1})_{\bm{\mu}\bm{\sigma}}^{\alpha\gamma}(x^0,z^0,\textbf{k})  D_{\bm{\sigma}\bm{\nu}}^{\gamma\beta}(z^0,y^0,\textbf{k})=\mathcal{I}_{\bm{\mu}\bm{\nu}}^{\alpha\beta}\delta_\mathcal{C}(x^0-y^0)\,,
\label{eq:Grid}
\end{align}
where
\begin{align}
 \mathcal{I}_{\bm{\mu}\bm{\nu}}^{\alpha\beta}\equiv
\begin{pmatrix}
\delta_{mn} &   \\
&  1 & 0 & 0\\
& 0  & \delta^{\alpha\beta} & 0\\
& 0 & 0 & 1
\end{pmatrix}
_{(2N+3)\times (2N+3)}
\end{align}
and $D_{\bm{\mu}\bm{\nu}}^{\alpha\beta}$ has components parametrized as
\begin{align}
D_{\bm{\mu}\bm{\nu}}^{\alpha\beta}(x^0,y^0,\textbf{k})\equiv
\begin{pmatrix}
G_{mn}(x^0,y^0,\textbf{k}) & \tilde{K}_{m s^\prime}^\beta(x^0,y^0,\textbf{k}) \\
K_{v^\prime n}^\alpha(x^0,y^0,\textbf{k}) &  \bar{B}_{v^\prime s^\prime}^{\alpha\beta}(x^0,y^0,\textbf{k})
\end{pmatrix}
\end{align}
with Eq.~(\ref{eq:Gtwo}) and
\begin{align}
(2\pi)^{d-1}\delta^{d-1}(\textbf{k}-\textbf{k}^\prime)\tilde{K}_{ms^\prime}^\beta(x^0,y^0,\textbf{k})=&\,
\begin{cases}
\big\langle\mathcal{T}_\mathcal{C}\,\hat{\Phi}_{\frac{m+1}{2}}(x^0,\textbf{k})\hat{\Xi}_{s^\prime}^\beta(y^0,-\textbf{k}^\prime)\big\rangle\,,\,\, m\,\,{\rm odd}\,,\\
\big\langle\mathcal{T}_\mathcal{C}\,\hat{\Phi}_{\frac{m}{2}}^*(x^0,-\textbf{k})\hat{\Xi}_{s^\prime}^\beta(y^0,-\textbf{k}^\prime)\big\rangle\,,\,\, m\,\,{\rm even}\,,
\label{eq:Ktwo}
\end{cases}
\\
(2\pi)^{d-1}\delta^{d-1}(\textbf{k}-\textbf{k}^\prime)K_{v^\prime n}^\alpha(x^0,y^0,\textbf{k})=&\,
\begin{cases}
\big\langle\mathcal{T}_\mathcal{C}\,\hat{\Xi}_{v^\prime}^\alpha(x^0,\textbf{k})\hat{\Phi}_{\frac{n+1}{2}}^*(y^0,\textbf{k}^\prime)\big\rangle\,,\,\, n\,\,{\rm odd}\,,\\
\big\langle\mathcal{T}_\mathcal{C}\,\hat{\Xi}_{v^\prime}^\alpha(x^0,\textbf{k})\hat{\Phi}_{\frac{n}{2}}(y^0,-\textbf{k}^\prime)\big\rangle\,,\,\, n\,\,{\rm even}\,,
\label{eq:Kttw}
\end{cases}
\\
(2\pi)^{d-1}\delta^{d-1}(\textbf{k}-\textbf{k}^\prime)\bar{B}_{v^\prime s^\prime}^{\alpha\beta}(x^0,y^0,\textbf{k})=&\,\big\langle
\mathcal{T}_\mathcal{C}\,\hat{\Xi}_{v^\prime}^\alpha(x^0,\textbf{k})\hat{\Xi}_{s^\prime}^\beta(y^0,-\textbf{k}^\prime)\big\rangle\,.
\label{eq:Btwo}
\end{align}
Four types of components extracted from Eq.~(\ref{eq:Grid}) can be characterized as diagonal parts
\begin{align}
\delta_{mn}\delta_\mathcal{C}(x^0-y^0)=&\,i\bigg[\Big(\partial_{x^0}^2+m^2\Big)\delta_{ml}+M_{ml}(x^0,x^0,\textbf{k})\bigg]G_{ln}(x^0,y^0,\textbf{k})\nonumber\\
&\,-\int_{z^0,\mathcal{C}}\Sigma_{ml}(x^0,z^0,\textbf{k})G_{ln}(z^0,y^0,\textbf{k})+i\tilde{\Omega}_{m s^\prime}^\alpha(x^0,\textbf{k})K_{s^\prime n}^\alpha(x^0,y^0,\textbf{k})\nonumber\\
&\,-\int_{z^0,\mathcal{C}}\tilde{\Lambda}_{m s^\prime}^\alpha(x^0,z^0,\textbf{k})K_{s^\prime n}^\alpha(z^0,y^0,\textbf{k})\,, \label{eq:nev1}\\
\mathcal{I}_{v^\prime s^\prime}^{\alpha\beta}\delta_\mathcal{C}(x^0-y^0)=&\,i\Omega_{v^\prime n}^\alpha(x^0,\textbf{k})\tilde{K}_{ns^\prime}^\beta(x^0,y^0,\textbf{k})-\int_{z^0,\mathcal{C}}\Lambda_{v^\prime n}^\alpha(x^0,z^0,\textbf{k})\tilde{K}_{ns^\prime}^\beta(z^0,y^0,\textbf{k})\nonumber\\
&\,+i\frac{N}{\bar{\lambda}}(A^{-1})_{v^\prime h^\prime}^{\alpha\gamma}(\textbf{k})\bar{B}_{h^\prime s^\prime}^{\gamma\beta}(x^0,y^0,\textbf{k})-\int_{z^0,\mathcal{C}}\Pi_{v^\prime h^\prime}^{\alpha\gamma}(x^0,z^0,\textbf{k})\bar{B}_{h^\prime s^\prime}^{\gamma\beta}(z^0,y^0,\textbf{k})\,,
\label{eq:nev2}
\end{align}
and the off-diagonal parts 
\begin{align}
0=&\,i\bigg[\Big(\partial_{x^0}^2+m^2\Big)\delta_{ml}+M_{ml}(x^0,x^0,\textbf{k})\bigg]K_{ls^\prime}^\alpha(x^0,y^0,\textbf{k})-\int_{z^0,\mathcal{C}}\Sigma_{ml}(x^0,z^0,\textbf{k})\tilde{K}_{ls^\prime}^\alpha(z^0,y^0,\textbf{k})\nonumber\\
&\,+i\tilde{\Omega}_{mv^\prime}^\beta(x^0,\textbf{k})\bar{B}_{v^\prime s^\prime}^{\beta\alpha}(x^0,y^0,\textbf{k})-\int_{z^0,\mathcal{C}}\tilde{\Lambda}_{mv^\prime}^\beta(x^0,z^0,\textbf{k})\bar{B}_{v^\prime s^\prime}^{\beta\alpha}(z^0,y^0,\textbf{k}) \,, \label{eq:nev3}\\
0=&\,i\Omega_{v^\prime m}^\alpha(x^0,\textbf{k})G_{mn}(x^0,y^0,\textbf{k})-\int_{z^0,\mathcal{C}}\Lambda_{v^\prime m}^\alpha(x^0,z^0,\textbf{k})G_{mn}(z^0,y^0,\textbf{k})\nonumber\\
&\,+i\frac{N}{\bar{\lambda}}(A^{-1})_{v^\prime s^\prime}^{\alpha\beta}(\textbf{k})K_{s^\prime n}^\beta(x^0,y^0,\textbf{k})-\int_{z^0,\mathcal{C}}\Pi_{v^\prime s^\prime}^{\alpha\beta}(x^0,z^0,\textbf{k})K_{s^\prime n}^\beta(z^0,y^0,\textbf{k}).
\label{eq:nev4}
\end{align}
The auxiliary field's two-point function is found to be decomposed into a local and a non-local part 
\begin{align}
\bar{B}_{v^\prime s^\prime}^{\alpha\beta}(x^0,y^0,\textbf{k})\equiv-i\frac{\bar{\lambda}}{N}A_{v^\prime s^\prime}^{\alpha\beta}(\textbf{k})\,\delta_\mathcal{C}(x^0-y^0)+\frac{\bar{\lambda}}{N}B_{v^\prime s^\prime}^{\alpha\beta}(x^0,y^0,\textbf{k})
\label{eq:Bfln}
\end{align}
in order to remove the singularity in Eq.~(\ref{eq:nev2}) such that
\begin{align}
0=&\,i\Omega_{v^\prime n}^\alpha(x^0,\textbf{k})\tilde{K}_{ns^\prime}^\beta(x^0,y^0,\textbf{k})-\int_{z^0,\mathcal{C}}\Lambda_{v^\prime n}^\alpha(x^0,z^0,\textbf{k})\tilde{K}_{ns^\prime}^\beta(z^0,y^0,\textbf{k})+i(A^{-1})_{v^\prime h^\prime}^{\alpha\gamma}(\textbf{k})B_{h^\prime s^\prime}^{\gamma\beta}(x^0,y^0,\textbf{k})\nonumber\\
&\,+i\frac{\bar{\lambda}}{N}\Pi_{v^\prime h^\prime}^{\alpha\gamma}(x^0,y^0)A_{h^\prime s^\prime}^{\gamma\beta}(\textbf{k})-\frac{\bar{\lambda}}{N}\int_{z^0,\mathcal{C}}\Pi_{v^\prime h^\prime}^{\alpha\gamma}(x^0,z^0,\textbf{k})B_{h^\prime s^\prime}^{\gamma\beta}(z^0,y^0,\textbf{k})
\label{eq:bbbb}
\end{align}
and $K_{s^\prime n}^\alpha, \tilde{K}_{ns^\prime}^\alpha, \bar{B}_{v^\prime s^\prime}^{\alpha\beta}$ are at the order of $\mathcal{O}(1/N)$.

Defining the step function $\theta_\mathcal{C}(x^0-y^0)$ and ${\rm sgn}_\mathcal{C}(x^0-y^0)\equiv\theta_\mathcal{C}(x^0-y^0)-\theta_\mathcal{C}(y^0-x^0)$ on $\mathcal{C}$ following the contour-ordering, the two-point functions of fluctuations obeying contour-ordering on $\mathcal{C}$ can be decomposed into the statistical two-point function, denoted by $F$, and the spectral function, denoted by $\rho$, as follows
\begin{align}
G_{nm}(x^0,y^0,\textbf{k})=&\,F_{G_{nm}}(x^0,y^0,\textbf{k})-\frac{i}{2}\rho_{G_{nm}}(x^0,y^0,\textbf{k})\,{\rm sgn}_\mathcal{C}(x^0-y^0)\,, \label{eq:Gdic}\\
\tilde{K}_{ns^\prime}^\alpha(x^0,y^0,\textbf{k})=&\,F_{\tilde{K}_{ns^\prime}}^\alpha(x^0,y^0,\textbf{k})-\frac{i}{2}\rho_{\tilde{K}_{ns^\prime}}^\alpha(x^0,y^0,\textbf{k})\,{\rm sgn}_\mathcal{C}(x^0-y^0)\,, \label{eq:Kdic}\\
K_{s^\prime n}^\alpha(x^0,y^0,\textbf{k})=&\,F_{K_{s^\prime n}}^\alpha(x^0,y^0,\textbf{k})-\frac{i}{2}\rho_{K_{s^\prime n}}^\alpha(x^0,y^0,\textbf{k})\,{\rm sgn}_\mathcal{C}(x^0-y^0)\,, \label{eq:tdic}\\
B_{s^\prime v^\prime}^{\alpha\beta}(x^0,y^0,\textbf{k})=&\,F_{B_{s^\prime v^\prime}}^{\alpha\beta}(x^0,y^0,\textbf{k})-\frac{i}{2}\rho_{B_{s^\prime v^\prime}}^{\alpha\beta}(x^0,y^0,\textbf{k})\,{\rm sgn}_\mathcal{C}(x^0-y^0)\label{eq:Bdic}\,,
\end{align}
in which explicit expressions are defined by several field operators
\begin{align}
(2\pi)^{d-1}\delta_\mathcal{C}^{d-1}(\textbf{k}-\textbf{k}^\prime)F_{G_{nm}}(x^0,y^0,\textbf{k})\equiv&\,\frac{1}{2}\Big\langle \Big\{\hat{L}_n(x^0,\textbf{k}),\hat{I}_m(y^0,\textbf{k}^\prime)\Big\}\Big\rangle\,,\\
(2\pi)^{d-1}\delta^{d-1}(\textbf{k}-\textbf{k}^\prime)\rho_{G_{nm}}(x^0,y^0,\textbf{k})\equiv&\, i\Big\langle \Big[\hat{L}_n(x^0,\textbf{k}),\hat{I}_m(y^0,\textbf{k}^\prime)\Big]\Big\rangle\,,\\
(2\pi)^{d-1}\delta^{d-1}(\textbf{k}-\textbf{k}^\prime)F_{B_{s^\prime v^\prime}}^{\alpha\beta}(x^0,y^0,\textbf{k})\equiv&\,\frac{1}{2}\Big\langle\Big\{\hat{\Xi}_{s^\prime}^\alpha(x^0,\textbf{k}),\hat{\Xi}_{v^\prime}^\beta(y^0,-\textbf{k}^\prime)\Big\}\Big\rangle\,,
\\
(2\pi)^{d-1}\delta^{d-1}(\textbf{k}-\textbf{k}^\prime)\rho_{B_{s^\prime v^\prime}}^{\alpha\beta}(x^0,y^0,\textbf{k})\equiv&\,i\Big\langle\Big[\hat{\Xi}_{s^\prime}^\alpha(x^0,\textbf{k}),\hat{\Xi}_{v^\prime}^\beta(y^0,-\textbf{k}^\prime)\Big]\Big\rangle\,,\\
(2\pi)^{d-1}\delta^{d-1}(\textbf{k}-\textbf{k}^\prime)F_{K_{s^\prime n}}^\alpha(x^0,y^0,\textbf{k})\equiv&\,\frac{1}{2}\Big\langle\Big\{\hat{\Xi}_{s^\prime}^\alpha(x^0,\textbf{k}),\hat{P}_n(y^0,\textbf{k}^\prime)\Big\}\Big\rangle\,,\\
(2\pi)^{d-1}\delta^{d-1}(\textbf{k}-\textbf{k}^\prime)\rho_{K_{s^\prime n}}^\alpha(x^0,y^0,\textbf{k})\equiv&\,i\Big\langle\Big[\hat{\Xi}_{s^\prime}^\alpha(x^0,\textbf{k}),\hat{P}_n(y^0,\textbf{k}^\prime)\Big]\Big\rangle\,, \\
(2\pi)^{d-1}\delta^{d-1}(\textbf{k}-\textbf{k}^\prime)F_{\tilde{K}_{ns^\prime }}^\alpha(x^0,y^0,\textbf{k})\equiv&\,\frac{1}{2}\Big\langle\Big\{\hat{\tilde{P}}_n(x^0,\textbf{k}),\hat{\Xi}_{s^\prime}^\alpha(y^0,\textbf{k}^\prime)\Big\}\Big\rangle\,,\\
(2\pi)^{d-1}\delta^{d-1}(\textbf{k}-\textbf{k}^\prime)\rho_{\tilde{K}_{ns^\prime }}^\alpha(x^0,y^0,\textbf{k})\equiv&\,i\Big\langle\Big[\hat{\tilde{P}}_n(x^0,\textbf{k}),\hat{\Xi}_{s^\prime}^\alpha(y^0,\textbf{k}^\prime)\Big\}\Big\rangle\,,
\end{align}
where components are defined in Table.~\ref{tab:FGrG} and \ref{tab:FKrK}. In the case of equal time $x^0=y^0$, the commutation relations are given by
\begin{align}
&\,\Big[\hat{\Phi}_\textbf{i}(x^0,\textbf{k}),\partial_{x^0}\hat{\Phi}_\textbf{j}^*(x^0,\textbf{k}^\prime)\Big]=i(2\pi)^{d-1}\delta^{d-1}(\textbf{k}-\textbf{k}^\prime)\delta_{\textbf{i}\textbf{j}}\,,\label{eq:coid}\\
&\,\Big[\hat{\Phi}_\textbf{i}(x^0,\textbf{k}),\hat{\Phi}_\textbf{j}^*(x^0,\textbf{k}^\prime)\Big]=\Big[\hat{\Phi}_\textbf{i}(x^0,\textbf{k}),\hat{\Phi}_\textbf{j}(x^0,-\textbf{k}^\prime)\Big]=\Big[\hat{\Phi}_\textbf{i}^*(x^0,-\textbf{k}),\hat{\Phi}_\textbf{j}^*(x^0,\textbf{k}^\prime)\Big]=0\,.
\label{eq:comu}
\end{align}
Therefore, the spectral function in Eq.~(\ref{eq:Gdic}) vanishes at the equal time.

\renewcommand{\arraystretch}{1.5}
\begin{table}[htbp]
  \centering
    \caption{Definitions of $F_{G_{nm}}(x^0,y^0,\textbf{k})$ and $\rho_{G_{nm}}(x^0,y^0,\textbf{k})$}
    \label{tab:FGrG}
\begin{center}
\begin{tabular}[c]{|p{11em}|@{\hspace{0.6em}}c@{\hspace{0.6em}}|@{\hspace{0.6em}}c@{\hspace{0.6em}}|}
    \hline
\centering $\Big(\hat{L}_n(x^0,\textbf{k}),\hat{I}_m(y^0,\textbf{k}^\prime)\Big)$
    & \centering $n$ odd &  $n$ even \\ 
    \hline
   \centering $m$ odd & $\Big(\hat{\Phi}_{\frac{n+1}{2}}(x^0,\textbf{k}),\hat{\Phi}_{\frac{m+1}{2}}^*(y^0,\textbf{k}^\prime)\Big)$ & $\Big(\hat{\Phi}_{\frac{n}{2}}^*(x^0,-\textbf{k}),\hat{\Phi}_{\frac{m+1}{2}}^*(y^0,\textbf{k}^\prime)\Big)$\\ 
    \hline
    \centering $m$ even & $\Big(\hat{\Phi}_{\frac{n+1}{2}}(x^0,\textbf{k}),\hat{\Phi}_{\frac{m}{2}}(y^0,-\textbf{k}^\prime)\Big)$ & $\Big(\hat{\Phi}_{\frac{n}{2}}^*(x^0,-\textbf{k}),\hat{\Phi}_{\frac{m}{2}}(y^0,-\textbf{k}^\prime)\Big)$\,, \\
    \hline
\end{tabular}
\end{center}
\end{table}

\renewcommand{\arraystretch}{1.5}
\begin{table}[htbp]
  \centering
    \caption{Definitions of $F_{K_{s^\prime n}}^\alpha(x^0,y^0,\textbf{k})$, $\rho_{K_{s^\prime n}}^\alpha(x^0,y^0,\textbf{k})$, $F_{\tilde{K}_{ns^\prime }}^\alpha(x^0,y^0,\textbf{k})$ and $\rho_{\tilde{K}_{ns^\prime }}^\alpha(x^0,y^0,\textbf{k})$}
    \label{tab:FKrK}
\begin{center}
\begin{tabular}[c]{|p{11em}|@{\hspace{0.6em}}c@{\hspace{0.6em}}|@{\hspace{0.6em}}c@{\hspace{0.6em}}|}
    \hline
    & \centering $n$ odd &  $n$ even \\ 
    \hline
   \centering $\hat{P}_n(y^0,\textbf{k}^\prime)$  & $\hat{\Phi}_{\frac{n+1}{2}}^*(y^0,\textbf{k}^\prime)$ & $\hat{\Phi}_{\frac{n}{2}}(y^0,-\textbf{k}^\prime)$\\ 
    \hline
    \centering $\hat{\tilde{P}}_{n}(x^0,\textbf{k})$ & $\hat{\Phi}_{\frac{n+1}{2}}(x^0,\textbf{k})$ & $\hat{\Phi}_{\frac{n}{2}}^*(x^0,-\textbf{k})$\,, \\
    \hline
\end{tabular}
\end{center}
\end{table}

Symmetry properties of these two-point functions are summarized in Table.~\ref{tab:FGre}, \ref{tab:FKre} and $B_{s^\prime v^\prime}^{\alpha\beta}(x^0,y^0,\textbf{k})$ follow
\begin{align}
F_{B_{s^\prime v^\prime}}^{\alpha\beta}(x^0,y^0,\textbf{k})=&\,F_{B_{v^\prime s^\prime}}^{\beta\alpha}(y^0,x^0,-\textbf{k})\,,\quad
\rho_{B_{s^\prime v^\prime}}^{\alpha\beta}(x^0,y^0,\textbf{k})=-\rho_{B_{v^\prime s^\prime}}^{\beta\alpha}(y^0,x^0,-\textbf{k})\,,
\end{align}
It can be seen that $K_{s^\prime n}^\alpha$ and $\tilde{K}_{ms^\prime}^\alpha$ are not independent due to the correlations shown in Table.~\ref{tab:FKre}. Hence, Eq.~(\ref{eq:nev2}) out of four identities from Eq.~(\ref{eq:Grid}) can be neglected.
\renewcommand{\arraystretch}{1.5}
\begin{table}[htbp]
  \centering
    \caption{Symmetry properties of $F_{G_{nm}}(x^0,y^0,\textbf{k})$ and $\rho_{G_{nm}}(x^0,y^0,\textbf{k})$}
    \label{tab:FGre}
\begin{center}
\begin{tabular}[c]{|p{15.5em}|@{\hspace{1em}}c@{\hspace{1em}}|@{\hspace{1em}}c@{\hspace{1em}}|}
    \hline
    $F_{G_{nm}}(x^0,y^0,\mathbf{k})=F_{G_{ij}}(y^0,x^0,-\mathbf{k}) \newline
 \rho_{G_{nm}}(x^0,y^0,\mathbf{k})=-\rho_{G_{ij}}(y^0,x^0,-\mathbf{k})$ 
    & \centering $n$ odd &  $n$ even \\ 
    \hline
   \centering $m$ odd & \shortstack{$i = m+1$\\$j = n+1$} & \shortstack{$i = m+1$\\$j = n-1$} \\ 
    \hline
    \centering $m$ even & \shortstack{$i = m-1$\\$j = n+1$} & \shortstack{$i = m-1$\\$j = n-1$} \\
    \hline
\end{tabular}
\end{center}
\end{table}

\renewcommand{\arraystretch}{1.5}
\begin{table}[htbp]
  \centering
    \caption{Symmetry properties of $F_{K_{s^\prime n}}^\alpha(x^0,y^0,\textbf{k})$ and $\rho_{K_{s^\prime n}}^\alpha(x^0,y^0,\textbf{k})$}
    \label{tab:FKre}
\begin{center}
\begin{tabular}[c]{|p{16em}|@{\hspace{1em}}c@{\hspace{1em}}|@{\hspace{1em}}c@{\hspace{1em}}|}
\hline
 $F_{K_{s^\prime n}}^\alpha(x^0,y^0,\textbf{k})= F_{\tilde{K}_{is^\prime }}^\alpha(x^0,y^0,-\textbf{k})\newline \rho_{K_{s^\prime n}}^\alpha(x^0,y^0,\textbf{k})=-\rho_{\tilde{K}_{is^\prime }}^\alpha(y^0,x^0,-\textbf{k})$& \centering $n$ odd & $n$ even \\
 \hline 
 &  $i=n+1$ &  $i=n-1$\\
 \hline
\end{tabular}
\end{center}
\end{table}

\section{2PI $1/N$ expansion in non-equilibrium dynamics}
\label{sec:effe}

\subsection{2PI effective action and non-equilibrium dynamics}
\label{sec:2PIe}
The 2PI effective action is built up including Feynman diagrams not be disconnected by cutting two internal propagator lines (two-particle irreducible); these diagrams consist of two-point functions in Eq.~(\ref{eq:Gtwo}), (\ref{eq:Ktwo}), (\ref{eq:Kttw}) and (\ref{eq:Btwo}) with non-local self-energies presented in Eq.~(\ref{eq:DiR2}) encoded by the resummation and field expectation values. In contrast to the 1PI effective action as a functional of field expectation values only, the 2PI effective action additionally contains the dressed propagators as constituents  \cite{Berges:2004yj,berges2015nonequilibriumquantumfieldscold}. Therefore, the 2PI effective action determines a set of equations of motion for field expectation values and two-point functions. The 2PI effective action can be constructed following Ref.~\cite{Cornwall:1974vz,Aarts:2002dj}
\begin{align}
\Gamma[\phi_{\textbf{i}},\phi_{\textbf{i}}^*,\bar{\Theta}_{s^\prime}^\alpha,D_{\bm{\mu}\bm{\nu}}^{\alpha\beta}]=&\,S[\phi_{\textbf{i}},\phi_{\textbf{i}}^*,\bar{\Theta}_{s^\prime}^\alpha]+\frac{i}{2}{\rm Tr}\ln D^{-1}+\frac{i}{2}{\rm Tr}D_0^{-1}D+\Gamma_I^{\rm NLO}[D_{\bm{\mu}\bm{\nu}}^{\alpha\beta}]+{\rm const}\,.
\label{eq:efac}
\end{align}
The trace in Eq.~(\ref{eq:efac}) sums over all indices, which generates a factor of $N$ by summing over field indices expressed as Latin letters shown in italics; it also implies integrations over time variables on $\mathcal{C}$ and spatial momentum variables, e.g. $\textbf{k}$, by $V\int_\textbf{k}$. Therefore, the first three terms in Eq.~(\ref{eq:efac}) are at the LO since $\phi_\textbf{i}^*\phi_\textbf{i}\sim N$. Since complex scalar fields and auxiliary fields are re-defined by shifting by their expectation values, interactions in the last three terms of the action in Eq.~(\ref{eq:SSma}) only give rise to cubic terms. Therefore, the NLO part of the effective action contained in $\Gamma_I^{\rm NLO}$ do not depend on $\phi_\textbf{i}$ and $\bar{\Theta}_{s^\prime}^\alpha$ in the auxiliary field method. Note that interactions induce NLO contributions in the effective action at the order of $\mathcal{O}(1)$ in the auxiliary field method.

The evolution equation for the field expectation value $\phi_\textbf{i}^*(x^0)$ is determined by the equation of motion
\begin{align}
\frac{\delta\Gamma[\phi_\textbf{j},\phi_\textbf{j}^*,\bar{\Theta}_{s^\prime}^\alpha,D_{\bm{\mu}\bm{\nu}}^{\alpha\beta}]}{\delta \phi_\textbf{i}(x^0)}=\frac{\delta\Gamma[\phi_\textbf{j},\phi_\textbf{j}^*,\bar{\Theta}_{s^\prime}^\alpha,D_{\bm{\mu}\bm{\nu}}^{\alpha\beta}]}{\delta \phi_\textbf{i}^*(x^0)}=0\,.
\label{eq:eqof}
\end{align}
Since $\Gamma_I^{\rm NLO}$ is independent of field expectation values in the auxiliary field method, the functional derivative in Eq.~(\ref{eq:eqof}) only acts on $S[\phi_{\textbf{i}},\phi_{\textbf{i}}^*,\bar{\Theta}_{s^\prime}^\alpha]$ and $(D_0^{-1})_{\bm{\mu}\bm{\nu}}^{\alpha\beta}$ in Eq.~(\ref{eq:efac}) given by
\begin{align}
&\,V\Big(\partial_{x^0}^2+m^2+\tilde{M}^2(x^0,x^0,0)\Big)\phi_{\textbf{i}}^*(x^0)\nonumber\\
=&\,-\frac{1}{2}{\rm Tr}\bigg[\frac{\delta \tilde{\Omega}(z^0,\textbf{q})}{\delta \phi_{\textbf{i}}(x^0)}K(z^0,z^0,\textbf{q})\bigg]-\frac{1}{2}{\rm Tr}\bigg[\frac{\delta \Omega(z^0,\textbf{q})}{\delta \phi_{\textbf{i}}(x^0)}\tilde{K}(z^0,z^0,\textbf{q})\bigg]\nonumber\\
=&\,-V\int_\textbf{k}\bigg(K_{1^\prime (2\textbf{i}-1)}^0(x^0,x^0,\textbf{k})+k^\alpha K_{2^\prime (2\textbf{i}-1)}^\alpha(x^0,x^0,\textbf{k})+\textbf{k}^2K_{3^\prime (2\textbf{i}-1)}^0(x^0,x^0,\textbf{k})\bigg)\,,
\label{eq:feau}\\
&\,V\Big(\partial_{x^0}^2+m^2+\tilde{M}^2(x^0,x^0,0)\Big)\phi_{\textbf{i}}(x^0)=-V\int_\textbf{k}K_{1^\prime(2\textbf{i})}^0(x^0,x^0,\textbf{k})\,,
\end{align}
where the symmetry properties in Table.~\ref{tab:FKre} are used. Equations of motion for field expectation values of auxiliary fields by the variation of the effective action with respect to $\bar{\Theta}_s^\alpha$ have been shown in Eq.~(\ref{eq:quau}), (\ref{eq:a1au}) and (\ref{eq:a2au}). On the other hand, self-energies from NLO corrections obey
\begin{align}
\Sigma_{mn}(x^0,y^0,\textbf{k})=&\,\frac{2i}{V}\frac{\delta\Gamma_I^{\rm NLO}[D_{\bm{\mu}\bm{\nu}}^{\alpha\beta}]}{\delta G_{nm}(y^0,x^0,\textbf{k})}\,,\label{eq:tPIS}\\
\Pi_{v^\prime s^\prime}^{\beta\alpha}(x^0,y^0,\textbf{k})=&\,\frac{2i}{V}\frac{\delta\Gamma_I^{\rm NLO}[D_{\bm{\mu}\bm{\nu}}^{\alpha\beta}]}{\delta \bar{B}_{s^\prime v^\prime}^{\alpha\beta}(y^0,x^0,\textbf{k})}\,,\label{eq:tPIPI}\\
\Lambda_{s^\prime n}^\alpha(x^0,y^0,\textbf{k})=&\,\frac{2i}{V}\frac{\delta \Gamma_I^{\rm NLO}[D_{\bm{\mu}\bm{\nu}}^{\alpha\beta}]}{\delta \tilde{K}_{ns^\prime}^\alpha(y^0,x^0,\textbf{k})}\,,\\
\tilde{\Lambda}_{n s^\prime }^\alpha(x^0,y^0,\textbf{k})=&\,\frac{2i}{V}\frac{\delta\Gamma_I^{\rm NLO}[D_{\bm{\mu}\bm{\nu}}^{\alpha\beta}]}{\delta K_{s^\prime n}^\alpha(y^0,x^0,\textbf{k})}
\end{align}
in order to satisfy equations of motion of two-point functions.

Given the cubic interactions in the last three terms of Eq.~(\ref{eq:SSma}), the allowed NLO contributions in $\Gamma_I^{\rm NLO}$ that are two-particle irreducible can be constructed by 
\begin{align}
&\,\Gamma_I^{\rm NLO}[D_{\bm{\mu}\bm{\nu}}^{\alpha\beta}]\nonumber\\
=&\,V\frac{i}{2}\int_{w^0,z^0,\mathcal{C}}\int_{\textbf{q},\bm{\ell}}\bigg\{\bigg[\bar{B}_{1^\prime 1^\prime}^{00}(w^0,z^0,-\textbf{q})+(\ell^\alpha-q^\alpha) \ell^\beta\bar{B}_{2^\prime 2^\prime}^{\alpha\beta}(w^0,z^0,-\textbf{q})\nonumber\\
&\,+(\bm{\ell}-\textbf{q})^2\bm{\ell}^2\bar{B}_{3^\prime 3^\prime}^{00}(w^0,z^0,-\textbf{q})
+\ell^\beta\bar{B}_{1^\prime 2^\prime}^{0 \beta}(w^0,z^0,-\textbf{q})+\bm{\ell}^2\bar{B}_{1^\prime 3^\prime}^{00}(w^0,z^0,-\textbf{q})
\nonumber\\
&\,+(\ell^\alpha-q^\alpha)\bm{\ell}^2\bar{B}_{2^\prime 3^\prime}^{\alpha 0}(w^0,z^0,-\textbf{q})
+(\ell^\alpha-q^\alpha)\bar{B}_{2^\prime 1^\prime}^{\alpha 0 }(w^0,z^0,-\textbf{q})
\nonumber\\
&\,+(\bm{\ell}-\textbf{q})^2\bar{B}_{3^\prime 1^\prime}^{00}(w^0,z^0,-\textbf{q})
+(\bm{\ell}-\textbf{q})^2\ell^\beta\bar{B}_{3^\prime 2^\prime}^{0\beta}(w^0,z^0,-\textbf{q})\bigg]\nonumber\\
&\,\times G_{(2\textbf{j}-1)(2\textbf{i}-1)}(w^0,z^0,\bm{\ell})G_{(2\textbf{j})(2\textbf{i})}(w^0,z^0,\textbf{q}-\bm{\ell})\nonumber\\
&\,+\bigg[\bar{B}_{1^\prime 1^\prime}^{00}(w^0,z^0,-\textbf{q})
-\ell^\alpha \ell^\beta\bar{B}_{2^\prime 2^\prime}^{\alpha\beta}(w^0,z^0,-\textbf{q})
+\bm{\ell}^2\bm{\ell}^2\bar{B}_{3^\prime 3^\prime}^{00}(w^0,z^0,-\textbf{q})  \nonumber\\
&\,+\ell^\beta\bar{B}_{1^\prime 2^\prime}^{0 \beta}(w^0,z^0,-\textbf{q})
+\bm{\ell}^2\bar{B}_{1^\prime 3^\prime}^{00}(w^0,z^0,-\textbf{q})
-\ell^\alpha\bm{\ell}^2\bar{B}_{2^\prime 3^\prime}^{\alpha 0}(w^0,z^0,-\textbf{q})\nonumber\\
&\,
-\ell^\alpha\bar{B}_{2^\prime 1^\prime}^{\alpha 0 }(w^0,z^0,-\textbf{q})
+\bm{\ell}^2\bar{B}_{3^\prime 1^\prime}^{00}(w^0,z^0,-\textbf{q})
+\bm{\ell}^2\ell^\beta\bar{B}_{3^\prime 2^\prime}^{0\beta}(w^0,z^0,-\textbf{q})\bigg]\nonumber\\
&\,\times G_{(2\textbf{j})(2\textbf{i}-1)}(w^0,z^0,\bm{\ell})G_{(2\textbf{j}-1)(2\textbf{i})}(w^0,z^0,\textbf{q}-\bm{\ell})\bigg\}\,,
\end{align}
where $\bar{B}\sim\mathcal{O}(1/N)$ together with a factor of $N$ from the contraction of field indices are identified at the order $\mathcal{O}(1)$. By introducing new functions $S_{s^\prime v^\prime}^{\alpha\beta}(\bm{\ell},\textbf{q})$ and $S_{s^\prime v^\prime}^{\alpha\beta}(\bm{\ell})$ that obey
\begin{align}
\alpha=0,\,\beta=0:&\quad 
S_{1^\prime 1^\prime}^{\alpha\beta}(\bm{\ell},\textbf{q})=1\,,\quad S_{3^\prime 3^\prime}^{\alpha\beta}(\bm{\ell},\textbf{q})=(\bm{\ell}-\textbf{q})^2\bm{\ell}^2\,,\quad S_{1^\prime 3^\prime}^{\alpha\beta}(\bm{\ell},\textbf{q})=\bm{\ell}^2 \nonumber\\
&\quad 
S_{3^\prime 1^\prime}^{\alpha\beta}(\bm{\ell},\textbf{q})=(\bm{\ell}-\textbf{q})^2 \,,\label{eq:Sre1}\\
\alpha\neq0,\,\beta=0:&\quad
S_{2^\prime 1^\prime}^{\alpha\beta}(\bm{\ell},\textbf{q})=\ell^\alpha-q^\alpha \,,\quad S_{2^\prime 3^\prime}^{\alpha\beta}(\bm{\ell},\textbf{q})=(\ell^\alpha-q^\alpha)\bm{\ell}^2\,,\label{eq:Sre2}\\
\alpha=0,\,\beta\neq0:&\quad 
S_{1^\prime 2^\prime}^{\alpha\beta}(\bm{\ell},\textbf{q})=\ell^\beta \,,\quad S_{3^\prime 2^\prime}^{\alpha\beta}(\bm{\ell},\textbf{q})=(\bm{\ell}-\textbf{q})^2\ell^\beta\,,\label{eq:Sre3}\\
\alpha\neq 0,\,\beta\neq 0:&\quad
S_{2^\prime 2^\prime}^{\alpha\beta}(\bm{\ell},\textbf{q})=(\ell^\alpha-q^\alpha)\ell^\beta \,,\label{eq:Sre4}
\end{align}
and
\begin{align}
\alpha=0,\,\beta=0:&\quad 
S_{1^\prime 1^\prime}^{\alpha\beta}(\bm{\ell})=1\,,\quad S_{3^\prime 3^\prime}^{\alpha\beta}(\bm{\ell})=\bm{\ell}^2\bm{\ell}^2\,,\quad S_{1^\prime 3^\prime}^{\alpha\beta}(\bm{\ell})=\bm{\ell}^2 \nonumber\\
&\quad 
S_{3^\prime 1^\prime}^{\alpha\beta}(\bm{\ell})=\bm{\ell}^2 \,,\label{eq:Sree1}\\
\alpha\neq0,\,\beta=0:&\quad
S_{2^\prime 1^\prime}^{\alpha\beta}(\bm{\ell})=-\ell^\alpha \,,\quad S_{2^\prime 3^\prime}^{\alpha\beta}(\bm{\ell})=-\ell^\alpha\bm{\ell}^2\,,\label{eq:Sree2}\\
\alpha=0,\,\beta\neq0:&\quad 
S_{1^\prime 2^\prime}^{\alpha\beta}(\bm{\ell})=\ell^\beta \,,\quad S_{3^\prime 2^\prime}^{\alpha\beta}(\bm{\ell})=\bm{\ell}^2\ell^\beta\,,\label{eq:Sree3}\\
\alpha\neq 0,\,\beta\neq 0:&\quad
S_{2^\prime 2^\prime}^{\alpha\beta}(\bm{\ell})=-\ell^\alpha\ell^\beta \,,\label{eq:Sree4}
\end{align}
otherwise, zero, self-energies in Eq.~(\ref{eq:tPIS}) can be expressed concisely as
\begin{align}
\Sigma_{(2\textbf{i}-1)(2\textbf{j}-1)}(x^0,y^0,\textbf{k})=&-\int_\textbf{q} S_{s^\prime v^\prime}^{\alpha\beta}(\textbf{k},\textbf{q})\bar{B}_{s^\prime v^\prime}^{\alpha\beta}(y^0,x^0,-\textbf{q})G_{(2\textbf{i}-1)(2\textbf{j}-1)}(x^0,y^0,\textbf{k}-\textbf{q})\,,\label{eq:sdjk}\\
\Sigma_{(2\textbf{i})(2\textbf{j})}(x^0,y^0,\textbf{k})=&-\int_\textbf{q} S_{s^\prime v^\prime}^{\alpha\beta}(\textbf{q}-\textbf{k},\textbf{q})\bar{B}_{s^\prime v^\prime}^{\alpha\beta}(y^0,x^0,-\textbf{q})G_{(2\textbf{i})(2\textbf{j})}(x^0,y^0,\textbf{k}-\textbf{q})\,,
\label{eq:Siev}\,,\\
\Sigma_{(2\textbf{i}-1)(2\textbf{j})}(x^0,y^0,\textbf{k})=&-\int_\textbf{q} S_{s^\prime v^\prime}^{\alpha\beta}(\textbf{k})\bar{B}_{s^\prime v^\prime}^{\alpha\beta}(y^0,x^0,-\textbf{q})G_{(2\textbf{i}-1)(2\textbf{j})}(x^0,y^0,\textbf{k}-\textbf{q})\,,
\label{eq:Sie2}
\\
\Sigma_{(2\textbf{i})(2\textbf{j}-1)}(x^0,y^0,\textbf{k})=&-\int_\textbf{q} S_{s^\prime v^\prime}^{\alpha\beta}(\textbf{q}-\textbf{k})\bar{B}_{s^\prime v^\prime}^{\alpha\beta}(y^0,x^0,-\textbf{q})G_{(2\textbf{i})(2\textbf{j}-1)}(x^0,y^0,\textbf{k}-\textbf{q})\,,
\label{eq:Sie3}
\end{align}
and Eq.~(\ref{eq:tPIPI}) as 
\begin{align}
\Pi_{v^\prime s^\prime}^{\beta\alpha}(x^0,y^0,\textbf{k})=&\,-\int_\textbf{q}\bigg(S_{s^\prime v^\prime}^{\alpha\beta}(\textbf{q},-\textbf{k})G_{(2\textbf{j}-1)(2\textbf{i}-1)}(y^0,x^0,\textbf{q})G_{(2\textbf{i}-1)(2\textbf{j}-1)}(x^0,y^0,\textbf{k}+\textbf{q})\nonumber\\
&\,+S_{s^\prime v^\prime}^{\alpha\beta}(\textbf{q})G_{(2\textbf{j})(2\textbf{i}-1)}(y^0,x^0,\textbf{q})G_{(2\textbf{i}-1)(2\textbf{j})}(x^0,y^0,\textbf{k}+\textbf{q})\bigg)\,.\label{eq:Pisi}
\end{align}
Two self-energies $\Lambda_{i v^\prime}^\alpha$ and $\Lambda_{v^\prime i}^\alpha$ are vanishing at the NLO in the 2PI since they come from the next-to-next-to-leading order (NNLO) contributions to the effective action. 

Analogous to Eq.~(\ref{eq:Gdic}), self-energies can also be decomposed into $F/\rho$ parts identified by sections with/without ${\rm sgn}_\mathcal{C}$ in the same form. For Eq.~(\ref{eq:sdjk}), (\ref{eq:Siev}), (\ref{eq:Sie2}), (\ref{eq:Sie3}), inserting (\ref{eq:Bfln}) leads to
\begin{align}
\Sigma_{nm}(x^0,z^0,\textbf{k})=&\,\tilde{\Sigma}_{0\,nm}(x^0,x^0,\textbf{k})\,\delta_\mathcal{C}(x^0-z^0)+\tilde{\Sigma}_{nm}(x^0,z^0,\textbf{k})\,,
\end{align}
where
\begin{align}
\tilde{\Sigma}_{0\,nm}(x^0,x^0,\textbf{k})\equiv&\,
\begin{cases}
i\frac{\bar{\lambda}}{N}\int_\textbf{q}S_{s^\prime v^\prime}^{\alpha\beta}(\textbf{k},\textbf{q})A_{s^\prime v^\prime}^{\alpha\beta}(-\textbf{q})G_{nm}(x^0,x^0,\textbf{k}-\textbf{q})\,, \quad n,\,m \,\,{\rm odd}\,,\label{eq:locs}\\
i\frac{\bar{\lambda}}{N}\int_\textbf{q}S_{s^\prime v^\prime}^{\alpha\beta}(\textbf{q}-\textbf{k},\textbf{q})A_{s^\prime v^\prime}^{\alpha\beta}(-\textbf{q})G_{nm}(x^0,x^0,\textbf{k}-\textbf{q})\,, \quad n,\,m \,\,{\rm even} \,,\\
i\frac{\bar{\lambda}}{N}\int_\textbf{q}S_{s^\prime v^\prime}^{\alpha\beta}(\textbf{k})A_{s^\prime v^\prime}^{\alpha\beta}(-\textbf{q})G_{(2\textbf{i}-1)(2\textbf{j})}(x^0,y^0,\textbf{k}-\textbf{q})\,, \quad  n\,\,{\rm odd}\,,\,m \,\,{\rm even}\,,\\
i\frac{\bar{\lambda}}{N}\int_\textbf{q}S_{s^\prime v^\prime}^{\alpha\beta}(\textbf{q}-\textbf{k})A_{s^\prime v^\prime}^{\alpha\beta}(-\textbf{q})G_{(2\textbf{i})(2\textbf{j}-1)}(x^0,y^0,\textbf{k}-\textbf{q})\,, \quad  n\,\,{\rm even}\,,\,m \,\,{\rm odd}\,,
\end{cases}
\\
\tilde{\Sigma}_{nm}(x^0,z^0,\textbf{k})\equiv &\,
\begin{cases}
-\frac{\bar{\lambda}}{N}\int_\textbf{q}S_{s^\prime v^\prime}^{\alpha\beta}(\textbf{k},\textbf{q})B_{s^\prime v^\prime}^{\alpha\beta}(z^0,x^0,-\textbf{q}) G_{nm}(x^0,z^0,\textbf{k}-\textbf{q})\,,\quad n,\,m\,\, {\rm odd}\,,\\
-\frac{\bar{\lambda}}{N}\int_\textbf{q}S_{s^\prime v^\prime}^{\alpha\beta}(\textbf{q}-\textbf{k},\textbf{q})B_{s^\prime v^\prime}^{\alpha\beta}(z^0,x^0,-\textbf{q})G_{nm}(x^0,z^0,\textbf{k}-\textbf{q})\,,\quad n,\,m\,\, {\rm even}\,,\\
-\frac{\bar{\lambda}}{N}\int_\textbf{q}S_{s^\prime v^\prime}^{\alpha\beta}(\textbf{k})B_{s^\prime v^\prime}^{\alpha\beta}(z^0,x^0,-\textbf{q})G_{nm}(x^0,z^0,\textbf{k}-\textbf{q})\,,\quad n\,\,{\rm odd}\,,\,m \,\,{\rm even}\,,\\
-\frac{\bar{\lambda}}{N}\int_\textbf{q}S_{s^\prime v^\prime}^{\alpha\beta}(\textbf{q}-\textbf{k})B_{s^\prime v^\prime}^{\alpha\beta}(z^0,x^0,-\textbf{q})G_{nm}(x^0,z^0,\textbf{k}-\textbf{q})\,,\quad n\,\,{\rm even}\,,\,m \,\,{\rm odd}\,.
\end{cases}
\label{eq:tisi}
\end{align}
Their further decomposition into the $F/\rho$ can be accomplished by decomposing $G_{nm}$ and the non-local part of Eq.~(\ref{eq:Bfln}). Detailed expressions of self-energies in a decomposed form by $F/\rho$ can be found in Appendix.~\ref{sec:fide}.

Consequently, this decomposition implemented in Eq.~(\ref{eq:nev1}) at NLO leads to two evolution equations for $F_G$ and $\rho_G$ given by
\begin{align}
&\,\bigg[\Big(\partial_{x^0}^2+m^2\Big)\delta_{ml}+M_{ml}(x^0,x^0,\textbf{k})+i\tilde{\Sigma}_{0\,ml}(x^0,x^0,\textbf{k})\bigg]F_{G_{ln}}(x^0,y^0,\textbf{k})\nonumber\\
=&\,\int_{t^0}^{y^0}dz^0\,\tilde{\Sigma}_{ml}^F(x^0,z^0,\textbf{k})\rho_{G_{ln}}(z^0,y^0,\textbf{k})-\int_{t^0}^{x^0}dz^0\,\tilde{\Sigma}_{ml}^\rho(x^0,z^0,\textbf{k})F_{G_{ln}}(z^0,y^0,\textbf{k})\nonumber\\
&\,-\tilde{\Omega}_{ms^\prime}^\alpha(x^0,\textbf{k})F_{K_{s^\prime n}}^\alpha(x^0,y^0,\textbf{k})\,,\label{eq:FKev}\\
&\,\bigg[\Big(\partial_{x^0}^2+m^2\Big)\delta_{ml}+M_{ml}(x^0,x^0,\textbf{k})+i\tilde{\Sigma}_{0\,ml}(x^0,x^0,\textbf{k})\bigg]\rho_{G_{ln}}(x^0,y^0,\textbf{k})\nonumber\\
=&\,-\int_{y^0}^{x^0}dz^0\,\tilde{\Sigma}_{ml}^\rho(x^0,z^0,\textbf{k})\rho_{G_{ln}}(z^0,y^0,\textbf{k})-\tilde{\Omega}_{ms^\prime}^\alpha(x^0,\textbf{k})\rho_{K_{s^\prime n}}^\alpha(x^0,y^0,\textbf{k})\,,\label{eq:rKev}
\end{align}
respectively, where calculations employ Eq.~(\ref{eq:coid}), (\ref{eq:comu}), identities of integral on $\mathcal{C}$
\begin{align}
&\,\int_{z^0,\mathcal{C}}{\rm sgn}_\mathcal{C}(z^0-y^0)=-2\int_{t^0}^{y^0}dz^0\,,\quad\int_{z^0,\mathcal{C}}{\rm sgn}_{\mathcal{C}}(x^0-z^0)=2\int_{t^0}^{x^0}dz^0\,,\label{eq:cosg}\\
&\,\int_{z^0,\mathcal{C}}{\rm sgn}_\mathcal{C}(x^0-z^0)\,{\rm sgn}_\mathcal{C}(z^0-y^0)=2\int_{y^0}^{x^0}dz^0\,{\rm sgn}_\mathcal{C}(x^0-y^0)\,,\label{eq:coto}
\end{align}
and $\delta_\mathcal{C}(x^0-y^0)$ is removed by $\partial_{x^0}{\rm sgn}_\mathcal{C}(x^0-y^0)=2\delta_\mathcal{C}(x^0-y^0)$. The field decomposition of $K_{s^\prime n}^\alpha$ can be found in Appendix.~\ref{sec:fide} by the symmetry properties in Table.~\ref{tab:FKre}. For more detailed derivations of equations from Eq.~(\ref{eq:FKev}) to (\ref{eq:coto}), see Ref.~\cite{Berges:2004yj,berges2015nonequilibriumquantumfieldscold}. These two differential equations are coupled by memory integrals that sum up all the information in the evolution from the initial time $t^0$ to $x^0$ or $y^0$. Given the complexity from the coupled terms, analytical solutions of Eq.~(\ref{eq:FKev}) and (\ref{eq:rKev}) are not approachable; the discretized time step along with the spatial lattice is a feasible set up to simulate evolutions of $F_G$ and $\rho_G$ numerically.

\subsection{Alternative formulation in non-equilibrium dynamics}

In Sec.~\ref{sec:2PIe}, the evolution equations are constructed in the auxiliary field method. They are a bit different from the standard Kadanoff-Baym equations but are more convenient in numerical calculations \cite{Aarts:2002dj}. In this section, I show alternative forms of evolution equations of $\phi_\textbf{i}(x^0)$ in Eq.~(\ref{eq:feau}) and two-point functions in Eq.~(\ref{eq:FKev}) and (\ref{eq:rKev}) by re-writing functions $\bar{B}_{s^\prime v^\prime}^{\alpha\beta}$, $K_{v^\prime n}^\alpha$, $\tilde{K}_{ns^\prime}^\alpha$ shown in Sec.~\ref{sec:2PIe}. The auxiliary fields $K_{v^\prime n}^\alpha$ in the evolution equations are absorbed and combined into self-energies $\Sigma_{mn}$ by the redefinition, by which the standard Kadanoff-Baym equations are restored. Consequently, the vertex-resummed kinetic theory is built up by reducing the Kadanoff-Baym equations at the late time.

It is noticed that by re-writing Eq.~(\ref{eq:nev4}) at the NLO as
\begin{align} &\,\int_{z^0,\mathcal{C}}\bigg[\mathcal{I}_{v^\prime s^\prime}^{\alpha\beta}\delta_\mathcal{C}(x^0-z^0)+i\frac{\bar{\lambda}}{N}A_{v^\prime h^\prime}^{\alpha\gamma}(\textbf{k})\Pi_{h^\prime s^\prime}^{\gamma\beta}(x^0,z^0,\textbf{k})\bigg]K_{s^\prime n}^\beta(z^0,y^0,\textbf{k})\nonumber\\
\quad =&\,-\frac{\bar{\lambda}}{N}A_{v^\prime h^\prime}^{\alpha\beta}(\textbf{k})\Omega_{h^\prime m}^\beta(x^0,\textbf{k})G_{mn}(x^0,y^0,\textbf{k})\,,
\label{eq:watK}
\end{align}
and defining a function $R_{v^\prime s^\prime}^{\alpha\beta}(x^0,y^0,\textbf{k})$ and its inverse
\begin{align}
R_{v^\prime s^\prime}^{\alpha\beta}(x^0,y^0,\textbf{k})\equiv&\,\mathcal{I}_{v^\prime s^\prime}^{\alpha\beta}\delta_\mathcal{C}(x^0-y^0)+i\frac{\bar{\lambda}}{N}A_{v^\prime h^\prime}^{\alpha\gamma}(\textbf{k})\Pi_{h^\prime s^\prime}^{\gamma\beta}(x^0,y^0,\textbf{k})\,,
\label{eq:Rdef} \\
(R^{-1})_{v^\prime s^\prime}^{\alpha\beta}(x^0,y^0,\textbf{k})\equiv&\,\mathcal{I}_{v^\prime s^\prime}^{\alpha\beta}\delta_\mathcal{C}(x^0-y^0)-i{J}_{v^\prime s^\prime}^{\alpha\beta}(x^0,y^0,\textbf{k})\,.
\label{eq:Rinv}
\end{align}
The two-point function $J_{v^\prime s^\prime}^{\alpha\beta}(x^0,y^0,\textbf{k})$ is obtained by the convolution of Eq.~(\ref{eq:Rdef}) with (\ref{eq:Rinv}) 
\begin{align}
J_{v^\prime s^\prime}^{\alpha\beta}(x^0,y^0,\textbf{k})=\frac{\bar{\lambda}}{N}A_{v^\prime h^\prime}^{\alpha\gamma}(\textbf{k})\Pi_{h^\prime s^\prime}^{\gamma\beta}(x^0,y^0,\textbf{k})-i\frac{\bar{\lambda}}{N}\int_{z^0,\mathcal{C}}J_{v^\prime h^\prime}^{\alpha\mu}(x^0,z^0)A_{h^\prime l^\prime}^{\mu\nu}(\textbf{k})\Pi_{l^\prime s^\prime}^{\nu\beta}(z^0,y^0,\textbf{k})\,.
\label{eq:JJeq}
\end{align}
Therefore, the two-point function $K_{v^\prime n}^\alpha(x^0,y^0,\textbf{k})$ can be expressed as
\begin{align}
K_{v^\prime n}^\alpha(x^0,y^0,\textbf{k})=-\frac{\bar{\lambda}}{N}\int_{z^0,\mathcal{C}}(R^{-1})_{v^\prime s^\prime}^{\alpha\beta}(x^0,z^0,\textbf{k})A_{s^\prime h^\prime}^{\beta\gamma}(\textbf{k})\Omega_{h^\prime m}^\gamma(z^0,\textbf{k})G_{mn}(z^0,y^0,\textbf{k})\,.
\label{eq:Krel}
\end{align}
by convoluting Eq.~(\ref{eq:watK}) with Eq.~(\ref{eq:Rinv}). Subsequently, Eq.~(\ref{eq:nev1}) at the NLO becomes
\begin{align}
&\,\delta_{mn}\delta_\mathcal{C}(x^0-y^0)\nonumber\\
=&\,i\bigg[\Big(\partial_{x^0}^2+m^2\Big)\delta_{ml}+M_{ml}(x^0,x^0,\textbf{k})\bigg]G_{ln}(x^0,y^0,\textbf{k})-\int_{z^0,\mathcal{C}}\Sigma_{ml}(x^0,z^0,\textbf{k})G_{ln}(z^0,y^0,\textbf{k})\nonumber\\
&\,-i\frac{\bar{\lambda}}{N}\tilde{\Omega}_{ms^\prime}^\alpha(x^0,\textbf{k})A_{s^\prime h^\prime}^{\alpha\beta}(\textbf{k})\Omega_{h^\prime l}^\beta(x^0,\textbf{k})G_{ln}(x^0,y^0,\textbf{k}) \nonumber\\
&\,-\frac{\bar{\lambda}}{N}\int_{u^0,\mathcal{C}}\tilde{\Omega}_{ms^\prime}^\alpha(x^0,\textbf{k})\bar{J}_{s^\prime v^\prime}^{\alpha\beta}(x^0,u^0,\textbf{k})\Omega_{v^\prime l}^\beta(u^0,\textbf{k})G_{ln}\,(u^0,y^0,\textbf{k})\,,
\label{eq:reKa}
\\
&\,\quad\quad\quad\quad\quad\quad\quad\quad\quad\bar{J}_{s^\prime v^\prime}^{\alpha\beta}(x^0,y^0,\textbf{k})\equiv J_{s^\prime c^\prime}^{\alpha\gamma}(x^0,y^0,\textbf{k})A_{c^\prime v^\prime}^{\gamma\beta}(\textbf{k})
\end{align}
by plugging in Eq.~(\ref{eq:Krel}) and (\ref{eq:Rinv}).

Similarly, it is able to get an expression of $\bar{B}_{v^\prime s^\prime }^{\alpha\beta}$ at the NLO by Eq.~(\ref{eq:nev2}) and (\ref{eq:Rinv}) given by
\begin{align}
&\,-i\frac{\bar{\lambda}}{N}(R^{-1})_{v^\prime h^\prime}^{\alpha\gamma}(x^0,y^0,\textbf{k})A_{h^\prime s^\prime}^{\gamma\beta}(\textbf{k})\nonumber\\
=&\,\frac{\bar{\lambda}}{N}\int_{z^0,\mathcal{C}}(R^{-1})_{v^\prime l^\prime}^{\alpha\mu}(x^0,z^0,\textbf{k})A_{l^\prime h^\prime}^{\mu\nu}(\textbf{k})\Omega_{h^\prime n}^\nu(z^0,\textbf{k})\tilde{K}_{n s^\prime}^\beta(z^0,y^0,\textbf{k})+\bar{B}_{v^\prime s^\prime}^{\alpha\beta}(x^0,y^0,\textbf{k})\,.
\label{eq:bste}
\end{align}
After inserting Eq.~(\ref{eq:Krel}) into Eq.~(\ref{eq:bste}), one obtains
\begin{align}
\bar{B}_{v^\prime s^\prime}^{\alpha\beta}(x^0,y^0,\textbf{k})
=&\,-i\frac{\bar{\lambda}}{N}A_{v^\prime s^\prime}^{\alpha\beta}(\textbf{k})\delta_\mathcal{C}(x^0-y^0)-\frac{\bar{\lambda}}{N}\bar{J}_{v^\prime s^\prime}^{\alpha\beta}(x^0,y^0,\textbf{k})\nonumber\\
&\,+\bigg(\frac{\bar{\lambda}}{N}\bigg)^2\int_{z^0,u^0,\mathcal{C}}(R^{-1})_{v^\prime l^\prime}^{\alpha\mu}(x^0,z^0,\textbf{k})A_{l^\prime h^\prime}^{\mu \nu}(\textbf{k})(R^{-1})_{s^\prime w^\prime}^{\beta \gamma}(y^0,u^0,-\textbf{k})A_{w^\prime c^\prime}^{\gamma \sigma}(-\textbf{k})\nonumber\\
&\,\times\Omega_{c^\prime m}^\sigma(u^0,-\textbf{k})G_{mn}(u^0,z^0,-\textbf{k})\tilde{\Omega}_{n h^\prime}^\nu(z^0,-\textbf{k})
\label{eq:Bbar}
\end{align}
by $\Omega_{h^\prime n}^\nu(z^0,\textbf{k})\tilde{K}_{n s^\prime}^\beta(z^0,y^0,\textbf{k})=K_{s^\prime n}^\beta(y^0,z^0,-\textbf{k})\tilde{\Omega}_{n h^\prime}^\nu(z^0,-\textbf{k})$ according to Eq.~(\ref{eq:Ome1}), (\ref{eq:Ome2}), (\ref{eq:Ome3}) and Table.~\ref{tab:FKre}. Furthermore, Eq.~(\ref{eq:Bbar}) can be re-written as
\begin{align}
\bar{B}_{v^\prime s^\prime}^{\alpha\beta}(x^0,y^0,\textbf{k})
=&\,-i\frac{\bar{\lambda}}{N}(R^{-1})_{v^\prime h^\prime}^{\alpha\gamma}(x^0,y^0,\textbf{k})A_{h^\prime s^\prime}^{\gamma\beta}(\textbf{k})+\bigg(\frac{\bar{\lambda}}{N}\bigg)^2\int_{z^0,u^0,\mathcal{C}}(R^{-1})_{v^\prime q^\prime}^{\alpha\mu}(x^0,z^0,\textbf{k})A_{q^\prime h^\prime}^{\mu\nu}(\textbf{k})\nonumber\\
&\,\times\Omega_{h^\prime n}^\nu(z^0,\textbf{k})G_{nm}(z^0,u^0,\textbf{k})\tilde{\Omega}_{m c^\prime}^\sigma(u^0,\textbf{k})(R^{-1})_{c^\prime w^\prime}^{\sigma\gamma}(u^0,y^0,\textbf{k})A_{w^\prime s^\prime}^{\gamma\beta}(\textbf{k})
\label{eq:bbdef}
\end{align}
by symmetry properties of $\bar{B}_{v^\prime s^\prime}^{\alpha\beta}$, $\tilde{\Omega}_{h^\prime n}^\nu$, $\Omega_{mc^\prime}^\gamma$ and $G_{nm}$. In addition, self-energies Eq.~(\ref{eq:sdjk}) and (\ref{eq:Siev}) can also be re-written by defining Eq.~(\ref{eq:bbdef}) as
\begin{align}
\bar{B}_{s^\prime v^\prime}^{\alpha\beta}(x^0,y^0,-\textbf{q})
\equiv-i\frac{\bar{\lambda}}{N}A_{s^\prime v^\prime}^{\alpha\beta}(-\textbf{q})\delta_\mathcal{C}(x^0-y^0)-\frac{\bar{\lambda}}{N}\Big(\bar{J}_{s^\prime v^\prime}^{\alpha\beta}(x^0,y^0,-\textbf{q})+W_{s^\prime v^\prime}^{\alpha\beta}(x^0,y^0,-\textbf{q})\Big)
\label{eq:bwjb}
\end{align}
with Eq.~(\ref{eq:Rinv}) and
\begin{align}
W_{s^\prime v^\prime}^{\alpha\beta}(x^0,y^0,-\textbf{q})
\equiv&\,-\frac{\bar{\lambda}}{N}\int_{v^0,u^0,\mathcal{C}}(R^{-1})_{s^\prime l^\prime}^{\alpha\mu}(x^0,v^0,-\textbf{q})A_{l^\prime h^\prime}^{\mu\nu}(-\textbf{q})Q_{h^\prime c^\prime}^{\nu\gamma}(v^0,u^0,-\textbf{q})\nonumber\\&\,\times(R^{-1})_{c^\prime w^\prime}^{\gamma\sigma}(u^0,y^0,-\textbf{q})A_{w^\prime v^\prime}^{\sigma\beta}(-\textbf{q})\,, \label{eq:WWde}\\
Q_{h^\prime c^\prime}^{\nu\gamma}(v^0,u^0,-\textbf{q})
\equiv&\,\Omega_{h^\prime n}^\nu(v^0,-\textbf{q})G_{nm}(v^0,u^0,-\textbf{q})\tilde{\Omega}_{mc^\prime}^\gamma(u^0,-\textbf{q})\label{eq:QQde}
\end{align}
with the symmetry properties given by
\begin{align}
Q_{h^\prime c^\prime}^{\nu\gamma}(v^0,u^0,-\textbf{q})=&\,Q_{c^\prime h^\prime}^{\gamma\nu}(u^0,v^0,\textbf{q})\,,\\
W_{s^\prime v^\prime}^{\alpha\beta}(x^0,y^0,-\textbf{q})=&\, W_{v^\prime s^\prime}^{\beta\alpha}(y^0,x^0,\textbf{q})\,.
\end{align}
The field decompositions of Eq.~(\ref{eq:WWde}) and (\ref{eq:QQde}) are provided in Appendix.~\ref{sec:fide}.

Accordingly, it is allowed to re-arrange Eq.~(\ref{eq:reKa}) as
\begin{align}
\delta_{mn}\delta_\mathcal{C}(x^0-y^0)=&\,i\bigg[\Big(\partial_{x^0}^2+m^2\Big)\delta_{ml}+M_{ml}(x^0,x^0,\textbf{k})+i\bar{\Sigma}_{0 \,ml}(x^0,x^0,\textbf{k})\bigg]G_{ln}(x^0,y^0,\textbf{k})\nonumber\\
&\,-\int_{z^0,\mathcal{C}}\bar{\Sigma}_{ml}(x^0,z^0,\textbf{k})G_{ln}(z^0,y^0,\textbf{k})\,.
\label{eq:beKa}
\end{align}
By comparing Eq.~(\ref{eq:reKa}) with (\ref{eq:beKa}), new self-energies in Eq.~(\ref{eq:beKa}) are defined as follows
\begin{align}
\bar{\Sigma}_{0 \,nm}(x^0,x^0,\textbf{k})
\equiv&\, i\frac{\bar{\lambda}}{N}\tilde{\Omega}_{ns^\prime}^\alpha(x^0,\textbf{k})A_{s^\prime h^\prime}^{\alpha\beta}(\textbf{k})\Omega_{h^\prime m}^\beta(x^0,\textbf{k})+\tilde{\Sigma}_{0\,nm}(x^0,x^0,\textbf{k})\,,\label{eq:Sib1}\\
\bar{\Sigma}_{nm}(x^0,z^0,\textbf{k})\equiv &\,\frac{\bar{\lambda}}{N}\tilde{\Omega}_{ns^\prime}^\alpha(x^0,\textbf{k})\bar{J}_{s^\prime v^\prime}^{\alpha\beta}(x^0,z^0,\textbf{k})\Omega_{v^\prime m}^\beta(z^0,\textbf{k})+\tilde{\Sigma}_{nm}(x^0,z^0,\textbf{k})\,,\label{eq:Sib2}
\end{align}
where
\begin{align}
\tilde{\Sigma}_{nm}(x^0,z^0,\textbf{k})\equiv 
\begin{cases}
&\frac{\bar{\lambda}}{N}\int_\textbf{q}S_{s^\prime v^\prime}^{\alpha\beta}(\textbf{k},\textbf{q})\Big(\bar{J}_{s^\prime v^\prime}^{\alpha\beta}(z^0,x^0,-\textbf{q})+W_{s^\prime v^\prime}^{\alpha\beta}(z^0,x^0,-\textbf{q})\Big)\label{eq:mosi}\\
&\times G_{nm}(x^0,z^0,\textbf{k}-\textbf{q})\,,\quad n,\,m\,\, {\rm odd}\,,\\
&\frac{\bar{\lambda}}{N}\int_\textbf{q}S_{s^\prime v^\prime}^{\alpha\beta}(\textbf{q}-\textbf{k},\textbf{q})\Big(\bar{J}_{s^\prime v^\prime}^{\alpha\beta}(z^0,x^0,-\textbf{q})+W_{s^\prime v^\prime}^{\alpha\beta}(z^0,x^0,-\textbf{q})\Big)\\
&\times G_{nm}(x^0,z^0,\textbf{k}-\textbf{q})\,,\quad n,\,m\,\, {\rm even}\,,\\
&\frac{\bar{\lambda}}{N}\int_\textbf{q}S_{s^\prime v^\prime}^{\alpha\beta}(\textbf{k})\Big(\bar{J}_{s^\prime v^\prime}^{\alpha\beta}(z^0,x^0,-\textbf{q})+W_{s^\prime v^\prime}^{\alpha\beta}(z^0,x^0,-\textbf{q})\Big)\\
&\times G_{nm}(x^0,z^0,\textbf{k}-\textbf{q})\,,\quad n\,\,{\rm odd}\,,\,\,m\,\,{\rm even}\,,\\
&\frac{\bar{\lambda}}{N}\int_\textbf{q}S_{s^\prime v^\prime}^{\alpha\beta}(\textbf{q}-\textbf{k})\Big(\bar{J}_{s^\prime v^\prime}^{\alpha\beta}(z^0,x^0,-\textbf{q})+W_{s^\prime v^\prime}^{\alpha\beta}(z^0,x^0,-\textbf{q})\Big)\\
&\times G_{nm}(x^0,z^0,\textbf{k}-\textbf{q})\,,\quad n\,\,{\rm even}\,,\,\,m\,\,{\rm odd}
\end{cases}
\end{align}
re-expresses Eq.~(\ref{eq:tisi}) in terms of the comparison between Eq.~(\ref{eq:Bfln}) and (\ref{eq:bwjb}). By plugging in Eq.~(\ref{eq:Gdic}),  $\bar{\Sigma}_{nm}^F$ and $\bar{\Sigma}_{nm}^\rho$ extracted from $\bar{\Sigma}_{nm}$ into Eq.~(\ref{eq:beKa}), the Kadanoff-Baym equations are derived 
\begin{align}
&\,\bigg[\Big(\partial_{x^0}^2+m^2\Big)\delta_{ml}+M_{ml}(x^0,x^0,\textbf{k})+i\bar{\Sigma}_{0\,ml}(x^0,x^0,\textbf{k})\bigg]F_{G_{ln}}(x^0,y^0,\textbf{k})\nonumber\\
=&\,\int_{t^0}^{y^0}dz^0\,\bar{\Sigma}_{ml}^F(x^0,z^0,\textbf{k})\rho_{G_{ln}}(z^0,y^0,\textbf{k})-\int_{t^0}^{x^0}dz^0\,\bar{\Sigma}_{ml}^\rho(x^0,z^0,\textbf{k})F_{G_{ln}}(z^0,y^0,\textbf{k})\,,\label{eq:KaFe}\\
&\,\bigg[\Big(\partial_{x^0}^2+m^2\Big)\delta_{ml}+M_{ml}(x^0,x^0,\textbf{k})+i\bar{\Sigma}_{0\,ml}(x^0,x^0,\textbf{k})\bigg]\rho_{G_{ln}}(x^0,y^0,\textbf{k})\nonumber\\
=&\,-\int_{y^0}^{x^0}dz^0\,\bar{\Sigma}_{ml}^\rho(x^0,z^0,\textbf{k})\rho_{G_{ln}}(z^0,y^0,\textbf{k})\,.
\label{eq:Kare}
\end{align}
Extra terms generated by auxiliary fields in Eq.~(\ref{eq:FKev}) and (\ref{eq:rKev}) compared to Eq.~(\ref{eq:KaFe}) and (\ref{eq:Kare}) are absorbed by self-energies $\bar{\Sigma}_{0\,nm}$ and $\bar{\Sigma}_{nm}$. The reminiscence of auxiliary fields is connected to non-zero $\phi_\textbf{i}(x^0)$ and encoded in the first term on the right-hand side of Eq.~(\ref{eq:Sib1}) and (\ref{eq:Sib2}). Therefore, the alternative formulation in the non-equilibrium dynamics in this section replaces the dependence of $\bar{B}_{s^\prime v^\prime}^{\alpha\beta}$, $K_{v^\prime n}^\alpha$, $\tilde{K}_{ns^\prime}^\alpha$ in Sec.~\ref{sec:2PIe} with $\bar{J}_{s^\prime v^\prime}^{\alpha\beta}$ defined by the inverse of $R_{s^\prime v^\prime}^{\alpha\beta}$ shown in Eq.~(\ref{eq:Rinv}) \cite{Berges:2001fi,Aarts:2002dj}. Non-local self-energies $\bar{\Sigma}_{nm}$ are more complicated in expressions due to $W_{s^\prime v^\prime}^{\alpha\beta}$ in its analytical expression.

Similarly, the evolution equation of $\phi_\textbf{i}^*(x^0)$ in Eq.~(\ref{eq:feau}) can also be converted to an alternative form without the auxiliary field $K_{v^\prime n}^\alpha$. I employ Eq.~(\ref{eq:Krel}) and (\ref{eq:bste})
\begin{align}
-K_{v^\prime (2\textbf{i}-1)}^\alpha(x^0,x^0,\textbf{k})
=&\,i\int_{z^0,\mathcal{C}}\bigg[\bar{B}_{v^\prime 1^\prime}^{\alpha 0}(x^0,z^0,\textbf{k};\phi_\textbf{i}=0)\phi_\textbf{j}^*(z^0)G_{(2\textbf{i})(2\textbf{j})}(x^0,z^0,-\textbf{k})\nonumber\\
&\,+\bigg(\bar{B}_{v^\prime1^\prime}^{\alpha0}(x^0,z^0,\textbf{k};\phi_\textbf{i}=0)-k^\beta\bar{B}_{v^\prime2^\prime}^{\alpha\beta}(x^0,z^0,\textbf{k};\phi_\textbf{i}=0)\nonumber\\
&\,+\textbf{k}^2\bar{B}_{v^\prime3^\prime}^{\alpha0}(x^0,z^0,\textbf{k};\phi_\textbf{i}=0)\bigg)\phi_\textbf{j}(z^0)G_{(2\textbf{i})(2\textbf{j}-1)}(x^0,z^0,-\textbf{k})\bigg]
\label{eq:Kevo}
\end{align}
in Eq.~(\ref{eq:feau}). Surprisingly, it is observed that Eq.~(\ref{eq:feau}) re-written by Eq.~(\ref{eq:Kevo}) shares a common expression with Eq.~(\ref{eq:Siev}) and (\ref{eq:Sie3})
\begin{align}
&\,-\Sigma_{(2\textbf{i})(2\textbf{j})}(x^0,z^0,0;\phi_\textbf{i}=0)\nonumber\\
=&\,\int_\textbf{k}\bigg(\bar{B}_{1^\prime 1^\prime}^{00}(x^0,z^0,\textbf{k})+k^\alpha\bar{B}_{2^\prime 1^\prime}^{\alpha0}(x^0,z^0,\textbf{k})+\textbf{k}^2\bar{B}_{3^\prime 1^\prime}^{00}(x^0,z^0,\textbf{k})\bigg)G_{(2\textbf{i})(2\textbf{j})}(x^0,z^0,-\textbf{k})\,, \\
&-\Sigma_{(2\textbf{i})(2\textbf{j}-1)}(x^0,z^0,0;\phi_\textbf{i}=0)\nonumber\\
=&\,\int_\textbf{k}\bigg(\bar{B}_{1^\prime 1^\prime}^{00}(x^0,z^0,\textbf{k})+k^\alpha\bar{B}_{2^\prime 1^\prime}^{\alpha0}(x^0,z^0,\textbf{k})+\textbf{k}^2\bar{B}_{3^\prime 1^\prime}^{00}(x^0,z^0,\textbf{k})\nonumber\\
&\,-k^\alpha \bar{B}_{1^\prime 2^\prime}^{0\alpha}(x^0,z^0,\textbf{k})-k^\alpha k^\beta\bar{B}_{2^\prime 2^\prime}^{\beta\alpha}(x^0,z^0,\textbf{k})-k^\alpha\textbf{k}^2 \bar{B}_{3^\prime 2^\prime}^{0\alpha}(x^0,z^0,\textbf{k})\nonumber\\
&\,+\textbf{k}^2\bar{B}_{1^\prime 3^\prime}^{00}(x^0,z^0,\textbf{k})+k^\alpha\textbf{k}^2\bar{B}_{2^\prime 3^\prime}^{\alpha0}(x^0,z^0,\textbf{k})+\textbf{k}^2\textbf{k}^2\bar{B}_{3^\prime 3^\prime}^{00}(x^0,z^0,\textbf{k})\bigg)G_{(2\textbf{i})(2\textbf{j}-1)}(x^0,z^0,-\textbf{k})
\end{align}
where definitions from Eq.~(\ref{eq:Sre1}) to (\ref{eq:Sree4}) are used. Therefore, the evolution equation of $\phi_\textbf{i}^*(x^0)$ in the presence of non-vanishing scatterings indicated by self-energies $\Sigma_{nm}$ is given by
\begin{align}
&\,\bigg[\bigg(\partial_{x^0}^2+m^2+\tilde{M}^2(x^0,x^0,0)\bigg)\delta_{\textbf{i}\textbf{j}}-\frac{2\bar{\lambda}}{N}\int_\textbf{k}\textbf{k}^2\Big(G_{(2\textbf{i})(2\textbf{j})}(x^0,x^0,\textbf{k})+2G_{(2\textbf{i})(2\textbf{j}-1)}(x^0,x^0,\textbf{k})\Big)\bigg]\phi_\textbf{j}^*(x^0)\nonumber\\
=&\,-\int_{t^0}^{x^0}dz^0\,\tilde{\Sigma}_{(2\textbf{i})(2\textbf{j})}^\rho(x^0,z^0,0;\phi_\textbf{i}=0)\phi_\textbf{j}^*(z^0)-\int_{t^0}^{x^0}dz^0\,\tilde{\Sigma}_{(2\textbf{i})(2\textbf{j}-1)}^\rho(x^0,z^0,0;\phi_\textbf{i}=0)\phi_\textbf{j}(z^0)
\end{align}
using Eq.~(\ref{eq:cosg}). Similarly, one can also derive the equation of motion of $\phi_\textbf{i}(x^0)$
\begin{align}
&\,\Bigg[\bigg(\partial_{X^0}^2+m^2+\tilde{M}(x^0,x^0,0)\bigg)\delta_{\textbf{i}\textbf{j}}-\frac{2\bar{\lambda}}{N}\int_\textbf{k}\textbf{k}^2\Big(G_{(2\textbf{i}-1)(2\textbf{j}-1)}(x^0,x^0,\textbf{k})+2G_{(2\textbf{i}-1)(2\textbf{j})}(x^0,x^0,\textbf{k})\Big)\bigg]\phi_\textbf{j}(x^0)\nonumber\\
=&\,-\int_{t_0}^{x^0}dz^0\,\tilde{\Sigma}_{(2\textbf{i}-1)(2\textbf{j}-1)}^\rho(x^0,z^0,0;\phi_\textbf{i}=0)\phi_\textbf{j}(z^0)-\int_{t_0}^{x^0}dz^0\,\tilde{\Sigma}_{(2\textbf{i}-1)(2\textbf{j})}^\rho(x^0,z^0,0;\phi_\textbf{i}=0)\phi_\textbf{j}^*(z^0)\,.
\end{align}

\section{Vertex-resummed kinetic theory} \label{sec:vere}

In this section, the late-time behavior of Kadanoff-Baym equations are considered to derive the kinetic equation. The NLO corrections in the $1/N$ expansion are resummed and encoded in the vertex function as an correction to the coupling, leading to the vertex-resummed kinetic equation different from the result in the weak-coupling expansion. The following calculations show in detail the derivations of kinetic equations with a resummed vertex.

\subsection{Transport equation}

The non-equilibrium evolution of two-point functions are described by the Kadanoff-Baym equations in Eq.~(\ref{eq:KaFe}) and (\ref{eq:Kare}). The NLO corrections of self-energies at the RHS play a role of damping oscillations of the two-point function in the time evolution. After a long time period, the amplitude of oscillations become damped and $X^0$-dependence turns out to dominate. It will be sufficient to keep only the leading order in the gradient expansion of $\partial_{X^0}$ to describe the transport phenomena at the late time. To formulate the transport equation from Eq.~(\ref{eq:KaFe}) and (\ref{eq:Kare}), two-point functions of the spatially homogeneous system should be first re-written by central and relative temporal coordinates defined as $X_{xy}^0\equiv(x^0+y^0)/2$ and $s_{xy}^0\equiv x^0-y^0$, respectively. Thus, the Wigner transform of the two-point function, i.e.  $F_{G_{nm}}(x^0,y^0,\textbf{k}),$ is converted to
\begin{align}
    F_{G_{nm}}(X_{xy}^0,k)= \int_{-\infty}^\infty ds_{xy}^0\, e^{ik^0s_{xy}^0}F_{G_{nm}}\bigg(X_{xy}^0+\frac{s_{xy}^0}{2},X_{xy}^0-\frac{s_{xy}^0}{2},\textbf{k}\bigg).
\end{align}
In the following, I will mostly express $X_{xy}^0\equiv X^0 (s_{xy}^0\equiv s^0)$ for the sake of notational simplicity.

Taking Eq.~(\ref{eq:KaFe}) into account and subtracting the same equation with $x\leftrightarrow y$ and $\textbf{k}\rightarrow -\textbf{k}$ at the late time by the limit $t^0\rightarrow-\infty$ as the initial time of the evolution, one of the transport equation emerges by using the Wigner transform and the trace at the LO in the gradient expansion of $\partial_{X^0}$ \cite{Berges:2004yj,berges2015nonequilibriumquantumfieldscold}
\begin{align}
2k^0\frac{\partial}{\partial X^0}{\rm Tr}\Big[F_G(X^0,k)\Big]={\rm Tr}\Big[F_G(X^0,k)\bar{\Sigma}^{\tilde{\rho}}(X^0,k)-\bar{\Sigma}^F(X^0,k)\tilde{\rho}_G(X^0,k)\Big]\,,
\label{eq:kinF}
\end{align}
where I define the retarded (denoted by $R$) and the advanced (denoted by $A$) two-point functions for the function $\tilde{\rho}_{G_{nm}}(X^0,k)\equiv-i\rho_{G_{nm}}(X^0,k)$ 
\begin{align}
(G_R^{\tilde{\rho}})_{mn}(x^0,y^0,\textbf{k})\equiv&\,\theta_\mathcal{C}(x^0-y^0)\tilde{\rho}_{G_{mn}}(x^0,y^0,\textbf{k})\,,\label{eq:Ratt}\\ (G_A^{\tilde{\rho}})_{mn}(x^0,y^0,\textbf{k})\equiv&\,-\theta_\mathcal{C}(y^0-x^0)\tilde{\rho}_{G_{mn}}(x^0,y^0,\textbf{k})\,,
\label{eq:RAde}
\end{align}
and $\bar{\Sigma}_{nm}^{\tilde{\rho}}(X^0,k)\equiv-i\bar{\Sigma}_{nm}^\rho(X^0,k)$
\begin{align}
(\bar{\Sigma}_R^{\tilde{\rho}})_{mn}(x^0,y^0,\textbf{k})\equiv&\,\theta_\mathcal{C}(x^0-y^0)\bar{\Sigma}_{mn}^{\tilde{\rho}}(x^0,y^0,\textbf{k})\,,\label{eq:RAd1}\\ (\bar{\Sigma}_A^{\tilde{\rho}})_{mn}(x^0,y^0,\textbf{k})\equiv&\,-\theta_\mathcal{C}(y^0-x^0)\bar{\Sigma}_{mn}^{\tilde{\rho}}(x^0,y^0,\textbf{k})\,,
\label{eq:RAd2}
\end{align}
respectively. In the derivation of Eq.~(\ref{eq:kinF}) from Eq.~(\ref{eq:KaFe}), for example, the trace of the first integral in the RHS of Eq.~(\ref{eq:KaFe}) after conversion to expressions by $X^0$ and $s^0$ turns out to be
\begin{align}
&\,-i\int_{-\infty}^{\infty} ds_{xy}^0\int_{-\infty}^{\infty}dz^0\int_{-\infty}^\infty\frac{dp^0}{2\pi}\int_{-\infty}^\infty\frac{dp^{\prime 0}}{2\pi}\,e^{ik^0 s_{xy}^0-ip^0 s_{xz}^0-ip^\prime s_{zy}^0}\bar{\Sigma}_{nl}^F(X_{xz}^0,p^0,\textbf{k})(G_A^{\tilde{\rho}})_{ln}(X_{zy}^0,p^{\prime 0},\textbf{k})\nonumber\\
=&\,-i\int_{-\infty}^{\infty}ds_{xy}^0\int_{-\infty}^\infty dz^0\int_{-\infty}^\infty\frac{dp^0}{2\pi}\int_{-\infty}^\infty\frac{dp^{\prime 0}}{2\pi} \,e^{ik^0 s_{xy}^0-ip^0 s_{xz}^0-ip^\prime s_{zy}^0}\bigg[{\rm exp}\bigg(\frac{s_{zy}^0}{2}\frac{\partial}{\partial X_{xy}^0}\bigg)\bar{\Sigma}_{nl}^F(X_{xy}^0,p^0,\textbf{k})\bigg]\nonumber\\
&\,\times\bigg[{\rm exp}\bigg(-\frac{s_{xz}^0}{2}\frac{\partial}{\partial X_{xy}^0}\bigg)(G_A)^{\tilde{\rho}}_{ln}(X_{xy}^0,p^{\prime0},\textbf{k})\bigg]
\label{eq:Foex}
\end{align}
by the Fourier transform with respect to $s_{xy}^0$. Keeping only the LO term in the gradient expansion embedded in the exponential function, Eq.~(\ref{eq:Foex}) reads
\begin{align}
-i\bar{\Sigma}_{nl}^F(X^0,k)(G_A^{\tilde{\rho}})_{ln}(X^0,k)\,.
\end{align}
Therefore, the Wigner transform of a convolution of two two-point functions simply reports a product without any differential operator. The above procedure can be applied for the second integral in the RHS of Eq.~(\ref{eq:KaFe}) as well.

On the other hand, the other transport equation is derived by considering Eq.~(\ref{eq:Kare}) and adding itself with $x^0\leftrightarrow y^0$, $\textbf{k}\rightarrow-\textbf{k}$, the LO term is given by \cite{Berges:2004yj,berges2015nonequilibriumquantumfieldscold}
\begin{align}
2k^0\frac{\partial}{\partial X^0}{\rm Tr}\Big[\tilde{\rho}_G(X^0,k)\Big]=0\,.\label{eq:rost}
\end{align}
Hence, the trace of the spectral function at the LO in the gradient expansion is stationary at the late time.

\subsection{Vertex function}

In contrast to the weak coupling expansion, the coupling in the $1/N$ expansion receives modifications by resumming an infinite number of scalar loops in the vertex. In general, the former fails in the strongly correlated (system that is weakly coupled but over-occupied \cite{Walz:2017ffj}) or the strongly coupled system \cite{Berges:2016nru}. In this section, I derive the effective coupling that encodes the resummation in cubic and quartic interactions by vertex functions.

The decompositions of Eq.~(\ref{eq:JJeq}) into $F$ and $\rho$ are provided by
\begin{align}
&\,(J^F)_{v^\prime s^\prime}^{\alpha\beta}(x^0,y^0,\textbf{k})\nonumber\\
=&\,\frac{\bar{\lambda}}{N}A_{v^\prime h^\prime}^{\alpha\gamma}(\textbf{k})(\Pi^F)_{h^\prime s^\prime}^{\gamma\beta}(x^0,y^0,\textbf{k})-\frac{\bar{\lambda}}{N}\int_{t^0}^{x^0}dz^0\,(J^\rho)_{v^\prime h^\prime}^{\alpha\mu}(x^0,z^0,\textbf{k})A_{h^\prime l^\prime}^{\mu\nu}(\textbf{k})(\Pi^F)_{l^\prime s^\prime}^{\nu\beta}(z^0,y^0,\textbf{k})\nonumber\\
&\,+\frac{\bar{\lambda}}{N}\int_{t^0}^{y^0}dz^0\,(J^F)_{v^\prime h^\prime}^{\alpha\mu}(x^0,z^0,\textbf{k})A_{h^\prime l^\prime}^{\mu\nu}(\textbf{k})(\Pi^\rho)_{l^\prime s^\prime}^{\nu\beta}(z^0,y^0,\textbf{k})\,,\label{eq:JFxy}\\
&\,(J^\rho)_{v^\prime s^\prime}^{\alpha\beta}(x^0,y^0,\textbf{k})\nonumber\\
=&\,\frac{\bar{\lambda}}{N}A_{v^\prime h^\prime}^{\alpha\gamma}(\textbf{k})(\Pi^\rho)_{h^\prime s^\prime}^{\gamma\beta}(x^0,y^0,\textbf{k})-\frac{\bar{\lambda}}{N}\int_{y^0}^{x^0}dz^0\,(J^\rho)_{v^\prime h^\prime}^{\alpha\mu}(x^0,z^0,\textbf{k})A_{h^\prime l^\prime}^{\mu\nu}(\textbf{k})(\Pi^\rho)_{l^\prime s^\prime}^{\nu\beta}(z^0,y^0,\textbf{k})\,,
\label{eq:Jrxy}
\end{align}
by using Eq.~(\ref{eq:cosg}) and (\ref{eq:coto}). These two expression can be simplified by defining the retarded and advanced two-point functions of $(J^\rho)_{v^\prime s^\prime}^{\alpha\beta}(x^0,y^0,\textbf{k})$ following notations in Eq.~(\ref{eq:RAd1}) and (\ref{eq:RAd2}) by inserting appropriate retarded or advanced step functions. Subsequently, the upper limits of integrals in Eq.~(\ref{eq:JFxy}) are extended to $\infty$ and the integration range $[y^0,x^0]$ in Eq.~(\ref{eq:Jrxy}) becomes $[-\infty,\infty]$ since the desired ranges are constrained by the step function. Using the Wigner transform after taking the limit $t^0\rightarrow-\infty$, the LO term in Eq.~(\ref{eq:Jrxy}) in the gradient expansion of $\partial_{X^0}$ gives
\begin{align}
&\,\frac{\bar{\lambda}}{N}A_{v^\prime h^\prime}^{\alpha\gamma}(\textbf{k})(\Pi_{R(A)}^\rho)_{h^\prime s^\prime}^{\gamma\beta}(X^0,k)\nonumber\\
=&\,(J_{R(A)}^\rho)_{v^\prime s^\prime}^{\alpha\beta}(X^0,k)+\frac{\bar{\lambda}}{N}(J_{R(A)}^\rho)_{v^\prime h^\prime}^{\alpha\mu}(X^0,k)A_{h^\prime l^\prime}^{\mu\nu}(\textbf{k})(\Pi_{R(A)}^\rho)_{l^\prime s^\prime}^{\nu\beta}(X^0,k)\,,\label{eq:JRA1}
\end{align}
by following analogous calculations in Eq.~(\ref{eq:Foex}). Therefore, Eq.~(\ref{eq:JRA1}) becomes
\begin{align}
\frac{\bar{\lambda}}{N}A_{v^\prime h^\prime}^{\alpha\gamma}(\textbf{k})(\Pi^\rho)_{h^\prime s^\prime}^{\gamma\beta}(X^0,k)=
&\,(J^\rho)_{v^\prime s^\prime}^{\alpha\beta}(X^0,k)+\frac{\bar{\lambda}}{N}(J_R^\rho)_{v^\prime h^\prime}^{\alpha\mu}(X^0,k)A_{h^\prime l^\prime}^{\mu\nu}(\textbf{k})(\Pi^\rho)_{l^\prime s^\prime}^{\nu\beta}(X^0,k)\nonumber\\
&\,+\frac{\bar{\lambda}}{N}(J^\rho)_{v^\prime h^\prime}^{\alpha\mu}(X^0,k)A_{h^\prime l^\prime}^{\mu\nu}(\textbf{k})(\Pi_A^\rho)_{l^\prime s^\prime}^{\nu\beta}(X^0,k).
\label{eq:APiJ}
\end{align}
by $J^\rho=J_R^\rho-J_A^\rho$. Furthermore, defining
\begin{align}
&\,(H_{R(A)}^{-1})_{v^\prime s^\prime}^{\alpha\beta}(X^0,k)=(A^{-1})_{v^\prime s^\prime}^{\alpha\beta}(\textbf{k})+\frac{\bar{\lambda}}{N}(\Pi_{R(A)}^\rho)_{v^\prime s^\prime}^{\alpha\beta}(X^0,k)\,,
\end{align}
Eq.~(\ref{eq:JRA1}) can be re-expressed as
\begin{align}
(\bar{J}_{R(A)}^\rho)_{v^\prime s^\prime}^{\alpha\beta}(X^0,k)=&\,\frac{\bar{\lambda}}{N}A_{v^\prime h^\prime}^{\alpha\mu}(\textbf{k})(\Pi_{R(A)}^\rho)_{h^\prime l^\prime}^{\mu\nu}(X^0,k)(H_{R(A)})_{l^\prime s^\prime}^{\nu\beta}(X^0,k)
\nonumber\\
=&\,A_{v^\prime s^\prime}^{\alpha\beta}(\textbf{k})-(H_{R(A)})_{v^\prime s^\prime}^{\alpha\beta}(X^0,k)\,,
\label{eq:JbHA}
\end{align}
and Eq.~(\ref{eq:APiJ}) can be re-written as
\begin{align}
&\,(\bar{J}^\rho)_{v^\prime s^\prime}^{\alpha\beta}(X^0,k)\nonumber\\
=&\,\frac{\bar{\lambda}}{N}\bigg(A_{v^\prime h^\prime}^{\alpha\mu}(\textbf{k})-\frac{\bar{\lambda}}{N}A_{v^\prime u^\prime}^{\alpha\nu}(\textbf{k})(\Pi_R^\rho)_{u^\prime l^\prime}^{\nu\sigma}(X^0,k) (H_R)_{l^\prime h^\prime}^{\sigma\mu}(X^0,k)\bigg)(\Pi^\rho)_{h^\prime r^\prime}^{\mu\kappa}(X^0,k)(H_A)_{r^\prime s^\prime}^{\kappa\beta}(X^0,k)\,\nonumber\\
=&\,\frac{\bar{\lambda}}{N}(H_R)_{v^\prime h^\prime}^{\alpha\mu}(X^0,k)(\Pi^\rho)_{h^\prime l^\prime}^{\mu\nu}(X^0,k)(H_A)_{l^\prime s^\prime}^{\nu\beta}(X^0,k)\label{eq:JrJr}
\end{align}
by Eq.~(\ref{eq:JbHA}).

For Eq.~(\ref{eq:JFxy}), its leading term in the gradient expansion after applying the Wigner transform is
\begin{align}
(J^F)_{v^\prime s^\prime}^{\alpha\beta}(X^0,k)
=&\,\frac{\bar{\lambda}}{N}A_{v^\prime h^\prime}^{\alpha\gamma}(\textbf{k})(\Pi^F)_{h^\prime s^\prime}^{\gamma\beta}(X^0,k)-\frac{\bar{\lambda}}{N}(J_R^\rho)_{v^\prime h^\prime}^{\alpha\mu}(X^0,k)A_{h^\prime l^\prime}^{\mu\nu}(\textbf{k})(\Pi^F)_{l^\prime s^\prime}^{\nu\beta}(X^0,k)\nonumber\\
&\,-\frac{\bar{\lambda}}{N}(J^F)_{v^\prime h^\prime}^{\alpha\mu}(X^0,k)A_{h^\prime l^\prime}^{\mu\nu}(\textbf{k})(\Pi_A^\rho)_{l^\prime s^\prime}^{\nu\beta}(X^0,k)\nonumber\\
\Rightarrow (\bar{J}^F)_{v^\prime s^\prime}^{\alpha\beta}(X^0,k)=&\,\frac{\bar{\lambda}}{N}\bigg(A_{v^\prime h^\prime}^{\alpha\mu}(\textbf{k})-(\bar{J}_R^\rho)_{v^\prime h^\prime}^{\alpha\mu}(X^0,k)\bigg)(\Pi^F)_{h^\prime w^\prime}^{\mu\nu}(X^0,k)(H_A)_{w^\prime s^\prime}^{\nu\beta}(X^0,k)\,.
\label{eq:JAfi}
\end{align}
After plugging in Eq~(\ref{eq:JbHA}), Eq.~(\ref{eq:JAfi}) reads
\begin{align}
(\bar{J}^F)_{v^\prime s^\prime}^{\alpha\beta}(X^0,k)
=&\,\frac{\bar{\lambda}}{N}(H_R)_{v^\prime h^\prime}^{\alpha\mu}(X^0,k)(\Pi^F)_{h^\prime l^\prime}^{\mu\nu}(X^0,k)(H_A)_{l^\prime s^\prime}^{\nu\beta}(X^0,k)\,.
\label{eq:JFJF}
\end{align}
Therefore, it is shown that $J^{F(\rho)}$ corresponds to $\Pi^{F(\rho)}$ sandwiched by $H_{R(A)}$, which contains non-trivial momentum-dependent corrections to the coupling $\bar{\lambda}$ by resumming an infinite number of scalar loops encoded in $\Pi_{R(A)}^\rho$. The complete expressions of the Wigner transform of Eq.~(\ref{eq:JrJr}) and (\ref{eq:JFJF}) can be written down by plugging in
\begin{align}
(\Pi^F)_{v^\prime s^\prime}^{\beta\alpha}(X^0,k)=&\,-\int_q \bigg[ S_{s^\prime v^\prime}^{\alpha\beta}(\textbf{q},-\textbf{k})\bigg(F_{G_{(2\textbf{j}-1)(2\textbf{i}-1)}}(X^0,q)F_{G_{(2\textbf{i}-1)(2\textbf{j}-1)}}(X^0,k+q)\nonumber\\
&\,+\frac{1}{4}\rho_{G_{(2\textbf{j}-1)(2\textbf{i}-1)}}(X^0,q)\rho_{G_{(2\textbf{i}-1)(2\textbf{j}-1)}}(X^0,k+q)\bigg)\nonumber\\
&\,+S_{s^\prime v^\prime}^{\alpha\beta}(\textbf{q})\bigg(F_{G_{(2\textbf{j})(2\textbf{i}-1)}}(X^0,q)F_{G_{(2\textbf{i}-1)(2\textbf{j})}}(X^0,k+q)\nonumber\\
&\,+\frac{1}{4}\rho_{G_{(2\textbf{j})(2\textbf{i}-1)}}(X^0,q)\rho_{G_{(2\textbf{i}-1)(2\textbf{j})}}(X^0,k+q)\bigg)\bigg]\,,\label{eq:PiFg}\\
(\Pi^\rho)_{v^\prime s^\prime}^{\beta\alpha}(X^0,k)=&\,-\int_q\bigg[S_{s^\prime v^\prime}^{\alpha\beta}(\textbf{q},-\textbf{k})\bigg(-\rho_{G_{(2\textbf{j}-1)(2\textbf{i}-1)}}(X^0,q)F_{G_{(2\textbf{i}-1)(2\textbf{j}-1)}}(X^0,k+q)\nonumber\\
&\,+F_{G_{(2\textbf{j}-1)(2\textbf{i}-1)}}(X^0,q)\rho_{G_{(2\textbf{i}-1)(2\textbf{j}-1)}}(X^0,k+q)\bigg)\nonumber\\
&\,+S_{s^\prime v^\prime}^{\alpha\beta}(\textbf{q})\bigg(-\rho_{G_{(2\textbf{j})(2\textbf{i}-1)}}(X^0,q)F_{G_{(2\textbf{i}-1)(2\textbf{j})}}(X^0,k+q)\nonumber\\
&\,+F_{G_{(2\textbf{j})(2\textbf{i}-1)}}(X^0,q)\rho_{G_{(2\textbf{i}-1)(2\textbf{j})}}(X^0,k+q)\bigg)\bigg]\,,\label{eq:Pirg}
\end{align}
where $\int_q\equiv\int_{-\infty}^\infty d^dq/(2\pi)^d$, given by the LO gradient expansion of the Wigner transform of Eq.~(\ref{eq:PiFF}) and (\ref{eq:Pirr}).

The function $\bar{J}^{F(\rho)}(X^0,k)$ accounts for resummation in the quartic vertex, which is the case of  vanishing field expectation values $\phi_\textbf{i}(x^0)=0$. However, the non-zero $\phi_\textbf{i}(x^0)$ can also scatter with fields, which embodies new corrections in the effective coupling embedded in the function $W_{s^\prime v^\prime}^{\alpha\beta}(X^0,k)$.

Starting from Eq.~(\ref{eq:WFre}) and (\ref{eq:Wrre}), LO terms in the gradient expansion of the Wigner transform are given by
\begin{align}
&\,(W^F)_{s^\prime v^\prime}^{\alpha\beta}(X^0,k)\nonumber\\
=&\,-\Omega_{a^\prime n}^\mu(X^0,\textbf{k})\tilde{\Omega}_{ml^\prime}^\nu(X^0,\textbf{k})\frac{\bar{\lambda}}{N}\bigg[F_{G_{nm}}(X^0,k)\bigg(A_{s^\prime a^\prime}^{\alpha \mu}(\textbf{k})A_{l^\prime v^\prime}^{\nu\beta}(\textbf{k})-A_{l^\prime v^\prime}^{\nu\beta}(\textbf{k})(\bar{J}_R^\rho)_{s^\prime a^\prime}^{\alpha\mu}(X^0,\textbf{k})\nonumber\\
&\,-A_{s^\prime a^\prime}^{\alpha \mu}(\textbf{k})(\bar{J}_A^\rho)_{l^\prime v^\prime}^{\nu\beta}(X^0,\textbf{k})+(\bar{J}_R^\rho)_{s^\prime a^\prime}^{\alpha \mu}(X^0,k)(\bar{J}_A^\rho)_{l^\prime v^\prime}^{\nu\beta}(X^0,k)\bigg)\nonumber\\
&\,+(\bar{J}^F)_{r^\prime h^\prime}^{\gamma \sigma}(X^0,k)\bigg(-A_{s^\prime a^\prime}^{\alpha \mu}(\textbf{k})(G_R^\rho)_{nm}(X^0,k)\mathcal{I}_{l^\prime r^\prime}^{\nu \gamma}\mathcal{I}_{h^\prime v^\prime}^{\sigma\beta}-(G_A^\rho)_{nm}(X^0,k)A_{l^\prime v^\prime}^{\nu\beta}(\textbf{k})\mathcal{I}_{s^\prime r^\prime}^{\alpha \gamma}\mathcal{I}_{a^\prime h^\prime}^{\mu \sigma}\nonumber\\
&\,+\mathcal{I}_{l^\prime r^\prime}^{\nu \gamma}\mathcal{I}_{h^\prime v^\prime}^{\sigma\beta}(\bar{J}_R^\rho)_{s^\prime a^\prime}^{\alpha \mu}(X^0,k)(G_R^\rho)_{nm}(X^0,k)+\mathcal{I}_{s^\prime r^\prime}^{\alpha \gamma}\mathcal{I}_{a^\prime h^\prime}^{\mu \sigma}(G_A^\rho)_{nm}(X^0,k)(\bar{J}_A^\rho)_{l^\prime v^\prime}^{\nu\beta}(X^0,k)\bigg)\bigg]\,,
\end{align}
\begin{align}
&\,(W^{\tilde{\rho}})_{s^\prime v^\prime}^{\alpha\beta}(X^0,k)\equiv-i(W^\rho)_{s^\prime v^\prime}^{\alpha\beta}(X^0,k)\nonumber\\
=&\,-\Omega_{a^\prime n}^\mu(X^0,\textbf{k})\tilde{\Omega}_{ml^\prime}^\nu(X^0,\textbf{k})\frac{\bar{\lambda}}{N}\bigg[\tilde{\rho}_{G_{nm}}(X^0,k)\bigg(A_{s^\prime a^\prime}^{\alpha \mu}(\textbf{k})A_{l^\prime v^\prime}^{\nu\beta}(\textbf{k})-A_{s^\prime a^\prime}^{\alpha \mu}(\textbf{k})(\bar{J}_A^\rho)_{l^\prime v^\prime}^{\nu\beta}(X^0,k)\nonumber\\
&\,-(\bar{J}_R^\rho)_{s^\prime a^\prime}^{\alpha \mu}(X^0,k) A_{l^\prime v^\prime}^{\nu\beta}(\textbf{k})+(\bar{J}_R^\rho)_{s^\prime a^\prime}^{\alpha \mu}(X^0,k)(\bar{J}_A^\rho)_{l^\prime v^\prime}^{\nu\beta}(X^0,k)\bigg)\nonumber\\
&\,+(\bar{J}^{\tilde{\rho}})_{r^\prime h^\prime}^{\gamma \sigma}(X^0,k)\bigg(-A_{s^\prime a^\prime}^{\alpha \mu}(\textbf{k})(G_R^\rho)_{nm}(X^0,k)\mathcal{I}_{l^\prime r^\prime}^{\nu \gamma}\mathcal{I}_{h^\prime v^\prime}^{\sigma\beta}-(G_A^\rho)_{nm}(X^0,k)A_{l^\prime v^\prime}^{\nu\beta}(\textbf{k})\mathcal{I}_{s^\prime r^\prime}^{\alpha \gamma}\mathcal{I}_{a^\prime h^\prime}^{\mu \sigma}\nonumber\\
&\,+(\bar{J}_R^\rho)_{s^\prime a^\prime}^{\alpha \mu}(X^0,k)(G_R^\rho)_{nm}(X^0,k)\mathcal{I}_{l^\prime r^\prime}^{\nu \gamma}\mathcal{I}_{h^\prime v^\prime}^{\sigma\beta}+(G_A^\rho)_{nm}(X^0,k)(\bar{J}_A^\rho)_{l^\prime v^\prime}^{\nu\beta}(X^0,k)\mathcal{I}_{s^\prime r^\prime}^{\alpha \gamma}\mathcal{I}_{a^\prime h^\prime}^{\mu \sigma}\bigg)\bigg]\,,
\end{align}
where several terms are grouped together by extracting non-retarded/non-advanced functions $F_G$, $\bar{J}^F$ in $W^F$ and $\rho_G$, $\bar{J}^{\tilde{\rho}}$ in $W^{\tilde{\rho}}$, respectively. In addition, these groups share the same form simplified by $H_{R(A)}$ according to Eq.~(\ref{eq:JbHA})
\begin{align}
(W^F)_{s^\prime v^\prime}^{\alpha\beta}(X^0,k)
=&\,-\Omega_{a^\prime n}^\mu(X^0,\textbf{k})\tilde{\Omega}_{ml^\prime}^\nu(X^0,\textbf{k})\frac{\bar{\lambda}}{N}\bigg[F_{G_{nm}}(X^0,k)(H_R)_{s^\prime a^\prime}^{\alpha\mu}(X^0,k)(H_A)_{l^\prime v^\prime}^{\nu\beta}(X^0,k)\nonumber\\
&\,-(\bar{J}^F)_{r^\prime h^\prime}^{\gamma \sigma}(X^0,k)\bigg(\mathcal{I}_{s^\prime r^\prime}^{\alpha\gamma}\mathcal{I}_{a^\prime h^\prime}^{\mu \sigma}(G_A^\rho)_{nm}(X^0,k)(H_A)_{l^\prime v^\prime}^{\nu\beta}(X^0,k)\nonumber\\
&\,+\mathcal{I}_{l^\prime r^\prime}^{\nu\gamma}\mathcal{I}_{h^\prime v^\prime}^{\sigma\beta}(H_R)_{s^\prime a^\prime}^{\alpha\mu}(X^0,k)(G_R^\rho)_{nm}(X^0,k)\bigg)\bigg]\,,\label{eq:WFgr}\\
(W^{\tilde{\rho}})_{s^\prime v^\prime}^{\alpha\beta}(X^0,k)
=&\,\Big( F_G\rightarrow\tilde{\rho}_G\,,\,\,  \bar{J}^F\rightarrow \bar{J}^{\tilde{\rho}}\,\,{\rm in}\,\,{\rm Eq.}~(\ref{eq:WFgr})\Big)\,.
\label{eq:Wsu2}
\end{align}
It is noticed that $F_G(\tilde\rho_G)$ and $\bar{J}^F(\bar{J}^{\tilde{\rho}})$ in $W^F(W^{\tilde{\rho}})$ contribute to two different processes, respectively. The cubic vertex arises from the first term of Eq.~(\ref{eq:WFgr}) and (\ref{eq:Wsu2}) due to the non-zero $\phi_\textbf{i}(x^0)$ while the second term is identified as corrections to the quartic vertex.

\subsection{Vertex-resummed kinetic equation}

The transport equation in Eq.~(\ref{eq:kinF}) can be re-written in a different basis 
\begin{align}
2k^0\frac{\partial}{\partial X^0}{\rm Tr}\Big[F_G(X^0,k)\Big]={\rm Tr}\Big[\bar{\Sigma}^+(X^0,k)G^-(X^0,k)-\bar{\Sigma}^-(X^0,k)G^+(X^0,k)\Big]
\label{eq:kipm}
\end{align}
by
\begin{align}
\bar{\Sigma}_{nm}^\pm(X^0,k)\equiv&\,\bar{\Sigma}_{nm}^F(X^0,k)\pm\frac{1}{2}\bar{\Sigma}_{nm}^{\tilde{\rho}}(X^0,k)\,,\label{eq:pmde}\\
G_{nm}^\pm(X^0,k)\equiv&\, F_{G_{nm}}(X^0,k)\pm\frac{1}{2}\tilde{\rho}_{G_{nm}}(X^0,k)\,.
\label{eq:pmpm}
\end{align}
The notation with an upper index $\pm$ is also applied to other functions by following the decomposition in Eq.~(\ref{eq:pmde}). In addition, the linear combination $\bar{\Sigma}_{nm}^\pm(X^0,k)$ expressed by equations from (\ref{eq:bSiF1}) and (\ref{eq:bSir1}) can also be re-written with $G^\pm$, $\bar{J}^\pm$ and $W^\pm$. In the following, a specific basis is studied where $G_{nm}$ becomes block-diagonal in the form of $G_{nm}\rightarrow G_{\parallel_{nm}}$ for $n,m\in\{1,2\}$ and 
\begin{align}
G_{nm}\rightarrow 
\begin{pmatrix}
G_{\perp_{11}} & 0 \\
0 & G_{\perp_{22}}
\end{pmatrix}
\oplus
\cdots \oplus
\begin{pmatrix}
G_{\perp_{11}} & 0 \\
0 & G_{\perp_{22}}
\end{pmatrix}
\,,\quad
n,m\in\{3,4,\cdots,2N\}\,,
\end{align}
which consists of direct sums of $N-1$ identical diagonal matrices ${\rm diag}(G_{\perp_{11}},G_{\perp_{22}})$ due to the residual global $U(N-1)$ symmetry. In addition, the $N$-component field expectation values $\phi_\textbf{i}(x^0)$ are aligned to $\big(\sqrt{N}\phi_\parallel(x^0),0,\cdots,0\big)$ where all $\perp$ sectors are vanishing. That is, the chosen basis specifies the situation when the non-zero time-dependent $\phi_\textbf{i}(x^0)$ is confined in a single component. Therefore, this specific decomposition of Kadanoff-Baym equations lead to
\begin{align}
2k^0\frac{\partial}{\partial X^0}  {\rm Tr}_2\big[F_{G_\parallel}(X^0,k)\big]=&\,{\rm Tr}_2\Big[\Sigma_\parallel^+(X^0,k)G_\parallel^-(X^0,k)-\Sigma_\parallel^-(X^0,k)G_\parallel^+(X^0,k)\Big]\,, \label{eq:depa}\\
2k^0\frac{\partial}{\partial X^0}  {\rm Tr}_2\big[F_{G_{\perp}}(X^0,k)\big]=&\,{\rm Tr}_2\Big[\Sigma_\perp^+(X^0,k)G_\perp^-(X^0,k)-\Sigma_\perp^-(X^0,k)G_\perp^+(X^0,k)\Big]\,.\label{eq:depe}
\end{align}
where ${\rm Tr}_2$ traces over indices $n\in\{1,2\}$ in $\parallel$ or $\perp$ sector. The right-hand side of Eq.~(\ref{eq:depa}) and (\ref{eq:depe}) thus read  
\begin{align}
&\,\frac{\bar{\lambda}}{N}\bigg[{\rm Tr}_2\Big[(\tilde{\Omega}_{\parallel})^\alpha(X^0,\textbf{k})(\bar{J}^+)^{\alpha\beta}(X^0,k)(\Omega_{\parallel})^\beta(X^0,\textbf{k})G_\parallel^-(X^0,k)\Big]\nonumber\\
&\,+\int_qS_{s^\prime v^\prime}^{\alpha\beta}(\textbf{k},\textbf{q})\Big((\bar{J}^-)_{s^\prime v^\prime}^{\alpha\beta}(X^0,-q)+(W^-)_{s^\prime v^\prime}^{\alpha\beta}(X^0,-q)\Big)G_{\parallel_{11}}^+(X^0,k-q)G_{\parallel_{11}}^-(X^0,k)\nonumber\\
&\,+\int_qS_{s^\prime v^\prime}^{\alpha\beta}(\textbf{q}-\textbf{k},\textbf{q})\Big((\bar{J}^-)_{s^\prime v^\prime}^{\alpha\beta}(X^0,-q)+(W^-)_{s^\prime v^\prime}^{\alpha\beta}(X^0,-q)\Big)G_{\parallel_{22}}^+(X^0,k-q)G_{\parallel_{22}}^-(X^0,k)\nonumber\\
&\,+\int_qS_{s^\prime v^\prime}^{\alpha\beta}(\textbf{k})\Big((\bar{J}^-)_{s^\prime v^\prime}^{\alpha\beta}(X^0,-q)+(W^-)_{s^\prime v^\prime}^{\alpha\beta}(X^0,-q)\Big)G_{\parallel_{12}}^+(X^0,k-q)G_{\parallel_{21}}^-(X^0,k)\nonumber\\
&\,+\int_qS_{s^\prime v^\prime}^{\alpha\beta}(\textbf{q}-\textbf{k})\Big((\bar{J}^-)_{s^\prime v^\prime}^{\alpha\beta}(X^0,-q)+(W^-)_{s^\prime v^\prime}^{\alpha\beta}(X^0,-q)\Big)G_{\parallel_{21}}^+(X^0,k-q)G_{\parallel_{12}}^-(X^0,k)\bigg]\nonumber\\
&\,-\frac{\bar{\lambda}}{N}\bigg[G^\pm, \bar{J}^\pm, W^\pm \rightarrow G^\mp, \bar{J}^\mp\,, W^\mp\bigg]
\label{eq:cokp}
\end{align}
and
\begin{align}
&\,\frac{\bar{\lambda}(N-1)}{N}
\int_q\bigg[S_{s^\prime v^\prime}^{\alpha\beta}(\textbf{k},\textbf{q})\Big((\bar{J}^-)_{s^\prime v^\prime}^{\alpha\beta}(X^0,-q)+(W^-)_{s^\prime v^\prime}^{\alpha\beta}(X^0,-q)\Big)G_{\perp_{11}}^+(X^0,k-q)G_{\perp_{11}}^-(X^0,k)\nonumber\\
&\,+S_{s^\prime v^\prime}^{\alpha\beta}(\textbf{q}-\textbf{k},\textbf{q})\Big((\bar{J}^-)_{s^\prime v^\prime}^{\alpha\beta}(X^0,-q)+(W^-)_{s^\prime v^\prime}^{\alpha\beta}(X^0,-q)\Big)G_{\perp_{22}}^+(X^0,k-q)G_{\perp_{22}}^-(X^0,k)\bigg]\nonumber\\
&\,-\frac{\bar{\lambda}(N-1)}{N}\bigg[G^\pm, \bar{J}^\pm, W^\pm \rightarrow G^\mp, \bar{J}^\mp\,, W^\mp\bigg]\,,
\label{eq:cokp1}
\end{align}
respectively.

Following the form in Eq.~(\ref{eq:pmpm}), the linear combination $(\bar{J}^\pm)_{s^\prime v^\prime}^{\alpha\beta}$  
given by Eq.~(\ref{eq:JrJr}), (\ref{eq:JFJF}), (\ref{eq:PiFg}) and (\ref{eq:Pirg}) is expressed as
\begin{align}
(\bar{J}^\pm)_{s^\prime v^\prime}^{\alpha\beta}(X^0,k)=&\,-\frac{\bar{\lambda}}{N}\int_qU_{s^\prime h^\prime, l^\prime v^\prime}^{\alpha\mu,\nu\beta}(X^0,k) \bigg[ S_{l^\prime h^\prime}^{\nu\mu}(\textbf{q},-\textbf{k}) \bigg(G_{\parallel_{11}}^\mp(X^0,q)G_{\parallel_{11}}^\pm(X^0,k+q)\nonumber\\
&\,+(N-1)G_{\perp_{11}}^\mp(X^0,q)G_{\perp_{11}}^\pm(X^0,k+q)\bigg)+S_{l^\prime h^\prime}^{\nu\mu}(\textbf{q})G_{\parallel_{21}}^\mp(X^0,q)G_{\parallel_{12}}^\pm(X^0,k+q)\bigg]\,,\label{eq:Jlin} \\
U_{v^\prime h^\prime,l^\prime s^\prime}^{\alpha\mu,\nu\beta}(X^0,k)\equiv&\, (H_R)_{v^\prime h^\prime}^{\alpha\mu}(X^0,k)(H_A)_{l^\prime s^\prime}^{\nu\beta}(X^0,k)\,.
\end{align}
The expression of $(W^\pm)_{s^\prime v^\prime}^{\alpha\beta}$ is constructed by Eq.~(\ref{eq:WFgr}),(\ref{eq:Wsu2}) and (\ref{eq:Jlin}) as
\begin{align}
&\,(W^\pm)_{s^\prime v^\prime}^{\alpha\beta}(X^0,k)\nonumber\\
=&\,-\frac{\bar{\lambda}}{N}(\Omega_\parallel)_{a^\prime n}^\mu(X^0,\textbf{k})(\tilde{\Omega}_\parallel)_{ml^\prime}^\nu(X^0,\textbf{k})\bigg\{U_{s^\prime a^\prime,l^\prime v^\prime}^{\alpha \mu,\nu\beta}(X^0,k)G_{\parallel_{nm}}^\pm(X^0,k)\nonumber\\
&\,+\frac{\bar{\lambda}}{N}\bigg(\mathcal{I}_{s^\prime w^\prime}^{\alpha \gamma}\mathcal{I}_{a^\prime h^\prime}^{\mu\sigma}(G_A^\rho)_{\parallel_{nm}}(X^0,k)(H_A)_{l^\prime v^\prime}^{\nu\beta}(X^0,k)+\mathcal{I}_{l^\prime w^\prime}^{\nu \gamma}\mathcal{I}_{h^\prime v^\prime}^{\sigma\beta}(H_R)_{s^\prime a^\prime}^{\alpha \mu}(X^0,k)(G_R^\rho)_{\parallel_{nm}}(X^0,k)\bigg)\nonumber\\
&\,\times\int_qU_{w^\prime b^\prime, c^\prime h^\prime}^{\alpha\tau,\kappa\beta}(X^0,k) \bigg[ S_{c^\prime b^\prime}^{\kappa\tau}(\textbf{q},-\textbf{k}) \bigg(G_{\parallel_{11}}^\mp(X^0,q)G_{\parallel_{11}}^\pm(X^0,k+q)\nonumber\\
&\,+(N-1)G_{\perp_{11}}^\mp(X^0,q)G_{\perp_{11}}^\pm(X^0,k+q)\bigg)+S_{c^\prime b^\prime}^{\kappa\tau}(\textbf{q})G_{\parallel_{21}}^\mp(X^0,q)G_{\parallel_{12}}^\pm(X^0,k+q)\bigg]\bigg\}\,.
\label{eq:Wepm} 
\end{align}

The transport equation becomes in the form analogous to the kinetic equation that contains components classified as the three and four-vertex scatterings labeled with $\mathcal{C}_{(3)}[f]$ and $\mathcal{C}_{(4)}[f]$, respectively 
\begin{align}
\int_0^\infty \frac{dk^0}{2\pi} 2k^0\frac{\partial }{\partial X^0}{\rm Tr}_2\big[F_{G_\parallel}(X^0,k)\big]=&\,\frac{\partial f_\parallel(X^0,\textbf{k})}{\partial X^0}=\mathcal{C}_{(3)}^\parallel[f](X^0,\textbf{k})+\mathcal{C}_{(4)}^\parallel[f](X^0,\textbf{k})\,,\label{eq:kik1}\\
\int_0^\infty \frac{dk^0}{2\pi} 2k^0\frac{\partial }{\partial X^0}{\rm Tr}_2\big[F_{G_\perp}(X^0,k)\big]=&\,\frac{\partial f_\perp(X^0,\textbf{k})}{\partial X^0}=\mathcal{C}_{(3)}^\perp[f](X^0,\textbf{k})+\mathcal{C}_{(4)}^\perp[f](X^0,\textbf{k})\,,
\label{eq:kik2}
\end{align}
where the Kadanoff-Baym ansatz \cite{10.1063/1.1704738,zubarev1997statistical2,imamovic2001quasiparticle} connects $F_G$ with $\tilde{\rho}_G$ by the occupation number distribution $f(X,k)$ obeying
\begin{align}
F_{G_{\parallel_{mn}}}(X,k)=&\,\bigg(f_\parallel(X,k)+\frac{1}{2}\bigg)\tilde{\rho}_{G_{\parallel_{mn}}}(X,k)\,,\\\quad F_{G_{\perp_{mn}}}(X,k)=&\,\bigg(f_\perp(X,k)+\frac{1}{2}\bigg)\tilde{\rho}_{G_{\perp_{mn}}}(X,k)
\label{eq:Fran}
\end{align}
and $f_{\parallel(\perp)}(X,-k)=-1-f_{\parallel(\perp)}(X,k)$ \cite{Berges:2004yj,berges2015nonequilibriumquantumfieldscold}. At the left-hand side of Eq.~(\ref{eq:kik1}) and (\ref{eq:kik2}), the effective number distribution is defined
\begin{align}
f_\parallel(X^0,\textbf{k})\equiv&\,\int_0^\infty \frac{dk^0}{2\pi} 2k^0f_\parallel(X^0,k){\rm Tr}_2\big[\tilde{\rho}_{G_\parallel}(X^0,k)\big]\,,\\
f_\perp(X^0,\textbf{k})\equiv&\,\int_0^\infty \frac{dk^0}{2\pi} 2k^0f_\perp(X^0,k){\rm Tr}_2\big[\tilde{\rho}_{G_\perp}(X^0,k)\big]
\end{align}
and diagonal elements of spectral function are stationary in the variation of $X^0$ according to Eq.~(\ref{eq:rost}). The classification of $\mathcal{C}_{(3)}[f]$ and $\mathcal{C}_{(4)}[f]$ are extracted and sorted from Eq.~(\ref{eq:cokp}) and (\ref{eq:cokp1}). In the following, I only show explicit expressions of $\mathcal{C}_{(3)}^\parallel[f]$ and $\mathcal{C}_{(4)}^\parallel[f]$   
\begin{align}
&\,\mathcal{C}_{(3)}^\parallel(X^0,\textbf{k})\nonumber\\=&\,-\bigg(\frac{\bar{\lambda}}{N}\bigg)^2\int_0^\infty\frac{dk^0}{2\pi}\int_{q,p}(2\pi)^d\delta^d(k-q-p)\bigg\{\bigg[S_{s^\prime v^\prime}^{\alpha\beta}(\textbf{k},\textbf{q})\tilde{\rho}_{G_{\parallel_{11}}}(X^0,p)\tilde{\rho}_{G_{\parallel_{11}}}(X^0,k)\nonumber\\
&\,+S_{s^\prime v^\prime}^{\alpha\beta}(\textbf{q}-\textbf{k},\textbf{q})\tilde{\rho}_{G_{\parallel_{22}}}(X^0,p)\tilde{\rho}_{G_{\parallel_{22}}}(X^0,k)+S_{s^\prime v^\prime}^{\alpha\beta}(\textbf{k})\tilde{\rho}_{G_{\parallel_{12}}}(X^0,p)\tilde{\rho}_{G_{\parallel_{21}}}(X^0,k)\nonumber\\
&\,+S_{s^\prime v^\prime}^{\alpha\beta}(\textbf{q}-\textbf{k})\tilde{\rho}_{G_{\parallel_{21}}}(X^0,p)\tilde{\rho}_{G_{\parallel_{12}}}(X^0,k)\bigg]\tilde{\rho}_{G_{\parallel_{nm}}}(X^0,q)(\Omega_\parallel)_{l^\prime m}^\nu(X^0,\textbf{q})(\tilde{\Omega}_\parallel)_{na^\prime}^\mu(X^0,\textbf{q})U_{s^\prime a^\prime ,l^\prime v^\prime}^{\alpha\mu,\nu\beta}(X^0,-q)\nonumber\\
&\,\times\Big(\bar{f}_\parallel(X^0,q)\bar{f}_\parallel(X^0,p)f_\parallel(X^0,k)-f_\parallel(X^0,q)f_\parallel(X^0,p)\bar{f}_\parallel(X^0,k)\Big)\nonumber\\
&\,+(\tilde{\Omega}_\parallel)_{ns^\prime}^\alpha(X^0,\textbf{k})U_{s^\prime h^\prime,l^\prime v^\prime}^{\alpha\mu,\nu\beta}(X^0,k)(\Omega_\parallel)_{v^\prime m}^\beta(X^0,\textbf{k})\tilde{\rho}_{G_{\parallel_{mn}}}(X^0,k)\nonumber\\
&\,\times\bigg[\Big(S_{l^\prime h^\prime}^{\nu\mu}(-\textbf{p},-\textbf{k})\tilde{\rho}_{G_{\parallel_{22}}}(X^0,p)\tilde{\rho}_{G_{\parallel_{11}}}(X^0,q)+S_{l^\prime h^\prime}^{\nu\mu}(-\textbf{p})\tilde{\rho}_{G_{\parallel_{21}}}(X^0,p)\tilde{\rho}_{G_{\parallel_{12}}}(X^0,q)\Big)\nonumber\\
&\,\times\Big(\bar{f}_\parallel(X^0,p)\bar{f}_\parallel(X^0,q)f_\parallel(X^0,k)-f_\parallel(X^0,p)f_\parallel(X^0,q)\bar{f}_\parallel(X^0,k)\Big)\nonumber\\
&\,+(N-1)S_{l^\prime h^\prime}^{\nu\mu}(-\textbf{p},-\textbf{k})\tilde{\rho}_{G_{\perp_{22}}}(X^0,p)\tilde{\rho}_{G_{\perp_{11}}}(X^0,q)\nonumber\\
&\,\times\Big(\bar{f}_\perp(X^0,p)\bar{f}_\perp(X^0,q)f_\parallel(X^0,k)-f_\perp(X^0,p)f_\perp(X^0,q)\bar{f}_\parallel(X^0,k)\Big)\bigg]\bigg\}\,,
\label{eq:C3pa}
\end{align}

\begin{align}
&\,\mathcal{C}_{(4)}^\parallel(X^0,\textbf{k})\nonumber\\
=&\,-\bigg(\frac{\bar{\lambda}}{N}\bigg)^2\int_0^\infty \frac{dk^0}{2\pi}\int_{q,p,w} (2\pi)^d\delta^d(p-k+q-w)\bar{U}_{s^\prime b^\prime,c^\prime v^\prime}^{\alpha\kappa,\tau\beta}(X^0,q-k)\nonumber\\
&\,\times \Big(S_{s^\prime v^\prime}^{\alpha\beta}(\textbf{k},\textbf{k}-\textbf{q})\tilde{\rho}_{G_{\parallel_{11}}}(X^0,q)\tilde{\rho}_{G_{\parallel_{11}}}(X^0,k)+S_{s^\prime v^\prime}^{\alpha\beta}(-\textbf{q},\textbf{k}-\textbf{q})\tilde{\rho}_{G_{\parallel_{22}}}(X^0,q)\tilde{\rho}_{G_{\parallel_{22}}}(X^0,k)\nonumber\\
&\,+S_{s^\prime v^\prime}^{\alpha\beta}(\textbf{k})\tilde{\rho}_{G_{\parallel_{12}}}(X^0,q)\tilde{\rho}_{G_{\parallel_{21}}}(X^0,k)+S_{s^\prime v^\prime}^{\alpha\beta}(-\textbf{q})\tilde{\rho}_{G_{\parallel_{21}}}(X^0,q)\tilde{\rho}_{G_{\parallel_{12}}}(X^0,k)\Big)\nonumber\\
&\,\times\Big[\Big(S_{c^\prime b^\prime}^{\tau\kappa}(\textbf{p},\textbf{k}-\textbf{q})\tilde{\rho}_{G_{\parallel_{11}}}(X^0,p)\tilde{\rho}_{\parallel_{11}}(X^0,w)+S_{c^\prime b^\prime}^{\tau\kappa}(\textbf{p})\tilde{\rho}_{G_{\parallel_{21}}}(X^0,p)\tilde{\rho}_{G_{\parallel_{12}}}(X^0,w)\Big)\nonumber\\
&\,\times\Big(\bar{f}_\parallel(X^0,q)f_\parallel(X^0,k)\bar{f}_\parallel(X^0,p)f_\parallel(X^0,w)-f_\parallel(X^0,q)\bar{f}_\parallel(X^0,k)f_\parallel(X^0,p)\bar{f}_\parallel(X^0,w)\Big)\nonumber\\
&\,+(N-1)S_{c^\prime b^\prime}^{\tau\kappa}(\textbf{p},\textbf{k}-\textbf{q})\tilde{\rho}_{G_{\perp_{11}}}(X^0,p)\tilde{\rho}_{G_{\perp_{11}}}(X^0,w)\nonumber\\
&\,\times\Big(\bar{f}_\parallel(X^0,q)f_\parallel(X^0,k)\bar{f}_\perp(X^0,p)f_\perp(X^0,w)-f_\parallel(X^0,q)\bar{f}_\parallel(X^0,k)f_\perp(X^0,p)\bar{f}_\perp(X^0,w)\Big)\Big]
\label{eq:C4pa}
\end{align}
where $\bar{f}(X^0,k)\equiv 1+f(X^0,k)$ and
\begin{align}
\bar{U}_{s^\prime b^\prime ,c^\prime v^\prime}^{\alpha\kappa,\tau\beta}(X^0,q)\equiv &\,\bigg[ U_{s^\prime b^\prime, c^\prime v^\prime}^{\alpha \kappa,\tau\beta}(X^0,q)+\frac{\bar{\lambda}}{N}(\Omega_\parallel)_{a^\prime n}^\mu(X^0,\textbf{q})(\tilde{\Omega}_\parallel)_{ml^\prime}^\nu(X^0,\textbf{q})U_{w^\prime b^\prime,c^\prime h^\prime}^{\gamma\kappa,\tau \sigma}(X^0,q)\nonumber\\
&\,\times\bigg(\mathcal{I}_{s^\prime w^\prime}^{\alpha \gamma}\mathcal{I}_{a^\prime h^\prime}^{\mu\sigma}(G_A^\rho)_{\parallel_{nm}}(X^0,q)(H_A)_{l^\prime v^\prime}^{\nu\beta}(X^0,q)\nonumber\\
&\,+\mathcal{I}_{l^\prime w^\prime}^{\nu\gamma}\mathcal{I}_{h^\prime v^\prime}^{\sigma\beta}(H_R)_{s^\prime a^\prime}^{\alpha \mu}(X^0,q)(G_R^\rho)_{\parallel_{nm}}(X^0,q)\bigg) \bigg]\,.
\end{align}
On the other hand, sectors of $\mathcal{C}_{(3)}^\perp$ and $\mathcal{C}_{(4)}^\perp$ can be obtained analogously. The non-vanishing $\phi_\textbf{i}(x^0)$ induce the cubic interaction, which is found to contribute not only to the three-vertex scatterings but also be embedded in $\bar{U}_{s^\prime b^\prime, c^\prime v^\prime}^{\alpha\kappa,\tau\beta}$ in the four-vertex scatterings. In the case of vanishing $\phi_\textbf{i}(x^0)$, the remaining terms in the collision kernel is from the four-vertex interaction and $\bar{U}_{s^\prime b^\prime, c^\prime v^\prime}^{\alpha\kappa,\tau\beta}=U_{s^\prime b^\prime, c^\prime v^\prime}^{\alpha\kappa,\tau\beta}$.


\section{Discussions}
The non-equilibrium evolution of the strongly coupled complex scalar field theory with a dipole symmetry beyond LO is constructed based on the 2PI $1/N$ expansion to the NLO.
The auxiliary field method is performed to transform quartic terms into quadratic terms, which creates three kinds of auxiliary fields. The index structure in Eq.~(\ref{eq:maaa}) makes the whole discussion in this work more general to the case of higher dimension. The 2PI $1/N$ expansion to the NLO corrections arises from fluctuations around field expectation values of complex scalar and auxiliary fields.

Two forms of time evolution equations of two-point functions and $\phi_\textbf{i}^*(x^0)$ are derived including non-trivial scattering effects from self-energies beyond the mean-field level. One is directly from the auxiliary field method. The other one takes the form of Kadanoff-Baym equations. At the late time of the evolution, the kinetic description in the $1/N$ expansion is modified and generalized to the vertex-resummed kinetic theory by encompassing a resummation in the vertex, which is encoded in the vertex functions $(H_{R(A)})_{v^\prime s^\prime}^{\alpha\beta}(X^0,k)$ as modifications to the momentum-dependent coupling. It manifests a non-perturbative nature beyond the weak coupling. Therefore, the kinetic description of a bosonic system with a dipole symmetry becomes valid by the $1/N$ expansion.

Multiple potential outlooks can be explored based on the framework in this study. First, the auxiliary field formulation and the 2PI $1/N$ expansion can be generalized to the full class of dipole-invariant $U(N)$ complex scalar field theory, including conventional local quartic and quartic interactions with four spatial derivatives allowed by the dipole symmetry \cite{Jensen:2022ibn}. In addition, the 2PI $1/N$ expansion provides a systematic way to calculate transport coefficients of a given microscopic model \cite{Aarts:2003bk,Aarts:2004sd,Aarts:2005vc}. Second, the evolution equation of $\phi_\textbf{i}^*(x^0)$ and Kadanoff-Baym equations with their alternative expressions provide a concrete starting point for real-time simulations of non-equilibrium dynamics on the lattice, where UV regularization is handled explicitly. Such simulations are able to quantify the lifetime of prethermal regimes and test the emergence of kinetic scaling behavior beyond the mean-field level, which can be compared with the vertex-resummed kinetic theory derived as an late-time effective theory. It is tempting to search for stationary or scaling solutions, e.g. non-thermal fixed points and cascades \cite{PineiroOrioli:2015cpb,Walz:2017ffj,Mikheev:2018adp}, analytically within controlled inertial windows, and determine how their existence and exponents depend on the dipole-induced momentum structure of the effective vertex.

\section*{Acknowledgment}
I am grateful for the hospitality of Shiang-Bin Chiu during my visit to the University of Washington, and for the support from Paul Romatschke in the course of my visit to the TU Wien prior to the completion of this work.

\appendix

\section{Field decomposition in self-energies} 

\label{sec:fide}

The appendix shows field decompositions of several functions present in self-energies by employing Eq.~(\ref{eq:cosg}) and (\ref{eq:coto}).

\subsection{\texorpdfstring
  {$\tilde{K}_{ns^\prime}^\alpha$}
  {K}}
According to Eq.~(\ref{eq:bbbb}), the field decomposition related to $\tilde{K}_{ns^\prime}^\alpha$ to the NLO is given by
\begin{align}
\Omega_{v^\prime b}^\alpha(x^0,\textbf{k})F_{\tilde{K}_{ns^\prime}}^\beta(x^0,y^0,\textbf{k})=&\,-(A^{-1})_{v^\prime h^\prime}^{\alpha\gamma}(\textbf{k})F_{B_{h^\prime s^\prime}}^{\gamma\beta}(x^0,y^0,\textbf{k})-\frac{\bar{\lambda}}{N}(\Pi^F)_{v^\prime h^\prime}^{\alpha\gamma}(x^0,y^0)A_{h^\prime s^\prime}^{\gamma\beta}(\textbf{k})\nonumber\\
&\,-\frac{\bar{\lambda}}{N}\int_{t^0}^{x^0}dz^0\,(\Pi^\rho)_{v^\prime h^\prime}^{\alpha\gamma}(x^0,z^0,\textbf{k})F_{B_{h^\prime s^\prime}}^{\gamma\beta}(z^0,y^0,\textbf{k})\nonumber\\
&\,+\frac{\bar{\lambda}}{N}\int_{t^0}^{y^0}dz^0\,(\Pi^F)_{v^\prime h^\prime}^{\alpha\gamma}(x^0,z^0,\textbf{k})\rho_{B_{h^\prime s^\prime}}^{\gamma\beta}(z^0,y^0,\textbf{k})\,,\\
\Omega_{v^\prime n}^\alpha(x^0,\textbf{k})\rho_{\tilde{K}_{ns^\prime}}^\beta(x^0,y^0,\textbf{k})=&\,-(A^{-1})_{v^\prime h^\prime}^{\alpha\gamma}(\textbf{k})\rho_{B_{h^\prime s^\prime}}^{\gamma\beta}(x^0,y^0,\textbf{k})\nonumber-\frac{\bar{\lambda}}{N}(\Pi^\rho)_{v^\prime h^\prime}^{\alpha\gamma}(x^0,y^0)A_{h^\prime s^\prime}^{\gamma\beta}(\textbf{k})\nonumber\\
&\,-\frac{\bar{\lambda}}{N}\int_{y^0}^{x^0}dz^0\,(\Pi^\rho)_{v^\prime h^\prime}^{\alpha\gamma}(x^0,z^0,\textbf{k})\rho_{B_{h^\prime s^\prime}}^{\gamma\beta}(z^0,y^0,\textbf{k})\,.
\end{align}
By relations in Eq.~(\ref{eq:Ome1}), (\ref{eq:Ome2}), (\ref{eq:Ome3}) and Table.~\ref{tab:FKre}, the above composition of $\tilde{K}_{ns^\prime}^\alpha$ can be converted to that of $K_{s^\prime n}^\alpha$.

\subsection{\texorpdfstring
  {$\tilde{\Sigma}_{nm}$}
  {Sigma-bar }}

The integrand of $\tilde{\Sigma}_{nm}$ excluding the momentum-dependent prefactor in Eq.~(\ref{eq:mosi}) is decomposed into $F$ part and $\rho$ part given by
\begin{align}
&\,(\bar{J}^F)_{s^\prime v^\prime}^{\alpha\beta}(z^0,x^0,-\textbf{q})F_{G_{nm}}(x^0,z^0,\textbf{k}-\textbf{q})+\frac{1}{4}(\bar{J}^\rho)_{s^\prime v^\prime}^{\alpha\beta}(z^0,x^0,-\textbf{q})\rho_{G_{nm}}(x^0,z^0,\textbf{k}-\textbf{q})\nonumber\\
+&\,(W^F)_{s^\prime v^\prime}^{\alpha\beta}(z^0,x^0,-\textbf{q})F_{G_{nm}}(x^0,z^0,,\textbf{k}-\textbf{q})+\frac{1}{4}(W^\rho)_{s^\prime v^\prime}^{\alpha\beta}(z^0,x^0,-\textbf{q})\rho_{G_{nm}}(x^0,z^0,\textbf{k}-\textbf{q})\,,
\label{eq:bSiF1}
\end{align}
and
\begin{align}
&\, -(\bar{J}^\rho)_{s^\prime v^\prime}^{\alpha\beta}(z^0,x^0,-\textbf{q})F_{G_{nm}}(x^0,z^0,\textbf{k}-\textbf{q})+(\bar{J}^F)_{s^\prime v^\prime}^{\alpha\beta}(z^0,x^0,-\textbf{q})\rho_{G_{nm}}(x^0,z^0,\textbf{k}-\textbf{q})\nonumber\\
&\,-(W^\rho)_{s^\prime v^\prime}^{\alpha\beta}(z^0,x^0,-\textbf{q})F_{G_{nm}}(x^0,z^0,\textbf{k}-\textbf{q})+(W^F)_{s^\prime v^\prime}^{\alpha\beta}(z^0,x^0,-\textbf{q})\rho_{G_{nm}}(x^0,z^0,\textbf{k}-\textbf{q})
\label{eq:bSir1}
\end{align}
where
\begin{align}
(W^F)_{s^\prime v^\prime}^{\alpha\beta}(x^0,y^0,\textbf{k})
=&\,-\frac{\bar{\lambda}}{N}\bigg[A_{s^\prime h^\prime}^{\alpha\mu}(\textbf{k})(Q^F)_{h^\prime l^\prime}^{\mu\nu}(x^0,y^0,\textbf{k})A_{l^\prime v^\prime}^{\nu\beta}(\textbf{k})\nonumber\\
&-\int_{t^0}^{x^0}dz^0\bigg(A_{s^\prime l^\prime}^{\alpha\sigma}(\textbf{k})(Q^\rho)_{l^\prime h^\prime}^{\sigma\gamma}(x^0,z^0,\textbf{k})(\bar{J}^F)_{h^\prime v^\prime}^{\gamma\beta}(z^0,y^0,\textbf{k})\nonumber\\
&\,+(\bar{J}^\rho)_{s^\prime h^\prime}^{\alpha\gamma}(x^0,z^0,\textbf{k})(Q^F)_{h^\prime l^\prime}^{\gamma\sigma}(z^0,y^0,\textbf{k})A_{l^\prime v^\prime}^{\sigma\beta}(\textbf{k})\bigg)\nonumber\\
&\,+\int_{t^0}^{y^0}dz^0\bigg(A_{s^\prime l^\prime}^{\alpha\sigma}(\textbf{k})(Q^F)_{l^\prime h^\prime}^{\sigma\gamma}(x^0,z^0,\textbf{k})(\bar{J}^\rho)_{h^\prime v^\prime}^{\gamma\beta}(z^0,y^0,\textbf{k})\nonumber\\
&\,+(\bar{J}^F)_{s^\prime h^\prime}^{\alpha\gamma}(x^0,z^0,\textbf{k})(Q^\rho)_{h^\prime l^\prime}^{\gamma\sigma}(z^0,y^0,\textbf{k})A_{l^\prime v^\prime}^{\sigma\beta}(\textbf{k})\bigg)\nonumber\\
&\,-\int_{t^0}^{x^0}dz^0\int_{t^0}^{y^0}du^0 (\bar{J}^\rho)_{s^\prime h^\prime}^{\alpha\mu}(x^0,z^0,\textbf{k})(Q^F)_{h^\prime c^\prime}^{\mu\nu}(z^0,u^0,\textbf{k})(\bar{J}^\rho)_{c^\prime v^\prime}^{\nu\beta}(u^0,y^0,\textbf{k})\nonumber\\
&\,+\int_{t^0}^{x^0}dz^0\int_{t^0}^{z^0}du^0 (\bar{J}^\rho)_{s^\prime h^\prime}^{\alpha\mu}(x^0,z^0,\textbf{k})(Q^\rho)_{h^\prime c^\prime}^{\mu\nu}(z^0,u^0,\textbf{k})(\bar{J}^F)_{c^\prime v^\prime}^{\nu\beta}(u^0,y^0,\textbf{k})\nonumber\\
&\,+\int_{t^0}^{y^0}dz^0\int_{z^0}^{y^0}du^0 (\bar{J}^F)_{s^\prime h^\prime}^{\alpha\mu}(x^0,z^0,\textbf{k})(Q^\rho)_{h^\prime c^\prime}^{\mu\nu}(z^0,u^0,\textbf{k})(\bar{J}^\rho)_{c^\prime v^\prime}^{\nu\beta}(u^0,y^0,\textbf{k})\bigg]\,,
\label{eq:WFex}
\end{align}
\begin{align}
(W^\rho)_{s^\prime v^\prime}^{\alpha\beta}(x^0,y^0,\textbf{k})
=&\,-\frac{\bar{\lambda}}{N}\bigg[A_{s^\prime h^\prime}^{\alpha\mu}(\textbf{k})(Q^\rho)_{h^\prime l^\prime}^{\mu\nu}(x^0,y^0,\textbf{k})A_{l^\prime s^\prime}^{\nu\beta}(\textbf{k})\nonumber\\
&\,-\int_{y^0}^{x^0}dz^0\bigg(A_{s^\prime l^\prime}^{\alpha\sigma}(\textbf{k})(Q^\rho)_{l^\prime h^\prime}^{\sigma\gamma}(x^0,z^0,\textbf{k})(\bar{J}^\rho)_{h^\prime v^\prime}^{\gamma\beta}(z^0,y^0,\textbf{k})\nonumber\\
&\,+(\bar{J}^\rho)_{s^\prime h^\prime}^{\alpha\gamma}(x^0,z^0,\textbf{k})(Q^\rho)_{h^\prime l^\prime}^{\gamma\sigma}(z^0,y^0,\textbf{k})A_{l^\prime v^\prime}^{\sigma\beta}(\textbf{k})\bigg)\nonumber\\
&\,+\int_{y^0}^{x^0}dz^0\int_{y^0}^{z^0}du^0(\bar{J}^\rho)_{s^\prime h^\prime}^{\alpha\mu}(x^0,z^0,\textbf{k})(Q^\rho)_{h^\prime c^\prime}^{\mu\nu}(z^0,u^0,\textbf{k})(\bar{J}^\rho)_{c^\prime v^\prime}^{\nu\beta}(u^0,y^0,\textbf{k})\bigg]\,.
\label{eq:Wrex}
\end{align}
After taking the limit $t^0\rightarrow-\infty$, upper (lower) bounds of integrals in Eq.~(\ref{eq:WFex}) and (\ref{eq:Wrex}) can be extended to $\infty$ ($-\infty$) by defining the retarded and advanced two-point function similar to definitions in Eq.~(\ref{eq:Ratt}) and (\ref{eq:RAde})
\begin{align}
(W^F)_{s^\prime v^\prime}^{\alpha\beta}(x^0,y^0,\textbf{k})=&\,-\frac{\bar{\lambda}}{N}\bigg[A_{s^\prime h^\prime}^{\alpha\mu}(\textbf{k})(Q^F)_{h^\prime l^\prime}^{\mu\nu}(x^0,y^0,\textbf{k})A_{l^\prime v^\prime}^{\nu\beta}(\textbf{k})\nonumber\\
&\,-\int_{-\infty}^\infty dz^0\,A_{s^\prime l^\prime}^{\alpha\sigma}(\textbf{k})(Q_R^\rho)_{l^\prime h^\prime}^{\sigma\gamma}(x^0,z^0,\textbf{k})(\bar{J}^F)_{h^\prime v^\prime}^{\gamma\beta}(z^0,y^0,\textbf{k})\nonumber\\
&\,-\int_{-\infty}^\infty dz^0\,(\bar{J}_R^\rho)_{s^\prime h^\prime}^{\alpha\gamma}(x^0,z^0,\textbf{k})(Q^F)_{h^\prime l^\prime}^{\gamma\sigma}(z^0,y^0,\textbf{k})A_{l^\prime v^\prime}^{\sigma\beta}(\textbf{k})\nonumber\\
&\,-\int_{-\infty}^\infty dz^0\,A_{s^\prime l^\prime}^{\alpha\sigma}(\textbf{k})(Q^F)_{l^\prime h^\prime}^{\sigma\gamma}(x^0,z^0,\textbf{k})(\bar{J}_A^\rho)_{h^\prime v^\prime}^{\gamma\beta}(z^0,y^0,\textbf{k})\nonumber\\
&\,-\int_{-\infty}^\infty dz^0\,(\bar{J}^F)_{s^\prime h^\prime}^{\alpha\gamma}(x^0,z^0,\textbf{k})(Q_A^\rho)_{h^\prime l^\prime}^{\gamma\sigma}(z^0,y^0,\textbf{k})A_{l^\prime v^\prime}^{\sigma\beta}(\textbf{k})\nonumber\\
&\,+\int_{-\infty}^\infty dz^0\int_{-\infty}^\infty du^0\,\bigg((\bar{J}_R^\rho)_{s^\prime h^\prime}^{\alpha\mu}(x^0,z^0,\textbf{k})(Q^F)_{h^\prime c^\prime}^{\mu\nu}(z^0,u^0,\textbf{k})(\bar{J}_A^\rho)_{c^\prime v^\prime}^{\nu\beta}(u^0,y^0,\textbf{k})\nonumber\\
&\,+(\bar{J}_R^\rho)_{s^\prime h^\prime}^{\alpha\mu}(x^0,z^0,\textbf{k})(Q_R^\rho)_{h^\prime c^\prime}^{\mu\nu}(z^0,u^0,\textbf{k})(\bar{J}^F)_{c^\prime v^\prime}^{\nu\beta}(u^0,y^0,\textbf{k})\nonumber\\
&\,+(\bar{J}^F)_{s^\prime h^\prime}^{\alpha\mu}(x^0,z^0,\textbf{k})(Q_A^\rho)_{h^\prime c^\prime}^{\mu\nu}(z^0,u^0,\textbf{k})(\bar{J}_A^\rho)_{c^\prime v^\prime}^{\nu\beta}(u^0,y^0,\textbf{k})\bigg)\bigg]\,,
\label{eq:WFre}
\end{align}

\begin{align}
(W^{\tilde{\rho}})_{s^\prime v^\prime}^{\alpha\beta}(x^0,y^0,\textbf{k})\equiv&\,-i(W^\rho)_{s^\prime v^\prime}^{\alpha\beta}(x^0,y^0,\textbf{k})\nonumber\\
=&\,-\frac{\bar{\lambda}}{N}\bigg[A_{s^\prime h^\prime}^{\alpha\mu}(\textbf{k})(Q^{\tilde{\rho}})_{h^\prime l^\prime}^{\mu\nu}(x^0,y^0,\textbf{k})A_{l^\prime v^\prime}^{\nu\beta}(\textbf{k})\nonumber\\
&\,-\int_{-\infty}^\infty dz^0\,A_{s^\prime l^\prime}^{\alpha\sigma}(\textbf{k})(Q_R^\rho)_{l^\prime h^\prime}^{\sigma\gamma}(x^0,z^0,\textbf{k})(\bar{J}^{\tilde{\rho}})_{h^\prime v^\prime}^{\gamma\beta}(z^0,y^0,\textbf{k})\nonumber\\
&\,-\int_{-\infty}^\infty dz^0\,A_{s^\prime l^\prime}^{\alpha\sigma}(\textbf{k})(Q^{\tilde{\rho}})_{l^\prime h^\prime}^{\sigma\gamma}(x^0,z^0,\textbf{k})(\bar{J}_A^\rho)_{h^\prime v^\prime}^{\gamma\beta}(z^0,y^0,\textbf{k})\nonumber\\
&\,-\int_{-\infty}^\infty dz^0\, (\bar{J}_R^\rho)_{s^\prime h^\prime}^{\alpha\gamma}(x^0,z^0,\textbf{k})(Q^{\tilde{\rho}})_{h^\prime l^\prime}^{\gamma\sigma}(z^0,y^0,\textbf{k}) A_{l^\prime v^\prime}^{\sigma\beta}(\textbf{k})\nonumber\\
&\,-\int_{-\infty}^\infty dz^0\, (\bar{J}^{\tilde{\rho}})_{s^\prime h^\prime}^{\alpha\gamma}(x^0,z^0,\textbf{k})(Q_A^\rho)_{h^\prime l^\prime}^{\gamma\sigma}(z^0,y^0,\textbf{k})A_{l^\prime v^\prime}^{\sigma\beta}(\textbf{k})\nonumber\\
&\,+\int_{-\infty}^\infty dz^0\int_{-\infty}^\infty du^0\,\bigg((\bar{J}_R^\rho)_{s^\prime h^\prime}^{\alpha\mu}(x^0,z^0,\textbf{k})(Q_R^\rho)_{h^\prime c^\prime}^{\mu\nu}(z^0,u^0,\textbf{k})(\bar{J}^{\tilde{\rho}})_{c^\prime v^\prime}^{\nu\beta}(u^0,y^0,\textbf{k})\nonumber\\
&\,+(\bar{J}_R^\rho)_{s^\prime h^\prime}^{\alpha\mu}(x^0,z^0,\textbf{k})(Q^{\tilde{\rho}})_{h^\prime c^\prime}^{\mu\nu}(z^0,u^0,\textbf{k})(\bar{J}_A^\rho)_{c^\prime v^\prime}^{\nu\beta}(u^0,y^0,\textbf{k})\nonumber\\
&\,+(\bar{J}^{\tilde{\rho}})_{s^\prime h^\prime}^{\alpha\mu}(x^0,z^0,\textbf{k})(Q_A^\rho)_{h^\prime c^\prime}^{\mu\nu}(z^0,u^0,\textbf{k})(\bar{J}_A^\rho)_{c^\prime v^\prime}^{\nu\beta}(u^0,y^0,\textbf{k})\bigg)\bigg]\,,
\label{eq:Wrre}
\end{align}
where the integration range from $-\infty$ to $\infty$ enables one to do the gradient expansion of $\partial_{X^0}$ after the Fourier transform with respect to relative coordinates. The relation
\begin{align}
&\,\int_{y^0}^{x^0} dz^0\, Y(x^0,z^0,\textbf{k})Z(z^0,y^0,\textbf{k})\nonumber\\
=&\,\int_{-\infty}^\infty dz^0\, Y_R(x^0,z^0,\textbf{k})Z_R(z^0,y^0,\textbf{k})\nonumber\\
=&\,\int_{-\infty}^\infty dz^0\, Y_R(x^0,z^0,\textbf{k}
)\Big(Z(z^0,y^0,\textbf{k})+Z_A(z^0,y^0,\textbf{k})\Big)\nonumber\\
=&\,\int_{-\infty}^\infty dz^0\, \Big(Y_R(x^0,z^0,\textbf{k}
)Z(z^0,y^0,\textbf{k})+Y(x^0,z^0,\textbf{k})Z_A(z^0,y^0,\textbf{k})\Big)\,.
\label{eq:rett}
\end{align} for the convolution of two retarded two-point functions $Y_R$ and $Z_R$ is applied above in Eq.~(\ref{eq:Wrre}). In addition, the other relation including the third retarded two-point function $C_R$ used in Eq.~(\ref{eq:Wrre})
\begin{align}
&\,\int_{y^0}^{x^0}dz^0\int_{y^0}^{z^0}du^0\,C(x^0,z^0,\textbf{k})Y(z^0,u^0,\textbf{k})Z(u^0,y^0,\textbf{k})
\equiv\int_{y^0}^{x^0}dz^0\,C(x^0,z^0,\textbf{k})E(z^0,y^0,\textbf{k})\nonumber\\
=&\,\int_{y^0}^{x^0}dz^0\,\Big(C_R(x^0,z^0,\textbf{k})E(z^0,y^0,\textbf{k})+C(x^0,z^0,\textbf{k})E_A(z^0,y^0,\textbf{k})\Big)
\label{eq:sasa}
\end{align}
can be derived by Eq.~(\ref{eq:rett}) as well. It is noticed that the advanced part of $E(z^0,y^0,\textbf{k})$ can be extracted using Eq.~(\ref{eq:rett})
\begin{align}
E(z^0,y^0,\textbf{k})=&\,\int_{y^0}^{z^0}du^0\,Y(z^0,u^0,\textbf{k})Z(u^0,y^0,\textbf{k})\nonumber\\
=&\,\int_{-\infty}^\infty du^0\,\bigg[Y_R(z^0,u^0,\textbf{k})\Big(Z_R(u^0,y^0,\textbf{k})-Z_A(u^0,y^0,\textbf{k})\Big)\nonumber\\
&\,+\Big(Y_R(z^0,u^0,\textbf{k})-Y_A(z^0,u^0,\textbf{k})\Big)Z_A(u^0,y^0,\textbf{k})\bigg]\nonumber\\
=&\,\int_{-\infty}^\infty du^0\,\Big(Y_R(z^0,u^0,\textbf{k})Z_R(u^0,y^0,\textbf{k})-Y_A(z^0,u^0,\textbf{k})Z_A(u^0,y^0,\textbf{k})\Big)\label{eq:prog}\\
\Rightarrow\quad\quad\quad&\,E_A(z^0,y^0,\textbf{k})=\int_{-\infty}^\infty du^0\,Y_A(z^0,u^0,\textbf{k})Z_A(u^0,y^0,\textbf{k})\,.
\label{eq:EaEa}
\end{align}
Therefore, Eq.~(\ref{eq:sasa}) becomes
\begin{align}
&\,\int_{y^0}^{x^0}dz^0\int_{y^0}^{z^0}du^0\,\Big(C_R(x^0,z^0,\textbf{k})Y_R(z^0,u^0,\textbf{k})Z_R(u^0,y^0,\textbf{k})\nonumber\\
&\,-C_R(x^0,z^0,\textbf{k})Y_A(z^0,u^0,\textbf{k})Z_A(u^0,y^0,\textbf{k})+C(x^0,z^0,\textbf{k})Y_A(z^0,u^0,\textbf{k})Z_A(u^0,y^0,\textbf{k})\Big)\nonumber\\
=&\,\int_{y^0}^{x^0}dz^0\int_{y^0}^{z^0}du^0\,\bigg[C_R(x^0,z^0,\textbf{k})Y_R(z^0,u^0,\textbf{k})Z_R(u^0,y^0,\textbf{k})\nonumber\\
&\,-C_R(x^0,z^0,\textbf{k})\Big(Y_R(z^0,u^0,\textbf{k})-Y(z^0,u^0,\textbf{k})\Big)Z_A(u^0,y^0,\textbf{k})\nonumber\\
&\,+C(x^0,z^0,\textbf{k})Y_A(z^0,u^0,\textbf{k})Z_A(u^0,y^0,\textbf{k})\bigg]\nonumber\\
=&\,\int_{y^0}^{x^0}dz^0\int_{y^0}^{z^0}du^0\,\bigg[C_R(x^0,z^0,\textbf{k})Y_R(z^0,u^0,\textbf{k})Z(u^0,y^0,\textbf{k})\nonumber\\
&\,+C_R(x^0,z^0,\textbf{k})Y(z^0,u^0,\textbf{k})Z_A(u^0,y^0,\textbf{k})+C(x^0,z^0,\textbf{k})Y_A(z^0,u^0,\textbf{k})Z_A(u^0,y^0,\textbf{k})\bigg]
\end{align}
by plugging in Eq.~(\ref{eq:prog}) and (\ref{eq:EaEa}).

\subsection{\texorpdfstring
  {$\Pi_{v's'}^{\beta\alpha}$ }
  {Pi-bar}}
Self-energies $(\Pi)_{v^\prime s^\prime}^{\beta\alpha}$ shown in Eq.~(\ref{eq:Pisi}) has a decomposition given by
\begin{align}
(\Pi^F)_{v^\prime s^\prime}^{\beta\alpha}(x^0,y^0,k)=&\,-\int_q \bigg[ S_{s^\prime v^\prime}^{\alpha\beta}(\textbf{q},-\textbf{k})\bigg(F_{G_{(2\textbf{j}-1)(2\textbf{i}-1)}}(y^0,x^0,\textbf{q})F_{G_{(2\textbf{i}-1)(2\textbf{j}-1)}}(x^0,y^0,\textbf{k}+\textbf{q})\nonumber\\
&\,+\frac{1}{4}\rho_{G_{(2\textbf{j}-1)(2\textbf{i}-1)}}(y^0,x^0,\textbf{q})\rho_{G_{(2\textbf{i}-1)(2\textbf{j}-1)}}(x^0,y^0,\textbf{k}+\textbf{q})\bigg)\nonumber\\
&\,+S_{s^\prime v^\prime}^{\alpha\beta}(\textbf{q})\bigg(F_{G_{(2\textbf{j})(2\textbf{i}-1)}}(y^0,x^0,\textbf{q})F_{G_{(2\textbf{i}-1)(2\textbf{j})}}(x^0,y^0,\textbf{k}+\textbf{q})\nonumber\\
&\,+\frac{1}{4}\rho_{G_{(2\textbf{j})(2\textbf{i}-1)}}(y^0,x^0,\textbf{q})\rho_{G_{(2\textbf{i}-1)(2\textbf{j})}}(x^0,y^0,\textbf{k}+\textbf{q})\bigg)\bigg]\,,\label{eq:PiFF}\\
(\Pi^\rho)_{v^\prime s^\prime}^{\beta\alpha}(x^0,y^0,k)=&\,-\int_q\bigg[S_{s^\prime v^\prime}^{\alpha\beta}(\textbf{q},-\textbf{k})\bigg(-\rho_{G_{(2\textbf{j}-1)(2\textbf{i}-1)}}(y^0,x^0,\textbf{q})F_{G_{(2\textbf{i}-1)(2\textbf{j}-1)}}(x^0,y^0,\textbf{k}+\textbf{q})\nonumber\\
&\,+F_{G_{(2\textbf{j}-1)(2\textbf{i}-1)}}(y^0,x^0,\textbf{q})\rho_{G_{(2\textbf{i}-1)(2\textbf{j}-1)}}(x^0,y^0,\textbf{k}+\textbf{q})\bigg)\nonumber\\
&\,+S_{s^\prime v^\prime}^{\alpha\beta}(\textbf{q})\bigg(-\rho_{G_{(2\textbf{j})(2\textbf{i}-1)}}(y^0,x^0,\textbf{q})F_{G_{(2\textbf{i}-1)(2\textbf{j})}}(x^0,y^0,\textbf{k}+\textbf{q})\nonumber\\
&\,+F_{G_{(2\textbf{j})(2\textbf{i}-1)}}(y^0,x^0,\textbf{q})\rho_{G_{(2\textbf{i}-1)(2\textbf{j})}}(x^0,y^0,\textbf{k}+\textbf{q})\bigg)\bigg]\,,\label{eq:Pirr}
\end{align}

\bibliography{fra}

@article{Jensen:2022ibn,
    author = "Jensen, Kristan and Raz, Amir",
    title = "{Large N Fractons}",
    eprint = "2205.01132",
    archivePrefix = "arXiv",
    primaryClass = "hep-th",
    doi = "10.1103/PhysRevLett.132.071603",
    journal = "Phys. Rev. Lett.",
    volume = "132",
    number = "7",
    pages = "071603",
    year = "2024"
}

@article{Pretko:2020cko,
    author = "Pretko, Michael and Chen, Xie and You, Yizhi",
    title = "{Fracton Phases of Matter}",
    eprint = "2001.01722",
    archivePrefix = "arXiv",
    primaryClass = "cond-mat.str-el",
    doi = "10.1142/S0217751X20300033",
    journal = "Int. J. Mod. Phys. A",
    volume = "35",
    number = "06",
    pages = "2030003",
    year = "2020"
}

@article{Ma:2017aog,
    author = "Ma, Han and Lake, Ethan and Chen, Xie and Hermele, Michael",
    title = "{Fracton topological order via coupled layers}",
    eprint = "1701.00747",
    archivePrefix = "arXiv",
    primaryClass = "cond-mat.str-el",
    reportNumber = "NSF-KITP-17-003, NSF-KITP-17-003",
    doi = "10.1103/PhysRevB.95.245126",
    journal = "Phys. Rev. B",
    volume = "95",
    number = "24",
    pages = "245126",
    year = "2017"
}

@article{Nandkishore:2018sel,
    author = "Nandkishore, Rahul M. and Hermele, Michael",
    title = "{Fractons}",
    eprint = "1803.11196",
    archivePrefix = "arXiv",
    primaryClass = "cond-mat.str-el",
    doi = "10.1146/annurev-conmatphys-031218-013604",
    journal = "Ann. Rev. Condensed Matter Phys.",
    volume = "10",
    pages = "295--313",
    year = "2019"
}

@article{Pretko:2016lgv,
    author = "Pretko, Michael",
    title = "{Generalized Electromagnetism of Subdimensional Particles: A Spin Liquid Story}",
    eprint = "1606.08857",
    archivePrefix = "arXiv",
    primaryClass = "cond-mat.str-el",
    doi = "10.1103/PhysRevB.96.035119",
    journal = "Phys. Rev. B",
    volume = "96",
    number = "3",
    pages = "035119",
    year = "2017"
}

@article{Pretko:2016kxt,
    author = "Pretko, Michael",
    title = "{Subdimensional Particle Structure of Higher Rank U(1) Spin Liquids}",
    eprint = "1604.05329",
    archivePrefix = "arXiv",
    primaryClass = "cond-mat.str-el",
    doi = "10.1103/PhysRevB.95.115139",
    journal = "Phys. Rev. B",
    volume = "95",
    number = "11",
    pages = "115139",
    year = "2017"
}

@article{Haah:2011drr,
    author = "Haah, Jeongwan",
    title = "{Local stabilizer codes in three dimensions without string logical operators}",
    eprint = "1101.1962",
    archivePrefix = "arXiv",
    primaryClass = "quant-ph",
    reportNumber = "CALT-68-2816, CALT-68-2816",
    doi = "10.1103/physreva.83.042330",
    journal = "Phys. Rev. A",
    volume = "83",
    number = "4",
    pages = "042330",
    year = "2011"
}

@article{Gromov:2017vir,
    author = "Gromov, Andrey",
    title = "{Chiral Topological Elasticity and Fracton Order}",
    eprint = "1712.06600",
    archivePrefix = "arXiv",
    primaryClass = "cond-mat.str-el",
    reportNumber = "EFI-17-27",
    doi = "10.1103/PhysRevLett.122.076403",
    journal = "Phys. Rev. Lett.",
    volume = "122",
    number = "7",
    pages = "076403",
    year = "2019"
}

@article{Vijay:2016phm,
    author = "Vijay, Sagar and Haah, Jeongwan and Fu, Liang",
    title = "{Fracton Topological Order, Generalized Lattice Gauge Theory and Duality}",
    eprint = "1603.04442",
    archivePrefix = "arXiv",
    primaryClass = "cond-mat.str-el",
    doi = "10.1103/PhysRevB.94.235157",
    journal = "Phys. Rev. B",
    volume = "94",
    number = "23",
    pages = "235157",
    year = "2016"
}

@article{Pretko_2018,
   title={Fracton-Elasticity Duality},
   volume={120},
   ISSN={1079-7114},
   url={http://dx.doi.org/10.1103/PhysRevLett.120.195301},
   DOI={10.1103/physrevlett.120.195301},
   number={19},
   journal={Physical Review Letters},
   publisher={American Physical Society (APS)},
   author={Pretko, Michael and Radzihovsky, Leo},
   year={2018},
   month=may }

@article{Pretko:2017xar,
    author = "Pretko, Michael",
    title = "{Higher-Spin Witten Effect and Two-Dimensional Fracton Phases}",
    eprint = "1707.03838",
    archivePrefix = "arXiv",
    primaryClass = "cond-mat.str-el",
    doi = "10.1103/PhysRevB.96.125151",
    journal = "Phys. Rev. B",
    volume = "96",
    number = "12",
    pages = "125151",
    year = "2017"
}

@article{Pretko:2017fbf,
    author = "Pretko, Michael",
    title = "{Emergent gravity of fractons: Mach{\textquoteright}s principle revisited}",
    eprint = "1702.07613",
    archivePrefix = "arXiv",
    primaryClass = "cond-mat.str-el",
    doi = "10.1103/PhysRevD.96.024051",
    journal = "Phys. Rev. D",
    volume = "96",
    number = "2",
    pages = "024051",
    year = "2017"
}

@article{Aarts:2002dj,
    author = "Aarts, Gert and Ahrensmeier, Daria and Baier, Rudolf and Berges, Juergen and Serreau, Julien",
    title = "{Far from equilibrium dynamics with broken symmetries from the 2PI - 1/N expansion}",
    eprint = "hep-ph/0201308",
    archivePrefix = "arXiv",
    reportNumber = "HD-THEP-02-5",
    doi = "10.1103/PhysRevD.66.045008",
    journal = "Phys. Rev. D",
    volume = "66",
    pages = "045008",
    year = "2002"
}

@article{Islam:2023cmm,
    author = "Islam, Md Mursalin and Sengupta, K. and Sensarma, Rajdeep",
    title = "{Nonequilibrium dynamics of bosons with dipole symmetry: Large-N Keldysh approach}",
    eprint = "2305.13372",
    archivePrefix = "arXiv",
    primaryClass = "cond-mat.quant-gas",
    reportNumber = "TIFR/TH/23-9",
    doi = "10.1103/PhysRevB.108.214314",
    journal = "Phys. Rev. B",
    volume = "108",
    number = "21",
    pages = "214314",
    year = "2023"
}

@article{Cornwall:1974vz,
    author = "Cornwall, John M. and Jackiw, R. and Tomboulis, E.",
    title = "{Effective Action for Composite Operators}",
    reportNumber = "MIT-CTP-419",
    doi = "10.1103/PhysRevD.10.2428",
    journal = "Phys. Rev. D",
    volume = "10",
    pages = "2428--2445",
    year = "1974"
}

@article{Berges:2004yj,
    author = "Berges, Juergen",
    editor = "Bracco, Mirian and Chiapparini, Marcelo and Ferreira, Erasmo and Kodama, Takeshi",
    title = "{Introduction to nonequilibrium quantum field theory}",
    eprint = "hep-ph/0409233",
    archivePrefix = "arXiv",
    doi = "10.1063/1.1843591",
    journal = "AIP Conf. Proc.",
    volume = "739",
    number = "1",
    pages = "3--62",
    year = "2004"
}

@misc{berges2015nonequilibriumquantumfieldscold,
      title={Nonequilibrium Quantum Fields: From Cold Atoms to Cosmology}, 
      author={J. Berges},
      year={2015},
      eprint={1503.02907},
      archivePrefix={arXiv},
      primaryClass={hep-ph},
      url={https://arxiv.org/abs/1503.02907}, 
}

@article{Pretko:2018jbi,
    author = "Pretko, Michael",
    title = "{The Fracton Gauge Principle}",
    eprint = "1807.11479",
    archivePrefix = "arXiv",
    primaryClass = "cond-mat.str-el",
    doi = "10.1103/PhysRevB.98.115134",
    journal = "Phys. Rev. B",
    volume = "98",
    number = "11",
    pages = "115134",
    year = "2018"
}

@article{Berges:2001fi,
    author = "Berges, Juergen",
    title = "{Controlled nonperturbative dynamics of quantum fields out-of-equilibrium}",
    eprint = "hep-ph/0105311",
    archivePrefix = "arXiv",
    reportNumber = "HD-THEP-01-26",
    doi = "10.1016/S0375-9474(01)01295-7",
    journal = "Nucl. Phys. A",
    volume = "699",
    pages = "847--886",
    year = "2002"
}

@article{Seiberg:2020bhn,
    author = "Seiberg, Nathan and Shao, Shu-Heng",
    title = "{Exotic Symmetries, Duality, and Fractons in 2+1-Dimensional Quantum Field Theory}",
    eprint = "2003.10466",
    archivePrefix = "arXiv",
    primaryClass = "cond-mat.str-el",
    doi = "10.21468/SciPostPhys.10.2.027",
    journal = "SciPost Phys.",
    volume = "10",
    number = "2",
    pages = "027",
    year = "2021"
}

@article{Seiberg:2020wsg,
    author = "Seiberg, Nathan and Shao, Shu-Heng",
    title = "{Exotic $U(1)$ Symmetries, Duality, and Fractons in 3+1-Dimensional Quantum Field Theory}",
    eprint = "2004.00015",
    archivePrefix = "arXiv",
    primaryClass = "cond-mat.str-el",
    doi = "10.21468/SciPostPhys.9.4.046",
    journal = "SciPost Phys.",
    volume = "9",
    number = "4",
    pages = "046",
    year = "2020"
}

@article{Gorantla:2021bda,
    author = "Gorantla, Pranay and Lam, Ho Tat and Seiberg, Nathan and Shao, Shu-Heng",
    title = "{Low-energy limit of some exotic lattice theories and UV/IR mixing}",
    eprint = "2108.00020",
    archivePrefix = "arXiv",
    primaryClass = "cond-mat.str-el",
    doi = "10.1103/PhysRevB.104.235116",
    journal = "Phys. Rev. B",
    volume = "104",
    number = "23",
    pages = "235116",
    year = "2021"
}

@article{Berges:2015ixa,
    author = "Berges, J. and Boguslavski, K. and Schlichting, S. and Venugopalan, R.",
    title = "{Nonequilibrium fixed points in longitudinally expanding scalar theories: infrared cascade, Bose condensation and a challenge for kinetic theory}",
    eprint = "1508.03073",
    archivePrefix = "arXiv",
    primaryClass = "hep-ph",
    reportNumber = "INT-PUB-15-043",
    doi = "10.1103/PhysRevD.92.096006",
    journal = "Phys. Rev. D",
    volume = "92",
    number = "9",
    pages = "096006",
    year = "2015"
}

@article{Walz:2017ffj,
    author = "Walz, R. and Boguslavski, K. and Berges, J.",
    title = "{Large-N kinetic theory for highly occupied systems}",
    eprint = "1710.11146",
    archivePrefix = "arXiv",
    primaryClass = "hep-ph",
    doi = "10.1103/PhysRevD.97.116011",
    journal = "Phys. Rev. D",
    volume = "97",
    number = "11",
    pages = "116011",
    year = "2018"
}

@article{PineiroOrioli:2015cpb,
    author = "Pi{\~n}eiro Orioli, A. and Boguslavski, K. and Berges, J.",
    title = "{Universal self-similar dynamics of relativistic and nonrelativistic field theories near nonthermal fixed points}",
    eprint = "1503.02498",
    archivePrefix = "arXiv",
    primaryClass = "hep-ph",
    doi = "10.1103/PhysRevD.92.025041",
    journal = "Phys. Rev. D",
    volume = "92",
    number = "2",
    pages = "025041",
    year = "2015"
}

@article{Khlebnikov:1996mc,
    author = "Khlebnikov, S. Yu. and Tkachev, I. I.",
    title = "{Classical decay of inflaton}",
    eprint = "hep-ph/9603378",
    archivePrefix = "arXiv",
    reportNumber = "PURD-TH-96-02, OSU-TA-08-96",
    doi = "10.1103/PhysRevLett.77.219",
    journal = "Phys. Rev. Lett.",
    volume = "77",
    pages = "219--222",
    year = "1996"
}

@article{Bettencourt:1997nf,
    author = "Bettencourt, Luis M. A. and Wetterich, Christof",
    title = "{Time evolution of correlation functions in nonequilibrium field theories}",
    eprint = "hep-ph/9712429",
    archivePrefix = "arXiv",
    reportNumber = "HD-THEP-97-59, LAUR-97-4974, LAUR---974974",
    doi = "10.1016/S0370-2693(98)00509-7",
    journal = "Phys. Lett. B",
    volume = "430",
    pages = "140--150",
    year = "1998"
}

@article{Kofman:1997yn,
    author = "Kofman, Lev and Linde, Andrei D. and Starobinsky, Alexei A.",
    title = "{Towards the theory of reheating after inflation}",
    eprint = "hep-ph/9704452",
    archivePrefix = "arXiv",
    reportNumber = "IFA-97-28, SU-ITP-97-18",
    doi = "10.1103/PhysRevD.56.3258",
    journal = "Phys. Rev. D",
    volume = "56",
    pages = "3258--3295",
    year = "1997"
}

@article{Mihaila:2000ib,
    author = "Mihaila, Bogdan and Athan, Tara and Cooper, Fred and Dawson, John and Habib, Salman",
    title = "{Exact and approximate dynamics of the quantum mechanical O(N) model}",
    eprint = "hep-ph/0003105",
    archivePrefix = "arXiv",
    doi = "10.1103/PhysRevD.62.125015",
    journal = "Phys. Rev. D",
    volume = "62",
    pages = "125015",
    year = "2000"
}

@article{Berges:2000ur,
    author = "Berges, Juergen and Cox, Jurgen",
    title = "{Thermalization of quantum fields from time reversal invariant evolution equations}",
    eprint = "hep-ph/0006160",
    archivePrefix = "arXiv",
    reportNumber = "MIT-CTP-2988",
    doi = "10.1016/S0370-2693(01)01004-8",
    journal = "Phys. Lett. B",
    volume = "517",
    pages = "369--374",
    year = "2001"
}

@article{Berges:2010ez,
    author = "Berges, Jurgen and Sexty, Denes",
    title = "{Strong versus weak wave-turbulence in relativistic field theory}",
    eprint = "1012.5944",
    archivePrefix = "arXiv",
    primaryClass = "hep-ph",
    doi = "10.1103/PhysRevD.83.085004",
    journal = "Phys. Rev. D",
    volume = "83",
    pages = "085004",
    year = "2011"
}

@article{Berges:2017ldx,
    author = "Berges, J. and Boguslavski, K. and Chatrchyan, A. and Jaeckel, J.",
    title = "{Attractive versus repulsive interactions in the Bose-Einstein condensation dynamics of relativistic field theories}",
    eprint = "1707.07696",
    archivePrefix = "arXiv",
    primaryClass = "hep-ph",
    doi = "10.1103/PhysRevD.96.076020",
    journal = "Phys. Rev. D",
    volume = "96",
    number = "7",
    pages = "076020",
    year = "2017"
}

@article{Berges:2012us,
    author = "Berges, J. and Sexty, D.",
    title = "{Bose condensation far from equilibrium}",
    eprint = "1201.0687",
    archivePrefix = "arXiv",
    primaryClass = "hep-ph",
    doi = "10.1103/PhysRevLett.108.161601",
    journal = "Phys. Rev. Lett.",
    volume = "108",
    pages = "161601",
    year = "2012"
}

@article{Berges:2002cz,
    author = "Berges, Juergen and Serreau, Julien",
    title = "{Parametric resonance in quantum field theory}",
    eprint = "hep-ph/0208070",
    archivePrefix = "arXiv",
    reportNumber = "HD-THEP-02-25",
    doi = "10.1103/PhysRevLett.91.111601",
    journal = "Phys. Rev. Lett.",
    volume = "91",
    pages = "111601",
    year = "2003"
}

@article{Berges:2016nru,
    author = {Berges, J{\"u}rgen and Wallisch, Benjamin},
    title = "{Nonthermal Fixed Points in Quantum Field Theory Beyond the Weak-Coupling Limit}",
    eprint = "1607.02160",
    archivePrefix = "arXiv",
    primaryClass = "hep-ph",
    doi = "10.1103/PhysRevD.95.036016",
    journal = "Phys. Rev. D",
    volume = "95",
    pages = "036016",
    year = "2017"
}

@article{Schwinger:1960qe,
    author = "Schwinger, Julian S.",
    title = "{Brownian motion of a quantum oscillator}",
    doi = "10.1063/1.1703727",
    journal = "J. Math. Phys.",
    volume = "2",
    pages = "407--432",
    year = "1961"
}

@article{Keldysh:1964ud,
    author = "Keldysh, L. V.",
    title = "{Diagram Technique for Nonequilibrium Processes}",
    doi = "10.1142/9789811279461_0007",
    journal = "Sov. Phys. JETP",
    volume = "20",
    pages = "1018--1026",
    year = "1965"
}

@book{kadanoff1962quantum,
  title={Quantum Statistical Mechanics: Green's Function Methods in Equilibrium and Nonequilibrium Problems},
  author={Kadanoff, L.P. and Baym, G.},
  isbn={9780805352016},
  lccn={lc62015644},
  series={Frontiers in physics},
  url={https://books.google.com/books?id=1-FEAAAAIAAJ},
  year={1962},
  publisher={W.A. Benjamin}
}

@article{PhysRevD.37.2878,
  title = {Nonequilibrium quantum fields: Closed-time-path effective action, Wigner function, and Boltzmann equation},
  author = {Calzetta, E. and Hu, B. L.},
  journal = {Phys. Rev. D},
  volume = {37},
  issue = {10},
  pages = {2878--2900},
  numpages = {0},
  year = {1988},
  month = {May},
  publisher = {American Physical Society},
  doi = {10.1103/PhysRevD.37.2878},
  url = {https://link.aps.org/doi/10.1103/PhysRevD.37.2878}
  }

@article{10.1063/1.1704738,
    author = {Kane, Joseph W. and Kadanoff, Leo P.},
    title = {Green's Functions and Superfluid Hydrodynamics},
    journal = {Journal of Mathematical Physics},
    volume = {6},
    number = {12},
    pages = {1902-1912},
    year = {1965},
    month = {12},
    issn = {0022-2488},
    doi = {10.1063/1.1704738},
    url = {https://doi.org/10.1063/1.1704738},
}

@book{zubarev1997statistical2,
  title={Statistical Mechanics of Nonequilibrium Processes, Volume 2: Relaxation and Hydrodynamic Processes},
  author={Zubarev, Dmitrii and Morozov, Vladimir and R{\"o}pke, Gerd},
  year={1997},
  publisher={Wiley-VCH},
  isbn={9783527400843}
}

@article{imamovic2001quasiparticle,
  title={Quasiparticle kinetic equation in a trapped Bose gas at low temperatures},
  author={Imamovic-Tomasovic, M and Griffin, A},
  journal={Journal of low temperature physics},
  volume={122},
  number={5},
  pages={617--655},
  year={2001},
  publisher={Springer}
}

@article{Aarts:2003bk,
    author = "Aarts, Gert and Martinez Resco, Jose M.",
    title = "{Transport coefficients from the 2PI effective action}",
    eprint = "hep-ph/0303216",
    archivePrefix = "arXiv",
    doi = "10.1103/PhysRevD.68.085009",
    journal = "Phys. Rev. D",
    volume = "68",
    pages = "085009",
    year = "2003"
}

@article{Aarts:2004sd,
    author = "Aarts, Gert and Martinez Resco, Jose M.",
    title = "{Shear viscosity in the O(N) model}",
    eprint = "hep-ph/0402192",
    archivePrefix = "arXiv",
    doi = "10.1088/1126-6708/2004/02/061",
    journal = "JHEP",
    volume = "02",
    pages = "061",
    year = "2004"
}

@article{Aarts:2005vc,
    author = "Aarts, Gert and Martinez Resco, Jose M.",
    title = "{Transport coefficients in large N(f) gauge theories with massive fermions}",
    eprint = "hep-ph/0503161",
    archivePrefix = "arXiv",
    reportNumber = "SWAT-05-427",
    doi = "10.1088/1126-6708/2005/03/074",
    journal = "JHEP",
    volume = "03",
    pages = "074",
    year = "2005"
}

@article{Mikheev:2018adp,
    author = "Mikheev, Aleksandr N. and Schmied, Christian-Marcel and Gasenzer, Thomas",
    title = "{Low-energy effective theory of nonthermal fixed points in a multicomponent Bose gas}",
    eprint = "1807.10228",
    archivePrefix = "arXiv",
    primaryClass = "cond-mat.quant-gas",
    doi = "10.1103/PhysRevA.99.063622",
    journal = "Phys. Rev. A",
    volume = "99",
    number = "6",
    pages = "063622",
    year = "2019"
}

\end{document}